\documentclass[fleqn, usegraphicx, useAMS, usenatbib]{mn2e}
\usepackage{amsmath, amssymb, aas_macros, subfigure}
\begin{document}

\title[Magnetic mountains: equation of state]{Quadrupole moment of a magnetically confined mountain on an accreting
neutron star: effect of the equation of state}
\author[M. Priymak et al.]
{M.~Priymak,$^1$\thanks{E-mail: m.priymak@pgrad.unimelb.edu.au} A.~Melatos,$^1$
\& D.~J.~B.~Payne$^1$ \\ $^1$School of Physics, University of Melbourne,
Parkville, VIC 3010, Australia}

\maketitle
\begin{abstract}
Magnetically confined mountains on accreting neutron stars are promising sources
of continuous-wave gravitational radiation and are currently the targets of
directed searches with long-baseline detectors like the Laser Interferometer
Gravitational Wave Observatory (LIGO). In this paper, previous
ideal-magnetohydrodynamic models of isothermal mountains are
generalized to a range of physically motivated, adiabatic equations of state. It
is found that the mass ellipticity $\epsilon$ drops substantially, from
$\epsilon \approx 3\times10^{-4}$ (isothermal) to $\epsilon \approx 9\times
10^{-7}$ (non-relativistic degenerate neutrons), $6\times10^{-8}$ (relativistic
degenerate electrons) and $1\times10^{-8}$ (non-relativistic degenerate
electrons) (assuming a magnetic field of $10^{12.5} \ \mathrm{G}$ at birth). The
characteristic mass $M_{\mathrm{c}}$ at which the magnetic dipole moment halves from its
initial value is also modified, from $M_{\mathrm{c}}/\mathrm{M}_{\sun} \approx 5\times10^{-4}$
(isothermal) to $M_{\mathrm{c}}/\mathrm{M}_{\sun} \approx 2\times10^{-6}$, $1\times 10^{-7}$, and
$3\times10^{-8}$ for the above three equations of state, respectively. Similar
results are obtained for a realistic, piecewise-polytropic nuclear equation of
state. The adiabatic models are consistent with current LIGO upper limits,
unlike the isothermal models. Updated estimates of gravitational-wave
detectability are made. Monte Carlo simulations of the spin distribution of
accreting millisecond pulsars including gravitational-wave stalling agree better
with observations for certain adiabatic equations of state, implying that X-ray
spin measurements can probe the equation of state when coupled with magnetic
mountain models.
\end{abstract}


\section{Introduction}
\label{section_1}

Neutron star spins in low-mass X-ray binaries (LMXBs), measured from X-ray
pulsations or thermonuclear burst oscillations, are found to lie in the range
$95 - 619 \ \mathrm{Hz}$ \citep{chakrabarty2008, galloway2008a, watts2008, galloway2010}. The
upper end of this range falls well short of the centrifugal breakup frequency
for most equations of state \citep{cook1994, haensel1999, chakrabarty2008}, even
though the objects accrete enough angular momentum during their X-ray lifetime
of $10^{7} - 10^{9}$ years \citep{podsiadlowski2002} to spin up to $1.5 \
\mathrm{kHz} \lesssim \nu_{\mathrm{s}} \lesssim 3 \ \mathrm{kHz}$ \citep{bildsten1998,
chakrabarty2003}. This discrepancy cannot be attributed to an observational
selection effect, because the Rossi X-ray Timing Explorer (\textit{RXTE}) remains
sensitive up to $2 \ \mathrm{kHz}$ \citep{chakrabarty2008, galloway2008a}. To
describe the apparent spin clustering and cut-off, \citet{bildsten1998} invoked
gravitational radiation torques to stall the spin-up process; see also
\citet{papaloizou1978} and \citet{wagoner1984}. To achieve this, a mass
quadrupole moment of order $\sim 2 \times 10^{38} \ \mathrm{g} \
\mathrm{cm}^{2}$ is required.

Quadrupoles on accreting neutron stars are of two kinds: (i) core deformations,
e.g. from r-modes \citep{brink2004, nayyar2006, bondarescu2007} and (ii)
permanent crustal deformations, e.g. supported by thermal \citep{bildsten1998, ushomirsky2000}
or magnetic \citep{brown1998, melatos2001, choudhuri2002, payne2004,
vigelius2008a} gradients. In the absence of a magnetic field, the maximum
crustal quadrupole depends on the breaking strain \citep{ushomirsky2000,
haskell2006} and can be as large as $\sim 10^{40} \ \mathrm{g} \
\mathrm{cm}^{2}$ in the light of recent molecular dynamics simulations
\citep{horowitz2009a}. When magnetic stresses are included, the quadrupole
increases, as matter is funnelled to the magnetic poles of the star and
compresses the magnetic field laterally \citep{hameury1983, melatos2001,
choudhuri2002, payne2004, vigelius2008a}.

\citet{payne2004}, hereafter PM04, calculated self-consistent, axisymmetric,
ideal-magnetohydrodynamic (ideal-MHD) equilibria of \textit{isothermal} magnetic
mountains as a function of accreted mass $M_{\mathrm{a}}$. They found that the magnetic
field distorts appreciably for $M_{\mathrm{a}} \geq M_{\mathrm{c}} \sim 10^{-5} \mathrm{M}_{\sun}$, in
accord with the phenomenological field decay relation of \citet{shibazaki1989}
and well above previous calculations, which predicted $M_{\mathrm{c}} \sim
10^{-10} \mathrm{M}_{\sun}$ without including the back reaction from the compressed
equatorial magnetic field \citep{hameury1983, brown1998, litwin2001}.
\citet{payne2007} showed that the mountain oscillates stably in a superposition
of Alfv\'en and acoustic modes when perturbed, following a transient adjustment
via the undular submode of the magnetic buoyancy instability
\citep{mouschovias1974, hughes1987, vigelius2008a}. \citet{vigelius2008a} found
that the equilibrium state remains mountain-like after this transient
instability, with the mass quadrupole moment decreasing by $\approx 30$ per cent.
Ohmic dissipation contributes to the decay of the mass quadrupole by allowing
slippage of accreted matter across magnetic field lines, with a resistive
relaxation timescale of $10^{5} - 10^{8} \ \mathrm{yr}$ depending on the
conductivity \citep{vigelius2009a}. \citet{wette2010} examined the subsidence
of mountains into a fluid crust, generalizing earlier calculations on a rigid
surface, and found that the quadrupole shrinks by up to $\approx 60$ per cent.

The existing literature on magnetic mountains, summarized above, suffers from
several limitations. First, the time-dependent feedback between the
magnetosphere and the accretion disc is neglected \citep{romanova2003,
romanova2004, kulkarni2008, long2008}. Secondly, the mountain should solidify into
a body-centred-cubic crystal as it sinks, when the ionic coupling parameter
exceeds the crystallization threshold \citep{farouki1993, horowitz2009b}. This
occurs at different depths, depending on the local composition, density and
temperature \citep{brown2000}. The sudden transition to a solid affects the
magnetic line-tying boundary condition, which now depends on the local magnetic
stresses and critical strain. Thirdly, a nuclear reaction network that follows
accreted matter elements as they descend has not yet been implemented
\citep{haensel1990b, haensel1990a, haensel2003, chamel2008}. Deep crustal
heating deposits $1.5 - 1.9 \ \mathrm{MeV}$ per accreted baryon
\citep{haensel2008}, reduces the Ohmic decay time-scale, and introduces thermal
and electrical conductivity gradients due to compositional variations
\citep{chamel2008}, all of which affect the mountain structure. Finally, the
equation of state (EOS) of the accreted matter needs to be modelled
realistically. The calculations cited in the previous paragraph all utilize an
isothermal EOS, an accurate model for very low mass mountains with maximum
density $\rho_{\mathrm{max}} \leq 10^{6} \ \textrm{g} \ \textrm{cm}^{-3}$
\citep{shapiro1983}. The isothermal EOS is too soft and does not accurately
represent all pressure components (e.g. degenerate neutron and electron
pressures in the inner crust) for realistically sized mountains with
$\rho_{\mathrm{max}} \lesssim 10^{14} \ \mathrm{g} \ \mathrm{cm}^{-3}$ or
equivalently $M_{\mathrm{a}} \lesssim 10^{-2} \mathrm{M}_{\sun}$ \citep{shapiro1983, brown2000,
chamel2008}.

This work aims to quantify how the EOS influences the structure of the magnetic
mountain and its mass quadrupole moment. It turns out that the effect is large.
In Section \ref{section_2}, we generalize the Grad--Shafranov framework for
solving numerically the MHD equilibrium problem to incorporate an adiabatic EOS.
The numerical algorithm is validated against published isothermal results in
Section \ref{section_3}. We directly compare the structure of adiabatic and
isothermal magnetic mountains in Section \ref{section_4}, quantifying the
relation between the accreted mass and measurable quantities such as dipole
moment and ellipticity. In Section \ref{section_5}, we approximate the realistic
EOS in the neutron star crust by an effective polytrope and calculate the
structure of the associated mountain. In Section \ref{section_6}, we examine the
implications of the theoretical models for gravitational-wave (GW) stalling of LMXB
spins. The detectability of magnetic mountains as GW sources is
assessed briefly in Section \ref{section_7}, revising the latest estimates in
\citet{vigelius2009b}.

\section{Hydromagnetic Equilibrium}
\label{section_2}

To compute the structure of a magnetic mountain with an adiabatic EOS, we
generalize the isothermal Grad--Shafranov solver described in PM04 to handle a
general, barotropic, pressure-density relation of the form $P(\rho) = K \rho^{1
+ 1/n} = K \rho^{\Gamma}$, where $n$ is the polytropic index and $\Gamma$ is the
adiabatic index \citep{paczynski1983, shapiro1983}.

\subsection{Grad--Shafranov equation}
\label{section_2:GS}

Let us define a spherical coordinate system $(r, \theta, \phi)$, where $\theta =
0$ is the magnetic symmetry axis before accretion begins and the neutron star
surface is situated at $r = R_{\mathrm{in}}$ (i.e. the inner boundary of the
simulation; see Appendix \ref{appendix:pseudocode}). Time-dependent ideal MHD
and resistive simulations of magnetic mountains in ZEUS-MP show that the
magnetic field relaxes to an almost axisymmetric configuration (deviation from
axisymmetry $\lesssim 1$ per cent) within a few Alfv\'en times, following a transient,
Parker-type instability \citep{vigelius2008a}. Hence, to a good approximation,
the magnetic field is given everywhere by
\begin{equation}
\bmath{B} = \frac{\nabla \psi}{r \sin\theta} \times \hat{e}_{\phi},
\label{b_axisymmetric}
\end{equation}
where $\psi(r, \theta)$ is a flux function. In the steady state, the MHD
equations reduce to 
\begin{equation}
\nabla P + \rho \nabla \phi + (\Delta^{2}\psi)\nabla \psi = 0,
\label{mhs_alternative}
\end{equation}
where $\Delta^{2}$ denotes the Grad--Shafranov operator,
\begin{equation}
\Delta^{2} = \frac{1}{4 \pi r^{2} \sin^{2}\theta} \Bigg[
\frac{\partial^{2}}{\partial r^{2}} + \frac{\sin \theta}{r^{2}}
\frac{\partial}{\partial \theta} \Bigg( \frac{1}{\sin \theta}
\frac{\partial}{\partial \theta} \Bigg) \Bigg].
\label{gs_operator}
\end{equation}

We solve the projection of equation (\ref{mhs_alternative}) along $\bmath{B}$
by the method of characteristics. The result depends critically on the EOS.
Under isothermal conditions, i.e. $P = c_{\mathrm{s}}^{2}\rho$, we find
\begin{equation}
\Delta^{2} \psi = - \frac{\mathrm{d}F(\psi)}{\mathrm{d}\psi} \exp[-(\phi - \phi_{0})/c_{\mathrm{s}}^{2}],
\label{gs_isothermal}
\end{equation}
where $\phi_{0}$ denotes the reference gravitational potential at the neutron
star surface, and $c_{\mathrm{s}}^{2}$ is the isothermal sound speed \citep{payne2004}.
Under adiabatic conditions, i.e. $P = K \rho^{\Gamma}$, we find 
\begin{equation}
\Delta^{2} \psi = - \frac{\mathrm{d}F(\psi)}{\mathrm{d}\psi} \Bigg\{1 - \frac{(\Gamma - 1) (\phi -
\phi_{0})}{\Gamma K^{1/\Gamma}
[F(\psi)]^{(\Gamma-1)/\Gamma}}\Bigg\}^{1/(\Gamma-1)}.
\label{gs_adiabatic}
\end{equation}
The pressure along a flux surface $\psi$ under isothermal and adiabatic
conditions is given by
\begin{equation}
P = F(\psi) \exp[-(\phi - \phi_{0})/c_{\mathrm{s}}^{2}],
\label{pressure_isothermal}
\end{equation}
and
\begin{equation}
P = F(\psi)\Bigg\{1 - \frac{(\Gamma - 1) (\phi - \phi_{0})}{\Gamma K^{1/\Gamma}
[F(\psi)]^{(\Gamma-1)/\Gamma}}\Bigg\}^{\Gamma/(\Gamma-1)},
\label{pressure_adiabatic}
\end{equation}
respectively. Formally speaking, $F(\psi)$ is an arbitrary function of the
magnetic flux in equations (\ref{gs_isothermal})--(\ref{pressure_adiabatic}).
Equation (\ref{pressure_isothermal}) is the usual barometric formula; the base
pressure $F(\psi)$ varies from field line to field line, and $P$ decreases with
arc length along any particular field line because $|\phi|$ is inversely
proportional to $r$. Equation (\ref{pressure_adiabatic}) behaves similarly, but
its form is not barometric, in the sense that $F(\psi)$ does not factorize out.

In order to establish a one-to-one mapping between the initial (pre-accretion)
and final (post-accretion) states that preserves the flux freezing encoded in
the mass-continuity and magnetic-induction equations of ideal MHD, we require that the final, steady-state, mass-flux distribution $\mathrm{d}M/\mathrm{d}\psi$, defined as the mass enclosed by the infinitesimally separated flux surfaces $\psi$ and $\psi +
\mathrm{d}\psi$, equals that of the initial state plus the accreted mass. This approach
uniquely determines $F(\psi)$ through
\begin{equation}
F(\psi) = \frac{c_{\mathrm{s}}^{2}}{2 \pi} \frac{\mathrm{d} M}{\mathrm{d} \psi} \Bigg[ \int_{C} \mathrm{d}s \ r 
\sin \theta | \nabla \psi |^{-1} \mathrm{e}^{-(\phi - \phi_{0})/c_{\mathrm{s}}^{2}} \Bigg]^{-1},
\label{F_isothermal}
\end{equation}
for the isothermal EOS and
\begin{align}
\label{F_adiabatic}
F(\psi) = & \frac{K}{(2 \pi)^{\Gamma}} \Bigg(\frac{\mathrm{d}M}{\mathrm{d} \psi}\Bigg)^{\Gamma} \\
 & \times \Bigg[ \int_{C} \mathrm{d}s \ r \sin \theta |\nabla \psi|^{-1} \nonumber \\ 
 & \quad \Bigg\{ 1 - \frac{(\Gamma - 1) (\phi - \phi_{0})}{\Gamma K^{1/\Gamma} [F(\psi)]^{(\Gamma-1)/\Gamma}} \Bigg\}^{1/(\Gamma-1)} \Bigg]^{-\Gamma} \nonumber,
\end{align}
for the adiabatic EOS. This approach is self-consistent and therefore preferable
to guessing $F(\psi)$ \citep{hameury1983, brown1998, melatos2001}, but it
renders the solution more difficult. [\citet{duez2010} also solved
self-consistently for $F(\psi)$ by minimizing the total energy while conserving
invariants like the helicity and mass-flux ratio.] The integrals in equations
(\ref{F_isothermal}) and (\ref{F_adiabatic}) are performed along the magnetic
field line $\psi = \mathrm{constant}$. In accordance with earlier work, we
prescribe the mass-flux distribution in one hemisphere to be 
\begin{equation}
M(\psi) = \frac{M_{\mathrm{a}}[1 - \exp(-\psi/\psi_{\mathrm{a}})]}{2[1 - \exp(-b)]},
\label{mass-flux} 
\end{equation}
where $M_{\mathrm{a}}$ is the accreted mass, $\psi_{\ast}$ labels the flux surface
emerging from the magnetic equator, $\psi_{\mathrm{a}}$ labels the field line that closes
just inside the inner edge of the accretion disc, and we write
$b=\psi_{\ast}/\psi_{\mathrm{a}}$. Equation (\ref{mass-flux}) ensures that $\approx 63$ per cent of the accreted mass accumulates within the polar cap $0 \leq \psi \leq
\psi_{\mathrm{a}}$ for $\psi_{\mathrm{a}} \ll \psi_{\ast}$.

The gravitational acceleration is assumed to be constant in this paper, with a
gravitational potential of the form $\phi(r) = G M_{\ast}r/R_{\mathrm{in}}^{2}$. This
assumption is justified, because the mountain never rises more than $\sim 10^{4}
\ \mathrm{cm}$ above the hard surface at $r = R_{\mathrm{in}}$ (see Section
\ref{section_4:hydromagnetic_structure}). A simple numerical check shows that
the altitude above $r = R_{\mathrm{in}}$ where the density distribution falls to zero
changes by $\approx 2$ per cent when $\phi(r) = G M_{\ast}r/R_{\mathrm{in}}^{2}$ is replaced by
$\phi(r) = -G M_{\ast}/r$. Self-gravity is also ignored, although the correction
$M_{\mathrm{a}}/M_{\ast}$ to the gravitational potential is significant in LMXBs with
$M_{\mathrm{a}} \gtrsim 10^{-1} \mathrm{M}_{\sun}$.

We conduct our numerical simulations as follows: a fixed dipolar magnetic field at the inner radial boundary of the numerical mesh is assumed, and a prescribed amount of accreted
matter $M_{\mathrm{a}}$ (described by one of the EOS in Table \ref{table:eos}) is added into the simulation volume according to the mass-flux relation (\ref{mass-flux}). We then allow the system to relax quasi-statically
to hydromagnetic equilibrium by solving equation (\ref{gs_isothermal}) or (\ref{gs_adiabatic}) simultaneously with equation (\ref{F_isothermal}) or (\ref{F_adiabatic}) for $\psi(r, \theta)$,
using an iterative under-relaxation algorithm combined with a finite-difference
Poisson solver. The details can be found in Appendix \ref{appendix:pseudocode}. We adopt the following boundary conditions, as in previous papers (e.g. PM04):
$\psi(R_{\mathrm{in}}, \theta) = \psi_{\ast} \sin^{2} \theta$ (surface dipole; magnetic
line tying), $\mathrm{d}\psi/\mathrm{d}r(R_{\mathrm{m}}, \theta) = 0$ (outflow), $\psi(r, 0) = 0$
(straight polar field line) and $\mathrm{d}\psi/\mathrm{d}\theta(r, \pi/2) = 0$ (north--south
symmetry), where $R_{\mathrm{in}} \leq r \leq R_{\mathrm{m}}$ and $0 \leq \theta \leq \pi/2$
delimit the computational volume. The outer radius $R_{\mathrm{m}}$ is chosen
large enough to encompass most of the screening currents (isothermal EOS) or the
outer edge of the accreted matter (adiabatic EOS).

\subsection{Inner boundary}

The nature of the rigid inner boundary at $R_{\mathrm{in}}$ deserves special mention. It is not the stellar surface; it is not meaningful to build a mountain $100 \ \mathrm{m}$ high and reaching neutron drip density at its base on top of a low-density ocean, using a realistic EOS. Instead, the outer layers of the neutron star are `constructed' from the accreted material of mass $M_{\mathrm{a}}$.
Thus, $R_{\mathrm{in}}$ does not correspond to the neutron star surface $R_{\ast}$, but to the depth in the neutron star crust above which lies the mass $M_{\mathrm{a}}$ (for a given EOS). 
Since $M_{\ast}$ and $R_{\mathrm{in}}$ are fixed, the total mass and radius vary slightly (few per cent) between models with different $M_{\mathrm{a}}$ but the 
same EOS (Table \ref{table:eos}). The inner boundary of our simulation volume $R_{\mathrm{in}}$ represents a solid surface at the corresponding base density. This simplification assumes that movement of matter below this depth is approximately radial due to compression and that the solid-surface prescription is valid. 
In reality, accreting matter is expected to displace both radially and laterally \citep{choudhuri2002}. The lateral flow would alter our computed results by decreasing the mass quadrupole moment slightly and increasing the magnetic dipole moment. We can eliminate this approximation by injecting the accreted matter according to the approach advocated by \citet{wette2010}, generalizing the latter paper to a realistic EOS. 
Such a procedure is feasible but technically difficult; we defer it to future work.

Referring to fig. 12 of \citet{wette2010}, the ellipticity of an isothermal mountain in the fluid-surface model appears to converge to the saturation ellipticity of the hard-surface model as $M_{\mathrm{a}}$ increases; 
the difference in ellipticities relative to the hard-surface model decreases from $\sim 60$ to $\sim 25$ per cent as $M_{\mathrm{a}}$ increases from $\sim 10^{-3}$ to $\sim 10^{-1} \mathrm{M}_{\sun}$. 
We expect similar convergent behaviour for adiabatic mountains at significantly lower $M_{\mathrm{a}}$, since saturation ellipticities of adiabatic mountains are attained at accreted masses $2$--$4$ orders of magnitude below that of the isothermal one (see Fig. \ref{fig:ellipticity_mass}).
Realistic accreted masses in LMXB systems of $M_{\mathrm{a}} \sim 10^{-1}$ $\mathrm{M}_{\sun}$ are $2$--$6$ orders of magnitude greater (depending on the EOS) than the accreted masses which we can reliably simulate. At realistic accreted masses, we expect the saturation ellipticities of mountains with and without sinking to approximately converge.
The population-synthesis results in Section \ref{section_6} and GW-detectability estimates in Section \ref{section_7} depend solely on the saturation ellipticity.

\subsection{Adiabatic index}
\label{section_2:adiabatic_index}

The realistic EOS of a neutron star crust is piecewise adiabatic, as discussed
in Section \ref{section_5}. However, before modelling the realistic EOS, we
conduct numerical experiments in Sections \ref{section_3} and \ref{section_4} to
see how the mountain structure depends on the adiabatic index $\Gamma$. In these
numerical experiments, we employ a purely adiabatic EOS with unique $K$ and
$\Gamma$. The values of $K$ and $\Gamma$ are chosen to correspond to density
regimes of interest in the crust, e.g. degenerate non-relativistic electron gas
$10^{5} \lesssim \rho/(\mathrm{g} \ \mathrm{cm}^{-3}) \lesssim 10^{7}$,
degenerate relativistic electron gas $10^{7} \lesssim \rho/(\mathrm{g} \
\mathrm{cm}^{-3}) \lesssim 10^{12}$ and degenerate neutron gas $10^{12}
\lesssim \rho/(\mathrm{g} \ \mathrm{cm}^{-3}) \lesssim 10^{16}$. In an ideal
electron gas, which is approximately isothermal, radiation and lattice pressures
dominate, but this occurs at much lower densities $\rho \lesssim 10^{4} \
\mathrm{g} \ \mathrm{cm}^{-3}$, which are irrelevant to the mountain problem.

Table \ref{table:eos} displays the magnetic mountain models we compute here,
with the details of their respective EOS. $K$ is a function of mean molecular
weight per electron, $\mu_{e} = m_{b}/(m_{u}Y_{e})$, according to the scaling $K
\propto \mu_{e}^{-4/3}$, where $m_{b}$ is the mean baryon rest mass, $m_{u}$ is
the atomic mass unit, and $Y_{e}$ is the mean number of electrons per baryon.
Under the assumption of symmetric nuclear matter, we take $\mu_{e} = 2$ and
$m_{b} = m_{u}$ and hence $K$ is a constant (i.e. independent of $\rho$). This
form of the EOS describes well a completely degenerate, ideal Fermi gas
\citep{shapiro1983}. Hence we use it to model degenerate relativistic electrons
($n=3, \ \Gamma=4/3, \ K=4.93\times10^{14} \ \mathrm{dyn} \ \mathrm{g}^{-4/3} \
\mathrm{cm}^{2}$), degenerate non-relativistic electrons ($n=3/2, \ \Gamma=5/3,
\ K=3.16\times10^{12} \ \mathrm{dyn} \ \mathrm{g}^{-5/3} \ \mathrm{cm}^{3}$)
and degenerate non-relativistic neutrons ($n=3/2, \ \Gamma=5/3, \
K=5.38\times10^{9} \ \mathrm{dyn} \ \mathrm{g}^{-5/3} \ \mathrm{cm}^{3}$).

\begin{table*}
\begin{minipage}{93mm}
\begin{tabular}{@{}cccc}
\hline
Model & $K$ (cgs) & $\Gamma$ & Equation of State \\
\hline
A & $1.0\times10^{8}$ & 1 & Isothermal \\
B & $3.2\times10^{12}$ & 5/3 & Non-relativistic degenerate electrons \\
C & $4.9\times10^{14}$ & 4/3 & Relativistic degenerate electrons \\
D & $5.4\times10^{9}$ & 5/3 & Non-relativistic degenerate neutrons \\
E & Variable & Variable & Piecewise polytropic \\
\hline
\end{tabular}
\caption{Numerical models of magnetic mountains with their associated EOS. In
models A--D, the EOS is polytropic, with $P(\rho) = K\rho^{\Gamma}$, where $K$
is measured in cgs units ($\mathrm{dyn} \ \mathrm{g}^{-\Gamma} \
\mathrm{cm}^{3\Gamma-2}$) \citep{shapiro1983}. In models A--D, $K$ and $\Gamma$
are held constant as $M_{\mathrm{a}}$ varies. In model E, $K$ and $\Gamma$ assume average
values, which depend on $M_{\mathrm{a}}$ (see Section \ref{section_5}).}
\label{table:eos}
\end{minipage}
\end{table*}

\section{Validation in the isothermal limit}
\label{section_3}

We assume the following neutron star parameters throughout this paper, except
where stipulated otherwise: $M_{\ast} = 1.4 \mathrm{M}_{\sun}, \ R_{\mathrm{in}} = 10^{6} \
\mathrm{cm}$, and $\psi_{\ast} = 1.6 \times 10^{24} \ \mathrm{G} \
\mathrm{cm}^{2}$ (with $\psi_{\ast} = B_{\ast} R_{\mathrm{in}}/2$, where $B_{\ast}$ is
the polar magnetic field strength before accretion begins). The fiducial value
of the magnetic field, $B_{\ast} = 10^{12.5} \ \mathrm{G}$, is chosen in accord
with population synthesis models, which predict natal magnetic fields of
$10^{12}-10^{13} \ \mathrm{G}$ \citep{hartman1997, arzoumanian2002,
faucher-giguere2006}.

The adiabatic Grad--Shafranov formalism in Section \ref{section_2}, and the
numerical solver described in Appendix \ref{appendix:pseudocode}, must reproduce
the results of PM04 in the isothermal limit (i.e. $n \to \infty, \ \Gamma \to 1,
\ K \to c_{\mathrm{s}}^{2}$). In this limit, the isodensity contours and magnetic field
lines of an adiabatic mountain with $\Gamma \to 1$ must converge to those
plotted in figs 4, 5 and 9 in PM04 for identical accreted masses. As there is
no unique way to continuously transform an adiabatic EOS into an isothermal EOS,
we test the adiabatic/isothermal correspondence by taking the limit $(K, \Gamma)
\to (c_{\mathrm{s}}^{2}, 1)$ in three different ways below.

\begin{enumerate}
\item\label{model_1} We set $K = c_{\mathrm{s}}^{2}$ and let $\Gamma$ tend to unity, such
that
\begin{equation}
P(\rho) = c_{\mathrm{s}}^{2} \rho^{\Gamma}.
\label{cK}
\end{equation}

\item Exploiting the tendency for the surface pressure
$P_{\mathrm{surf}}(\theta)$ and density $\rho_{\mathrm{surf}}(\theta)$ to be
roughly EOS-independent for $\Gamma \approx 1$, we write
$P_{\mathrm{surf}} \approx c_{\mathrm{s}}^{2} \rho_{\mathrm{surf}} \approx K
\rho_{\mathrm{surf}}^{\Gamma}$ to eliminate $K$, and hence obtain
\begin{equation}
P(\rho) = c_{\mathrm{s}}^{2 \Gamma} \ P_{\mathrm{surf}}^{1 - \Gamma} \ \rho^{\Gamma},
\label{cRho}
\end{equation}
where $P_{\mathrm{surf}}$ is a function of $M_{\mathrm{a}}$.

\item We take $K \propto \Gamma$ and interpolate between a selected polytrope
$(K_{0}, \Gamma_{0})$ and the isothermal target according to
\begin{equation}
P(\rho) = \frac{(\Gamma - 1)K_{0} + (\Gamma_{0} - \Gamma)c_{\mathrm{s}}^{2}}{\Gamma_{0} -
1} \rho^{\Gamma},
\label{fK}
\end{equation}
with $\Gamma \to 1$.
\end{enumerate}

\begin{figure}
\begin{minipage}{0.5\linewidth}
\includegraphics[width=84mm]{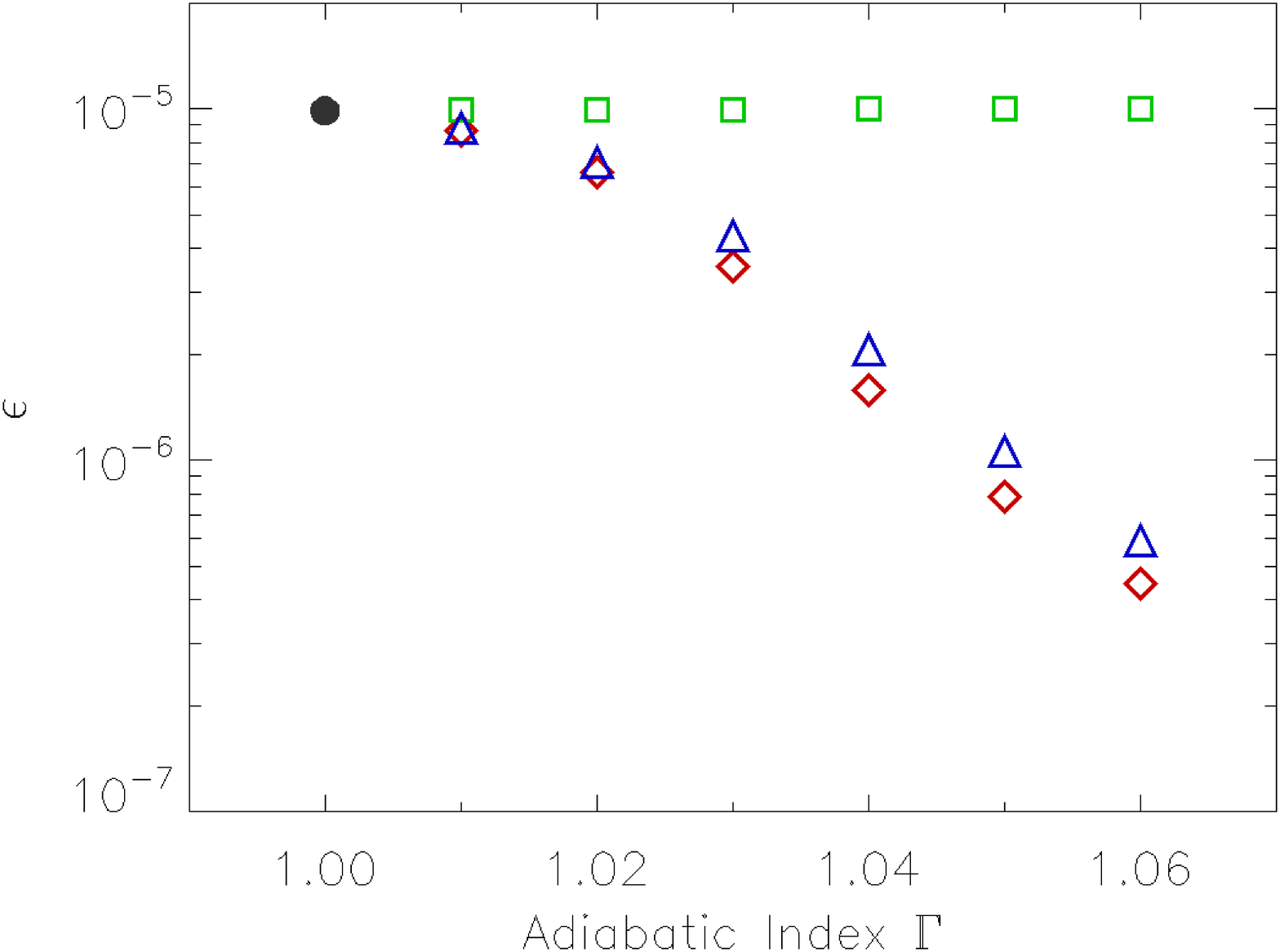}
\end{minipage}
\begin{minipage}{0.5\linewidth}
\includegraphics[width=84mm]{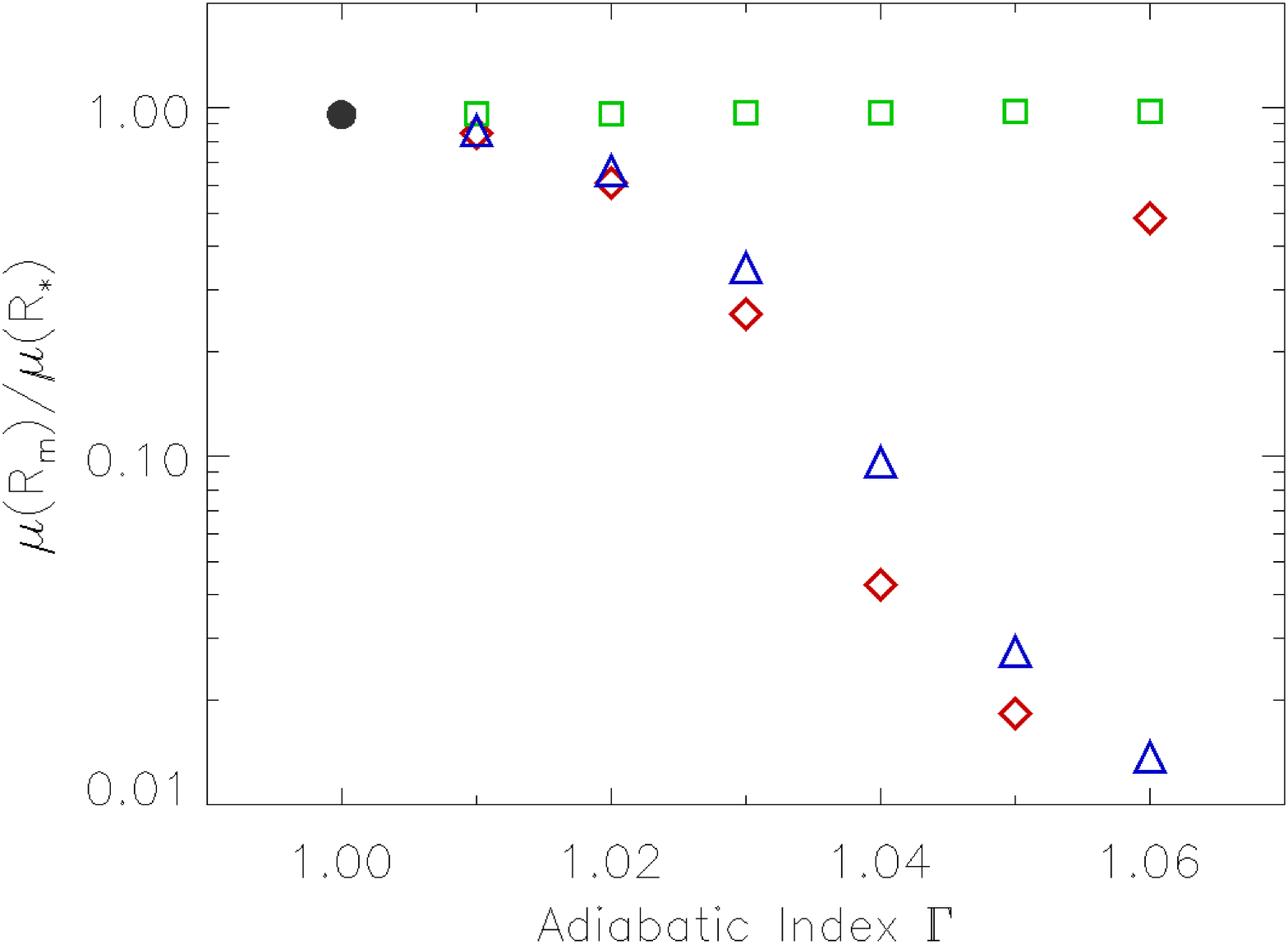}
\end{minipage}
\begin{minipage}{0.5\linewidth}
\includegraphics[width=84mm]{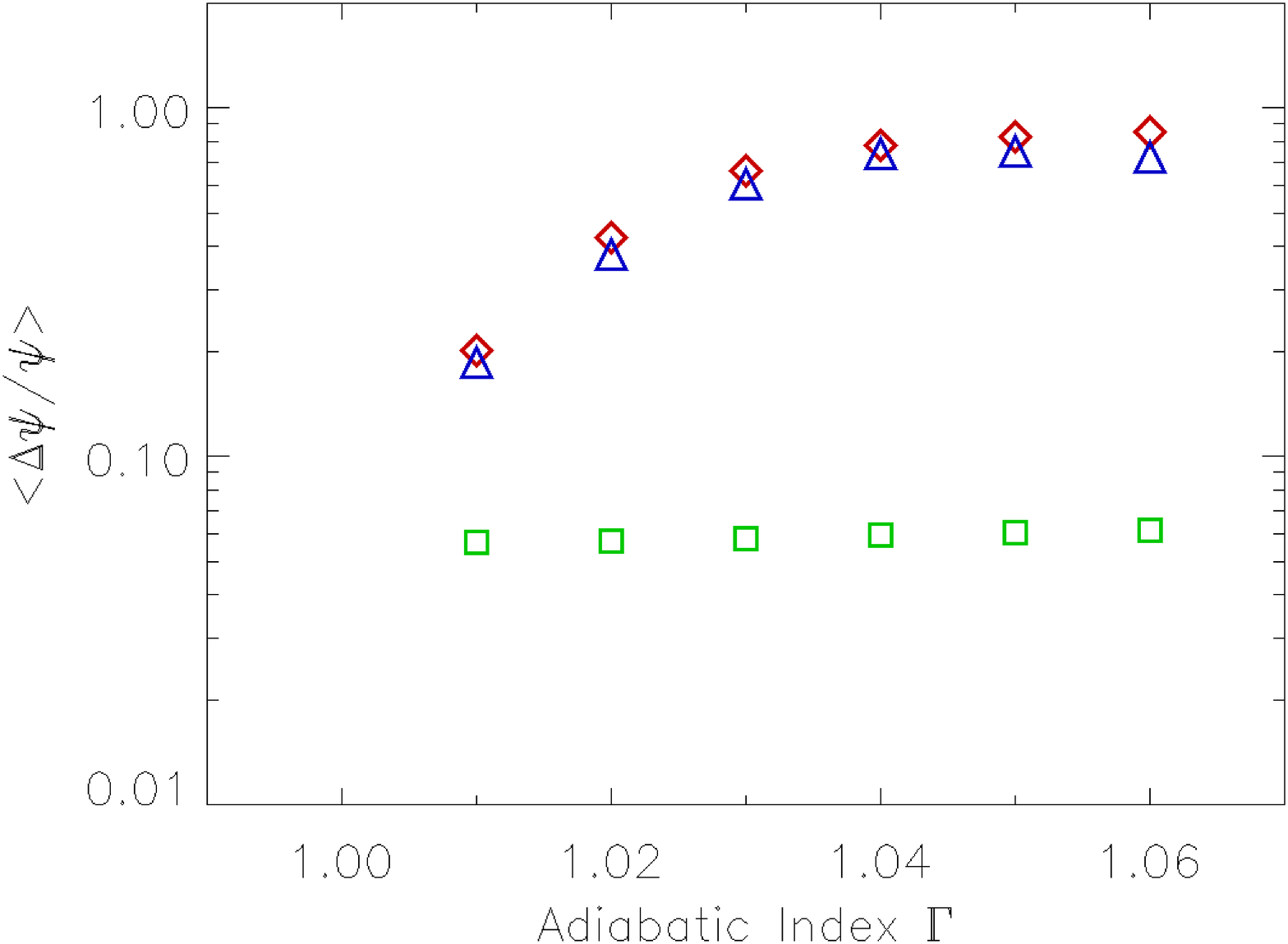}
\end{minipage}
\caption{Convergence between the isothermal ($\Gamma = 1$) and adiabatic
($\Gamma \to 1$) magnetic mountain models at $M_{\mathrm{a}} = 10^{-5} \mathrm{M}_{\sun}$ for the
mass ellipticity $\epsilon$ (top panel), magnetic dipole moment $\mu$ (middle panel) and grid-averaged residuals (bottom panel) as a function of
adiabatic index $\Gamma$. The data points represent magnetic mountains modelled
with the EOS in equations (\ref{cK}) (red diamonds), (\ref{cRho}) (green
rectangles) and (\ref{fK}) (blue triangles). Isothermal results are denoted by filled black circles in the top two panels.}
\label{fig:convergence}
\end{figure}

We apply these three approaches to the case $M_{\mathrm{a}} = 1.0 \times 10^{-5}
\mathrm{M}_{\sun}$, starting from a relativistic degenerate electron EOS (model C in
Table \ref{table:eos}). This EOS prevails over a large logarithmic range of
densities in a realistic stellar crust [$10^{7} \lesssim \rho/(\mathrm{g} \
\mathrm{cm}^{-3}) \lesssim 10^{12}$; see Section \ref{section_5}] and gives way
to an isothermal EOS in the upper atmosphere ($\rho < 10^{4} \ \mathrm{g} \
\mathrm{cm}^{-3}$). We find that all three approaches converge correctly to the
$\Gamma = 1$ results of PM04 after $\sim 3\times10^{3}$ iterations. Fig.
\ref{fig:convergence} displays the mass ellipticity, magnetic dipole moment, and
grid-averaged $\psi$ residual (relative to the $\Gamma = 1$ result) as a function of $\Gamma$ 
for approaches (i) (red diamonds), (ii) (green rectangles) and (iii) (blue triangles). 
As indicated by Fig. \ref{fig:convergence}, the rate of convergence towards the isothermal 
results differs between models. The abnormally high dipole moment for case (i) at $\Gamma = 1.06$ in Fig.
\ref{fig:convergence} is caused by insufficient resolution in $\theta$ and
can be prevented by scaling the grid logarithmically in $\theta$ to handle the
steep magnetic field gradients at the equator. We defer this project to future
work.

We compute the mass enclosed within the computational grid as a function of
iteration number, to track the mass lost through the outer boundary. In every
converged equilibrium, the total mass in the final state is always within $4$ per cent
(and typically within $1$ per cent) of the initial mass. The iterative solver also
preserves the divergence-free nature of the magnetic field, with $|\nabla \cdot
\bmath{B}| = 0$ to machine precision everywhere on the grid.

\section{Adiabatic mountains}
\label{section_4}

In this section, we compute Grad--Shafranov equilibria for several adiabatic EOS
using the method described in Section \ref{section_2} and validated in Section
\ref{section_3}. Table \ref{table:eos} lists the parameters of each EOS,
corresponding to different depth intervals within the stellar crust (see Section
\ref{section_5}). The scalings of the magnetic dipole moment $\mu$ and mass
ellipticity $\epsilon$ with accreted mass $M_{\mathrm{a}}$ are studied in Sections
\ref{section_4:dipole_moment} and \ref{section_4:ellipticity}, respectively. The
maximum density and local magnetic field strength are computed in Sections
\ref{section_4:maximum_magnetic_field} and \ref{section_4:maximum_density}, 
respectively. In Section \ref{section_4:hydromagnetic_structure}, we compare the
equilibrium density and magnetic field distributions for adiabatic and
isothermal magnetic mountains. For each model in Table \ref{table:eos}, we stop 
our simulations once $|\Delta \psi/\psi|$ is less than $5$ per cent averaged over the
grid (see Appendix \ref{appendix:pseudocode}).

\subsection{Magnetic burial: $\mu$ versus $M_{\mathrm{a}}$}
\label{section_4:dipole_moment}

\begin{figure}
\includegraphics[width=84mm]{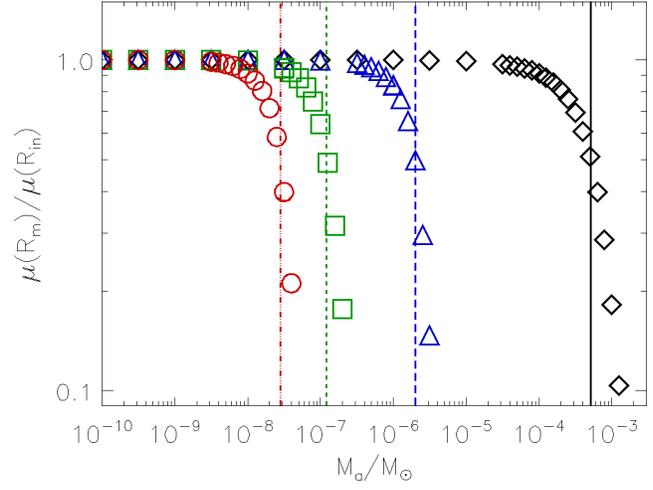}
\caption{Magnetic dipole moment $\mu$, computed at the outer edge of the grid and normalized to its surface value, as a function of accreted mass,
$M_{\mathrm{a}}$, measured in solar masses, for models A (black diamonds), B (red
circles), C (green squares) and D (blue triangles). Values of the
characteristic masses $M_{\mathrm{c}}$ are plotted as vertical lines for models A (solid line), B (triple-dot--dashed line), C (short-dashed line), D (long-dashed line), and coloured accordingly.}
\label{fig:dipole_mass}
\end{figure}

As accretion proceeds and the initial dipolar magnetic field lines are
distorted, magnetic energy is transferred from the dipole to higher order
multipole moments. The north--south antisymmetry of $B_{r}$ precludes the
existence of even multipoles. Fig. \ref{fig:dipole_mass} displays the magnetic dipole moment $\mu$ (normalized by its initial, or surface, value) 
as a function of the accreted mass $M_{\mathrm{a}}$ for models A--E in Table \ref{table:eos}.
The maximum accreted mass for which the
iterative solver converges reliably (grid-averaged residual $\leq 5$ per cent) depends
on the EOS, with $M_{\mathrm{a},\mathrm{max}} \approx 1\times10^{-3}, \
3\times10^{-8}, \ 2\times10^{-7}, \ 3\times10^{-6} \mathrm{M}_{\sun}$
for models A--D, respectively. (As a corollary, the gradient $\mathrm{d}\mu/\mathrm{d}M_{\mathrm{a}}$ in the
vicinity of the rightmost data point for each model in Fig.
\ref{fig:dipole_mass} is unphysically steep.) The method we use to calculate the
dipole moment differs slightly from that in PM04; we integrate $\psi$ directly
rather than $B_{r}$, according to 
\begin{equation}
\mu_{l} = \frac{l (2l + 1) R_{\mathrm{m}}^{l}}{2 (l + 1)} \int^{1}_{-1} \mathrm{d}(\cos
\theta) \ \psi(R_{\mathrm{m}}, \cos\theta) \frac{\mathrm{d}P_{l}(\cos \theta)}{\mathrm{d}(\cos
\theta)}
\label{multipole_moments}
\end{equation}
for the \textit{l}th multipole moment, circumventing one set of numerical derivatives
and improving the accuracy of the results. Equation (\ref{multipole_moments}) is
$\sim 10$ per cent more accurate than equation (34) in PM04 for a $64 \times 64$ grid.
The discrepancy shrinks to $ < 1$ per cent for a $256 \times 256$ grid.

\begin{figure}
\includegraphics[width=84mm]{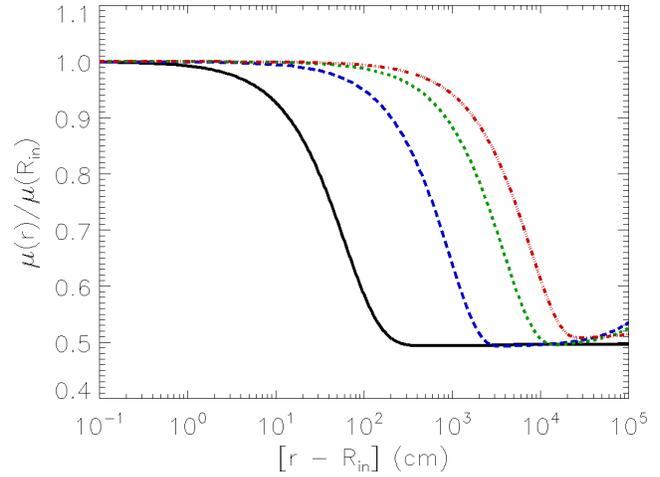}
\caption{Magnetic dipole moment $\mu$, calculated as a function of altitude (in
centimetres) from the inner boundary of the computational mesh $R_{\mathrm{in}}$ and
normalized to its surface value, for the characteristic masses $M_{\mathrm{c}}$, measured
in solar masses, for models A (solid black curve), B (triple-dot--dashed red curve), C (short-dashed green curve) and
D (long-dashed blue curve). The altitude where the dipole moment reaches a minimum denotes
where the screening currents end.}
\label{fig:dipole_radius}
\end{figure}

It is clear from Fig. \ref{fig:dipole_mass} that the characteristic mass $M_{\mathrm{c}}$
required to significantly distort the initial configuration varies with the EOS.
If we define $M_{\mathrm{c}}$ to be the accreted mass that halves $\mu$ from its initial value $\mu_{\mathrm{i}}$, 
to be consistent with the empirical scaling introduced by \citet{shibazaki1989}, viz.
\begin{equation}
\mu = \mu_{\mathrm{i}}(1 + M_{\mathrm{a}}/M_{\mathrm{c}})^{-1},
\label{burial_expression}
\end{equation}
then Fig. \ref{fig:dipole_mass} yields $M_{\mathrm{c},\mathrm{A}} \approx 5 \times
10^{-4} \mathrm{M}_{\sun}$, $M_{\mathrm{c},\mathrm{B}} \approx 3 \times 10^{-8} \mathrm{M}_{\sun}$,
$M_{\mathrm{c},\mathrm{C}} \approx 1 \times 10^{-7} \mathrm{M}_{\sun}$ and $M_{\mathrm{c},\mathrm{D}}
\approx 2 \times 10^{-6} \mathrm{M}_{\sun}$ for models A--D in Table \ref{table:eos}.
Plainly, varying the EOS makes a big difference. $M_{\mathrm{c}}$ is reduced by a factor
of between $3 \times 10^{2}$ (model D) and $2 \times 10^{4}$ (model B) relative
to an isothermal mountain. This is because adiabatic mountains are up to $\sim
10^{2}$ times taller than isothermal ones for $M_{\mathrm{a}} = M_{\mathrm{c}}$ (see Fig.
\ref{fig:dipole_radius} below and Section
\ref{section_4:hydromagnetic_structure}). At higher altitudes, the magnetic
stress ($\propto r^{-6}$) is weaker and hence the pressure gradient pushes the
magnetic field sideways more than in an isothermal mountain.

In the limit of small $M_{\mathrm{a}}$, one can show (see Appendix \ref{appendix:gs_analytic}) that the scaling of the 
characteristic mass $M_{\mathrm{c}}$ for adiabatic mountains is proportional to the square of the magnetic field strength, as for isothermal magnetic mountains (see Section 3.2 in PM04).
Additionally, $M_{\mathrm{c}}$ is also inversely proportional to an extra factor $I(\Lambda_{0}, \Gamma)$ (evaluated as a contour plot on the $\Lambda_{0}$--$\Gamma$ plane in Fig. \ref{fig:lambda_gamma}), which depends 
only on the EOS parameters and the accreted mass through equations (\ref{integral_plotted}) and (\ref{lambda_general}). In this limit, one finds the following scalings of the magnetic dipole moment: $\mu \propto (1-k_{\mathrm{A}}M_{\mathrm{a}} B^{-2})$ 
(model A), $\mu \propto (1-k_{\mathrm{B,D}}M_{\mathrm{a}}^{9/5} B^{-2})$ (models B and D) and $\mu \propto (1-k_{\mathrm{C}}M_{\mathrm{a}}^{3/2} B^{-2})$ (model C), where $k_{\mathrm{A,B,C,D}}$ are constants. We confirm in Section \ref{section_5} that the realistic 
EOS (model E) is well approximated by model C and hence follows the same scaling. It is important to note that these $\mu(M_{\mathrm{a}})$ scalings are only valid in the small-$M_{\mathrm{a}}$ limit (i.e. $M_{\mathrm{a}} \leq M_{\mathrm{c}}$, where $M_{\mathrm{c}}$ is EOS-dependent). 
For $M_{\mathrm{a}} > M_{\mathrm{c}}$, the analytical solution no longer applies and numerical results have to be used.

In Fig. \ref{fig:dipole_radius}, we plot $\mu/\mu_{\mathrm{i}}$ as a function of altitude
above the surface for models A--D by replacing $R_{\mathrm{m}}$ with $r$ in
equation (\ref{multipole_moments}). The purpose is to illustrate how the
screening currents are distributed radially for different EOS. The accreted
masses are chosen to be the characteristic masses $M_{\mathrm{c}}$ of each model in Table
\ref{table:eos}. The dipole moment turns up by $\approx 5$ per cent at $r \approx
R_{\mathrm{m}}$ because the Neumann boundary condition $\partial \psi/\partial
\theta = 0$, which holds the field lines perpendicular to the outer grid
boundary, does not apply strictly to a dipole field. For the isothermal mountain
(model A), the screening currents are located $10^{1} - 10^{2}$ times closer to
the neutron star surface than in models B--D, and the isodensity contours
contract towards the surface by the same factor (see Section
\ref{section_4:hydromagnetic_structure}).

\subsection{Mass quadrupole: $\epsilon$ versus $M_{\mathrm{a}}$}
\label{section_4:ellipticity}

\begin{figure}
\includegraphics[width=84mm]{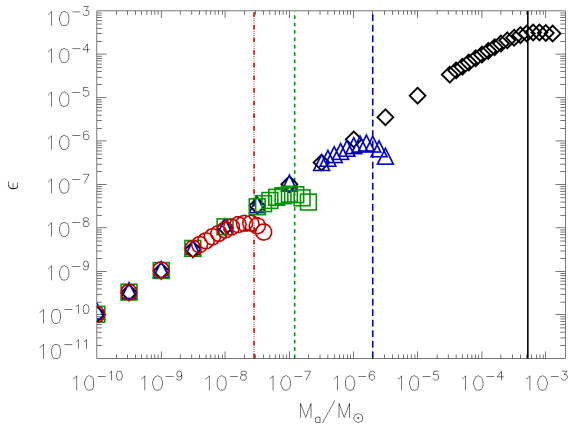}
\caption{Mass ellipticity $\epsilon$, as a function of accreted mass, $M_{\mathrm{a}}$,
measured in solar masses, for models A (black diamonds), B (red circles), C
(green squares) and D (blue triangles). Values of the
characteristic masses $M_{\mathrm{c}}$ are plotted as vertical lines for models A (solid line), B (triple-dot--dashed line), C (short-dashed line), D (long-dashed line), and coloured accordingly.}
\label{fig:ellipticity_mass}
\end{figure}

Fig. \ref{fig:ellipticity_mass} displays the mass quadrupole moment of the
mountain, expressed in terms of the mass ellipticity $\epsilon$, as a function
of $M_{\mathrm{a}}$. The ellipticity is given by $\epsilon = |I_{zz} - I_{yy}|/I_{0}$,
where $I_{ij}$ denotes the moment-of-inertia tensor, the $z$-axis lies along the
magnetic axis of symmetry, and we define $I_{0} =
(2/5)M_{\ast}R_{\mathrm{in}}^{2}$. To zeroth order, both $M_{\mathrm{a}}$ and $\epsilon$
are proportional to the surface density $\rho_{\mathrm{surf}}$. Hence, the
ellipticity is proportional to accreted mass for $M_{\mathrm{a}} < M_{\mathrm{c}}$. At $M_{\mathrm{a}}
\approx M_{\mathrm{c}}$, the hydrostatic pressure overwhelms the Lorentz force and the
mountain spreads laterally, distributing the extra accreted mass evenly over a
larger area (the enlarged magnetic polar cap) and moderating the growth of the
ellipticity such that $\mathrm{d}\epsilon/\mathrm{d}M_{\mathrm{a}} < 1/\mathrm{M}_{\sun}$.

The apparent turnover in $\epsilon$ after it peaks in Fig.
\ref{fig:ellipticity_mass} is a numerical artefact, which sets in as the
convergence of the numerical algorithm worsens (see Section
\ref{section_4:dipole_moment}). In reality, for $M_{\mathrm{a}} > M_{\mathrm{c}}$, the ellipticity
saturates at the value where $\mathrm{d}\epsilon/\mathrm{d}M_{\mathrm{a}} = 0$ in a hard-surface model.
\citet{wette2010} examined accretion on to a non-rigid neutron star crust,
thereby allowing the accreted matter to sink, and showed that the ellipticity
does not saturate (i.e. $\mathrm{d}\epsilon/\mathrm{d}M_{\mathrm{a}} > 0$) up to $M_{\mathrm{a}} \lesssim 0.12
\mathrm{M}_{\sun}$. Despite this monotonic increase, the ellipticity of soft-surface
mountains is always less than that of hard-surface mountains by $25$--$60$ per cent in
the mass range $1.2 \times 10^{-4} < M_{\mathrm{a}}/\mathrm{M}_{\sun} < 1.2\times10^{-1}$.

\subsection{Equatorial magnetic compression}
\label{section_4:maximum_magnetic_field}

\begin{figure}
\includegraphics[width=84mm]{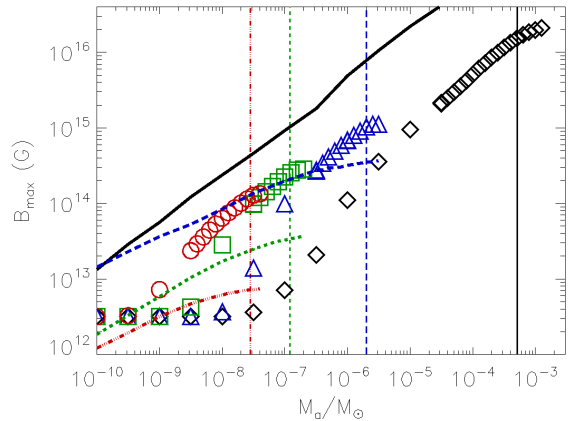}
\caption{Maximum field strength in the equatorial magnetic belt at hydromagnetic
equilibrium, $|\bmath{B}|_{\mathrm{max}}$ (in gauss), as a function of accreted mass,
$M_{\mathrm{a}}$ (in solar masses), for models A (black diamonds), B (red circles), C
(green squares) and D (blue triangles). Values of the
characteristic masses $M_{\mathrm{c}}$ are plotted as vertical lines for models A (solid line), B (triple-dot--dashed line), C (short-dashed line), D (long-dashed line), and coloured accordingly. 
Overplotted are curves of the maximum yield magnetic field
strength, $B_{\mathrm{yield}}$ (in gauss), at the base of a mountain of mass
$M_{\mathrm{a}}$ for models A (solid curve), B (triple-dot--dashed curve), C (short-dashed curve), D (long-dashed curve), and coloured accordingly. Adiabatic magnetic mountains
(models B--D) surpass $|\bmath{B}|_{\mathrm{max}}$, deforming plastically, while the
isothermal mountain (model A) does not exceed this threshold, and hence does not
crack.}
\label{fig:mag_mass}
\end{figure}

\begin{figure}
\subfigure
{
	\includegraphics[width=84mm]{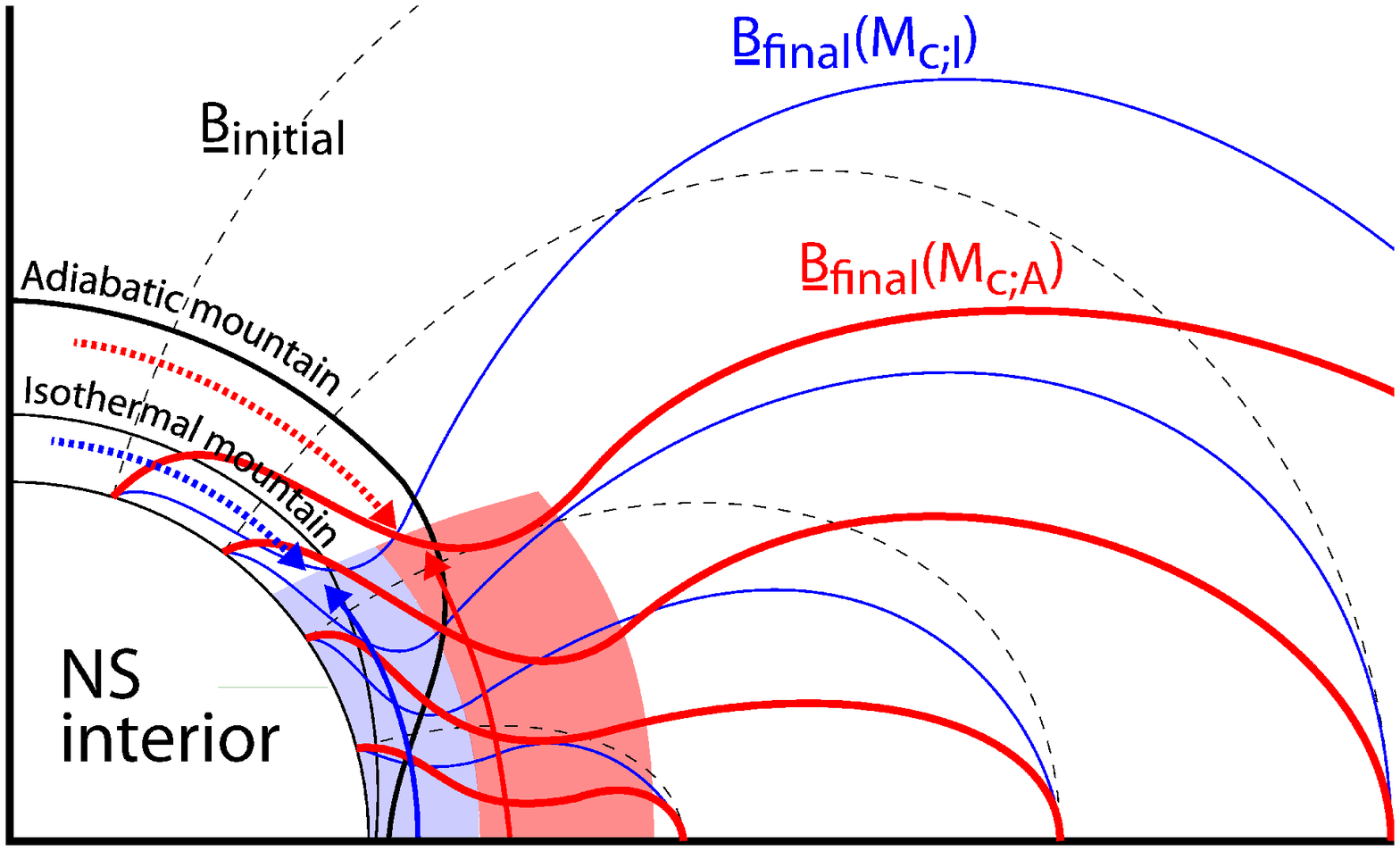}
}
\subfigure
{
	\includegraphics[width=84mm]{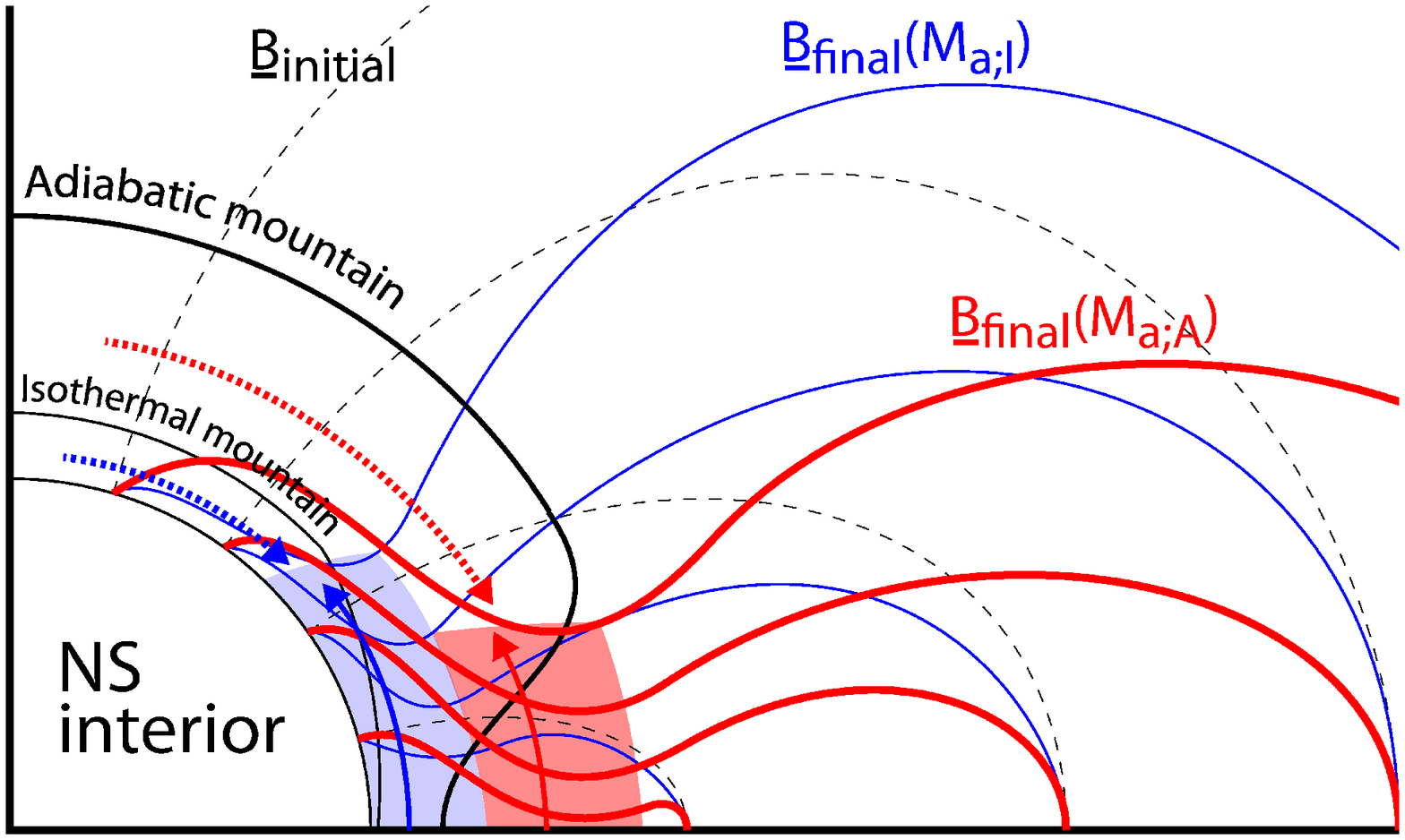}
}

\caption{Schematic diagram (not to scale) of the magnetic field of a neutron star during magnetic burial in the case of an adiabatic (thick black curve) and isothermal (thin black curve) magnetic mountain. An adiabatic magnetic mountain extends further above the surface. 
Undistorted dipole magnetic field lines before accretion are shown as dashed curves. Adiabatic and isothermal mountains of characteristic mass $M_{\mathrm{c}}$ (top panel) and equivalent accreted mass $M_{\mathrm{a}}$ (bottom panel) are represented [$M_{\textrm{c}}$ is the EOS-dependent characteristic mass defined in equation (\ref{burial_expression})]. 
The distorted magnetic field lines during magnetic burial are shown for both the adiabatic and isothermal EOS of the accreted matter (thick red curves and thin blue curves, respectively). Subscripts I and A denote isothermal and adiabatic EOS, respectively.
The Lorentz force (red/blue arrows for adiabatic/isothermal EOS) of the compressed magnetic field in the equatorial belt (the extent of the belt is denoted by red/blue shaded regions for adiabatic/isothermal EOS) balances the hydrostatic pressure gradient (red/blue dotted arrow for adiabatic/isothermal EOS) at the base of the mountain. 
As more matter accretes, the increasing hydrostatic pressure gradient at the base of the mountain is compensated by the enlarged Lorentz force in the magnetic belt. This compresses the magnetic belt further towards the equator.}
\label{fig:schematic}
\end{figure}

The accreted matter transports frozen-in magnetic flux equatorward as it spreads
sideways under its own weight. As a result, the magnetic field lines are
`pinched' near the surface at the equator and flare outwards at higher altitudes
like a `tutu' \citep{melatos2001, payne2006b}. The maximum magnetic field
strength $|\bmath{B}|_{\mathrm{max}}$ in the equatorial belt is computed as a
function of $M_{\mathrm{a}}$ and graphed in Fig. \ref{fig:mag_mass} for models A--D in
Table \ref{table:eos}. Naturally, the latitude where $|\bmath{B}|$ maximizes
moves towards the equator as $M_{\mathrm{a}}$ increases, and the equatorial belt narrows.
From Fig. \ref{fig:mag_mass}, we see that adiabatic magnetic mountains produce a
larger $|\bmath{B}|_{\mathrm{max}}$ (and hence a narrower belt, by flux
conservation) than isothermal ones with the same $M_{\mathrm{a}}$. Referring to Fig. \ref{fig:schematic}, this can be understood
as follows. The top panel of Fig. \ref{fig:schematic} shows the equilibrium magnetic field configuration of an adiabatic and an isothermal mountain, 
at their characteristic masses $M_{\mathrm{c}}$ (these masses are different since $M_{\mathrm{c}}$ is EOS-dependent). At equilibrium, the hydrostatic pressure gradient at 
the base of the mountain (dotted red/blue arrow for adiabatic/isothermal mountain in Fig. \ref{fig:schematic}) is
balanced by magnetic stresses (red/blue arrow for adiabatic/isothermal mountain in Fig. \ref{fig:schematic}) within the equatorial magnetic belt 
(the extent of the magnetic belt is denoted by red/blue shaded regions for adiabatic/isothermal mountains in Fig. \ref{fig:schematic}).
The hydrostatic pressure gradients for both EOSs are comparable at characteristic accreted masses, because the magnetic
field lines are bent by a similar angle for all models at $M_{\mathrm{a}} \approx M_{\mathrm{c}}$ [since $\mu(M_{\textrm{c}})$ is EOS-independent].
This can be expressed equivalently in terms of the comparable width of the equatorial magnetic belt of both mountains, since comparable deformation angles 
of the magnetic field lines result in corresponding widths of the magnetic belt. 
Referring to the bottom panel of Fig. \ref{fig:schematic}, the hydrostatic pressure gradient at the base of the accreted layer is greater for adiabatic
mountains than isothermal ones at an equivalent $M_{\mathrm{a}}$, because $M_{\mathrm{c}; \mathrm{A}} >
M_{\mathrm{c}; \mathrm{D}} > M_{\mathrm{c}; \mathrm{C}} > M_{\mathrm{c}; \mathrm{B}}$, where the subscripts A--D denote the models in Table \ref{table:eos} (see Section \ref{section_4:dipole_moment}). 
Hence, magnetic-field lines of an adiabatic mountain are more deformed than those of an isothermal one to counteract this. This decreases the lateral extent of the magnetic belt and, by magnetic flux
conservation, $|\bmath{B}|_{\mathrm{max}}$ increases as the belt shrinks. This explains
why the point where $|\bmath{B}|_{\mathrm{max}}$ is reached moves equatorward as $M_{\mathrm{a}}$
increases, and why $|\bmath{B}|_{\mathrm{max}}$ is greater for an adiabatic rather than an
isothermal mountain for the same $M_{\mathrm{a}}$.

The compressed magnetic field can surpass the yield strength of the crust, at
which point the magnetic stresses break the Coulomb lattice as the field
deforms. Taking the breaking strain of the neutron star crust to be $\approx
0.1$ from recent molecular dynamics simulations \citep{horowitz2009a}, the
magnetic field strength at which the crustal matter yields \citep{romani1990} is
\begin{equation}
B_{\mathrm{yield}} = 1.2 \times 10^{14} ZA^{-2/3} (\rho/10^{11} \ \mathrm{g} \
\mathrm{cm}^{-3})^{2/3} \ \mathrm{G},
\end{equation}
where $Z$ and $A$ are the mean atomic and mass numbers, respectively. We evaluate
$B_{\mathrm{yield}}$ at the base of a mountain of mass $M_{\mathrm{a}}$ from the nuclides
present at base pressure \citep{haensel1990b, haensel1990a, chamel2008}. The
results are plotted as curves in Fig. \ref{fig:mag_mass} for the models in Table
\ref{table:eos}. In an isothermal mountain, we find $|\bmath{B}|_{\mathrm{max}}
< B_{\mathrm{yield}}$, so that the accreted matter does not crack and remains
polycrystalline, with a frozen-in magnetic field. As the substrate of an
isothermal mountain does not spread significantly, $\epsilon$ and $M_{\mathrm{c}}$ are
larger. Indeed, strictly speaking, crustal freezing should be included in the
boundary conditions of an isothermal mountain calculation (implemented
dynamically at the depth where it first occurs). On the other hand, adiabatic
mountains compress the magnetic field in excess of $B_{\mathrm{yield}}$ for
$M_{\mathrm{a}} \gtrsim 3 \times 10^{-7} \mathrm{M}_{\sun}$ and $M_{\mathrm{a}} \gtrsim 6 \times 10^{-9}
\mathrm{M}_{\sun}$ for models D and C respectively, while $B_{\mathrm{yield}}$ is
surpassed for all accreted masses in the case of model B. This suggests that the
accreted matter continuously cracks or flows plastically at most depths
\citep{horowitz2009a}, validating the fluid approximation for models with
$\Gamma \geq 4/3$.

\subsection{Maximum Density}
\label{section_4:maximum_density}

\begin{figure}
\includegraphics[width=84mm]{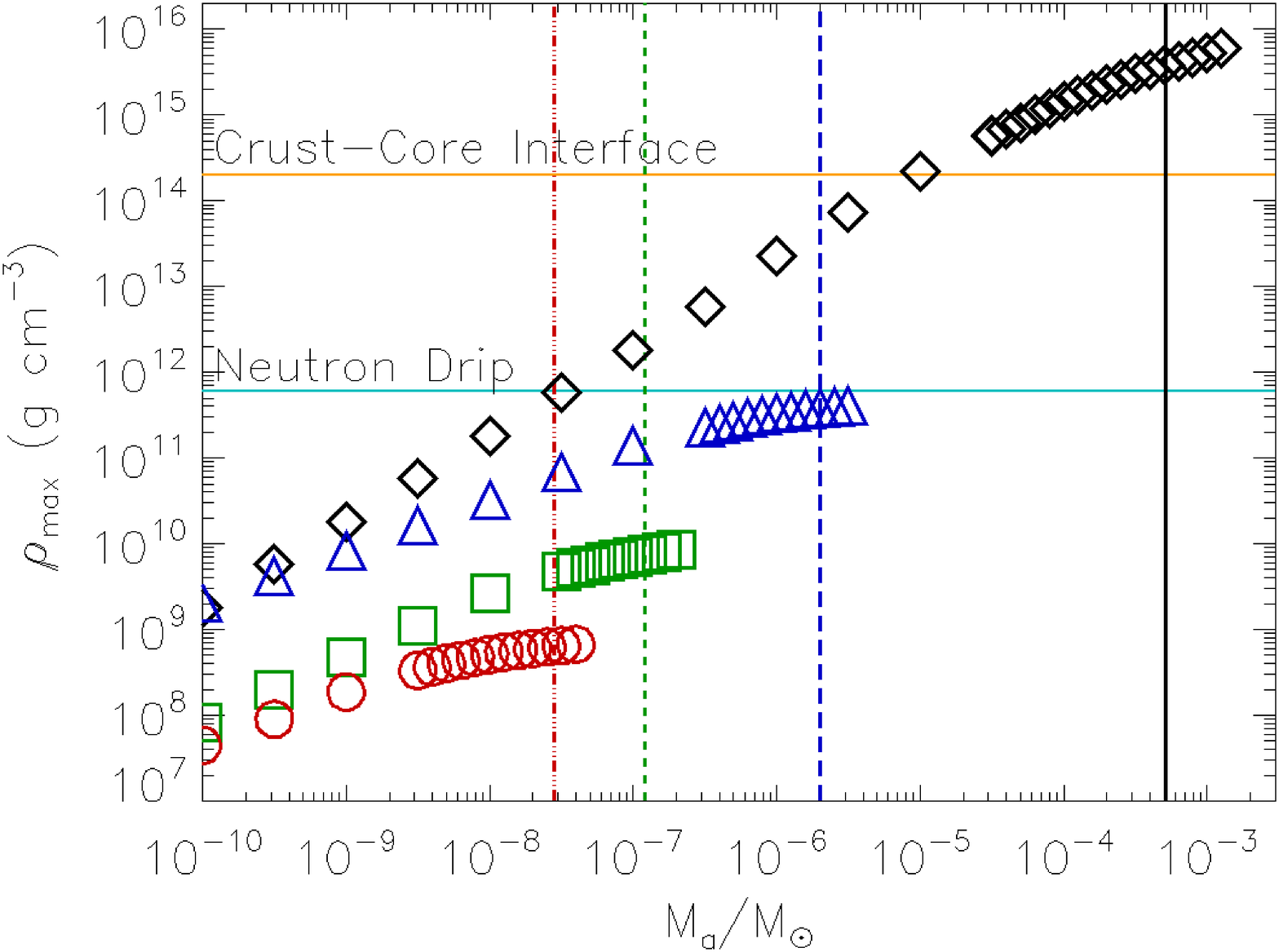}
\caption{Maximum density at hydromagnetic equilibrium, $\rho_{\mathrm{max}}$
($\mathrm{g} \ \mathrm{cm}^{-3}$), as a function of accreted mass, $M_{\mathrm{a}}$ (in
solar masses), for models A (black diamonds), B (red circles), C (green
squares) and D (blue triangles). Values of the
characteristic masses $M_{\mathrm{c}}$ are plotted as vertical lines for models A (solid line), B (triple-dot--dashed line), C (short-dashed line), D (long-dashed line), and coloured accordingly. 
Overplotted are lines of the neutron drip density $\rho_{\mathrm{ND}} = 6 \times
10^{11} \ \mathrm{g} \ \mathrm{cm}^{-3}$ and the density at the crust--core
interface $\rho_{\mathrm{CC}} = 2 \times 10^{14} \ \mathrm{g} \
\mathrm{cm}^{-3}$. Adiabatic magnetic mountains (models B--D) satisfy
$\rho_{\mathrm{max}} < \rho_{\mathrm{ND}}$, while isothermal magnetic mountains
exceed $\rho_{\mathrm{ND}}$ and $\rho_{\mathrm{CC}}$ for $M_{\mathrm{a}} > 3 \times
10^{-8} \mathrm{M}_{\sun}$ and $M_{\mathrm{a}} > 1\times10^{-5} \mathrm{M}_{\sun}$, respectively.}
\label{fig:rho_mass}
\end{figure}

The maximum density at the base of a magnetic mountain is reached at the
magnetic pole (see Section \ref{section_4:hydromagnetic_structure}). We extract
the maximum density $\rho_{\mathrm{max}}(R_{\mathrm{in}}, 0)$ as a function of $M_{\mathrm{a}}$
from the simulated models listed in Table \ref{table:eos} and graph the results
in Fig. \ref{fig:rho_mass}.

A deficiency of isothermal mountains, noted by PM04, is the unrealistically high
density at the base, which exceeds the neutron drip
point\footnote{$\rho_{\mathrm{ND}}$ depends on whether the crust is
cold-catalyzed or accreted, as well as the exact EOS \citep{chamel2008}. Here we
consider the compressible liquid drop model for the EOS of the accreted crust
\citep{haensel1990b, haensel1990a, chamel2008}.} $\rho_{\mathrm{ND}} \approx 6
\times 10^{11} \ \mathrm{g} \ \mathrm{cm}^{-3}$ at relatively small accreted
masses of $\sim 10^{-8} \mathrm{M}_{\sun}$ (cf. $M_{\mathrm{a}} \sim 10^{-1} \mathrm{M}_{\sun}$ in a
typical LMXB). In contrast, Fig. \ref{fig:rho_mass} shows that
$\rho_{\mathrm{max}}$ is several orders of magnitude lower for an adiabatic EOS;
none of the adiabatic mountains surpass $\rho_{\mathrm{ND}}$ for $M_{\mathrm{a}} \lesssim
M_{\mathrm{c}}$.

At accreted masses approaching $M_{\mathrm{a}} \sim 10^{-2} \mathrm{M}_{\sun}$, models C and D attain
crust--core densities\footnote{We choose $\rho_{\mathrm{CC}}$ large enough to
contain the crust--core transitions of both the Friedman--Pandharipande--Skyrme and
Skyrme--Lyon EOS \citep{pethick1995, haensel2007, chamel2008}.}
$\rho_{\mathrm{CC}} \approx 2 \times 10^{14} \ \mathrm{g} \ \mathrm{cm}^{-3}$ at
their bases. These models are good approximations to the EOS of the neutron star
crust at $\rho \gtrsim 10^{9} \ \mathrm{g} \ \mathrm{cm}^{-3}$ and $\rho \gtrsim
10^{13} \ \mathrm{g} \ \mathrm{cm}^{-3}$, respectively (see Section
\ref{section_5}). On the other hand, model B does not reach the
crust--core interface because it is too stiff and approximates the true crustal
EOS only at low densities $10^{5} < \rho/(\mathrm{g} \ \mathrm{cm}^{-3}) <
10^{7}$. In the isothermal mountain (model A), $\rho_{\mathrm{max}}$ exceeds
$\rho_{\mathrm{CC}}$ for $M_{\mathrm{a}} \gtrsim 10^{-5} \mathrm{M}_{\sun}$.

\subsection{Hydromagnetic structure}
\label{section_4:hydromagnetic_structure}

\begin{figure*}
\begin{minipage}{170mm}
\subfigure
{
	\includegraphics[width=84mm]{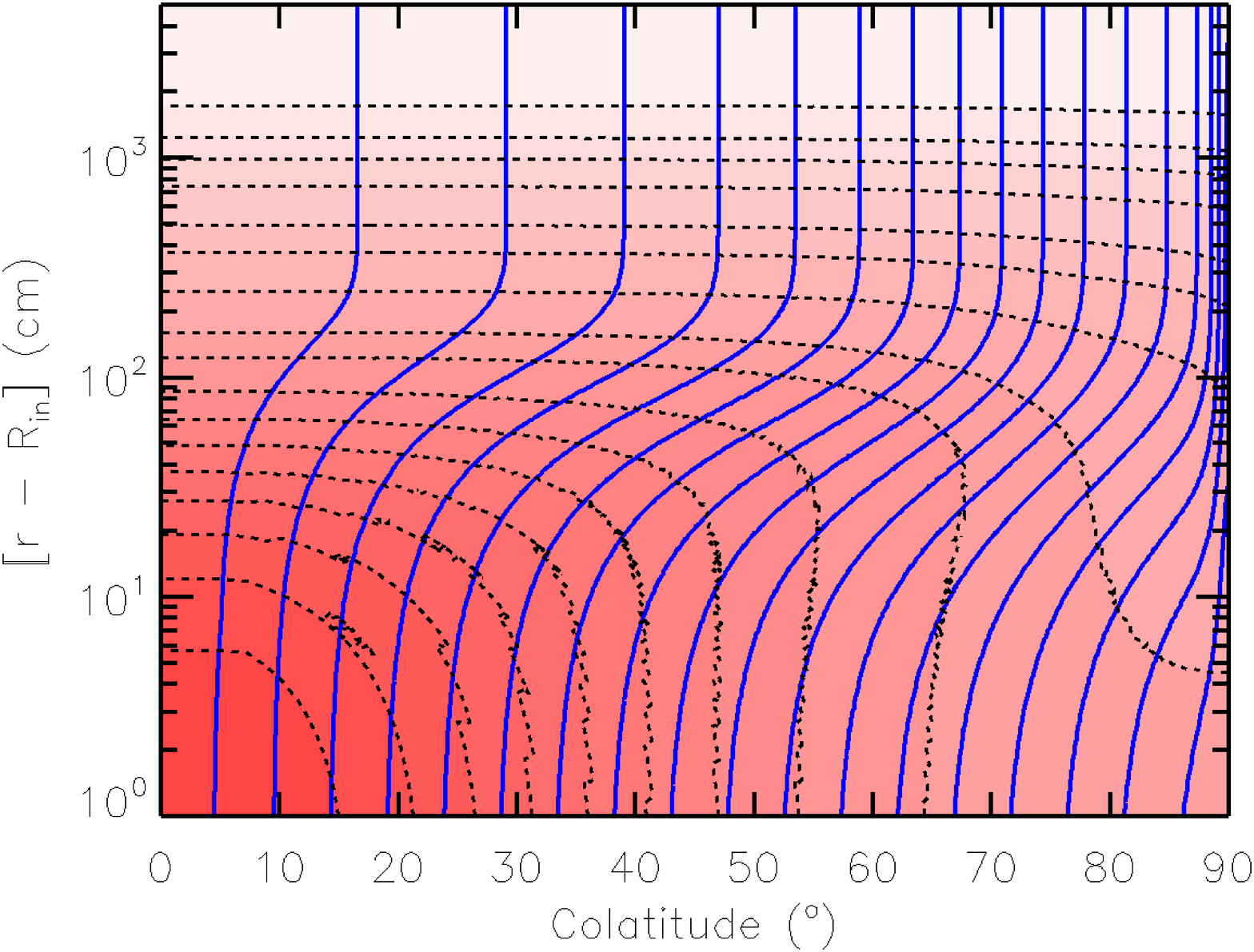}
}
\subfigure
{
	\includegraphics[width=84mm]{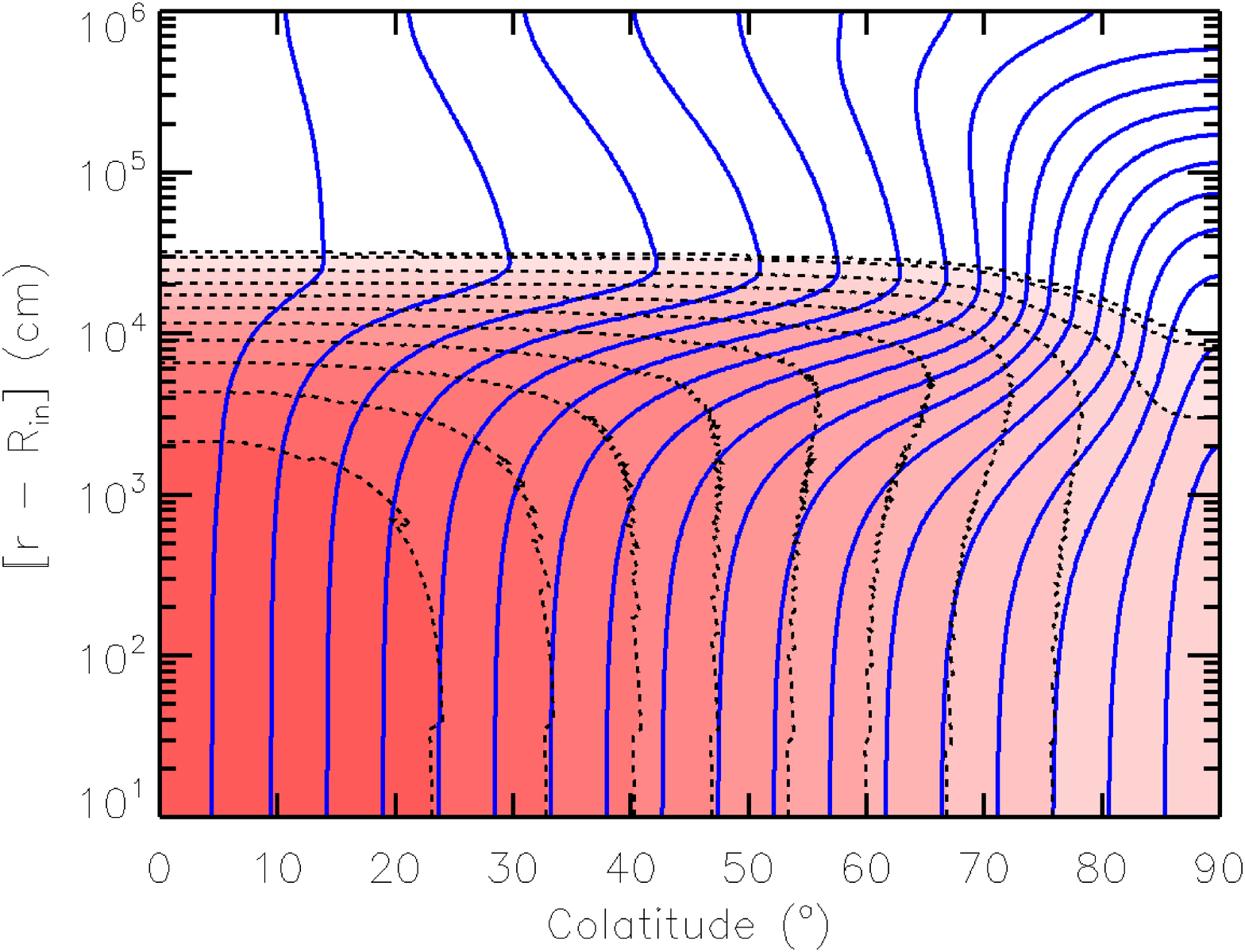}
}
\subfigure
{
	\includegraphics[width=84mm]{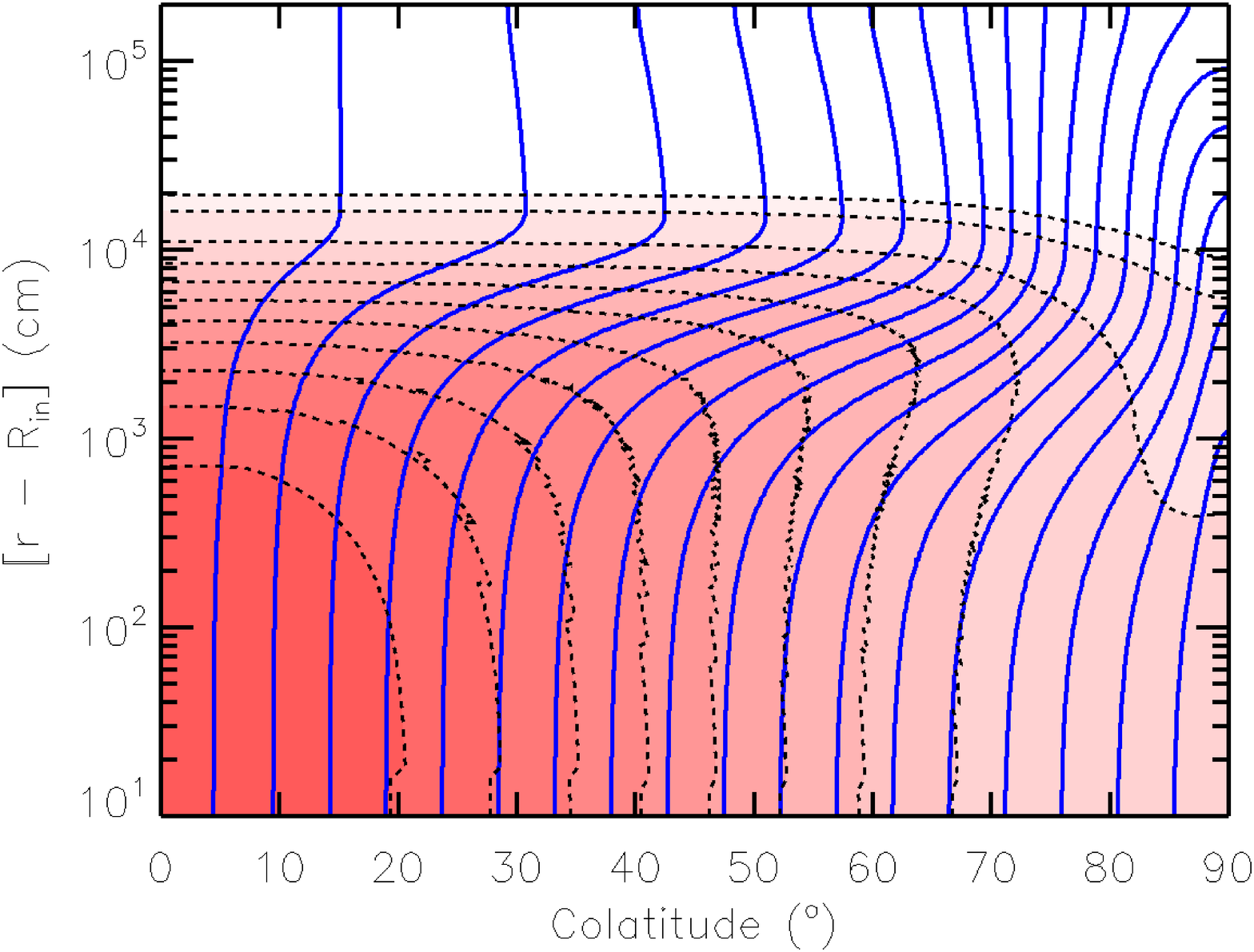}
}
\subfigure
{
	\includegraphics[width=84mm]{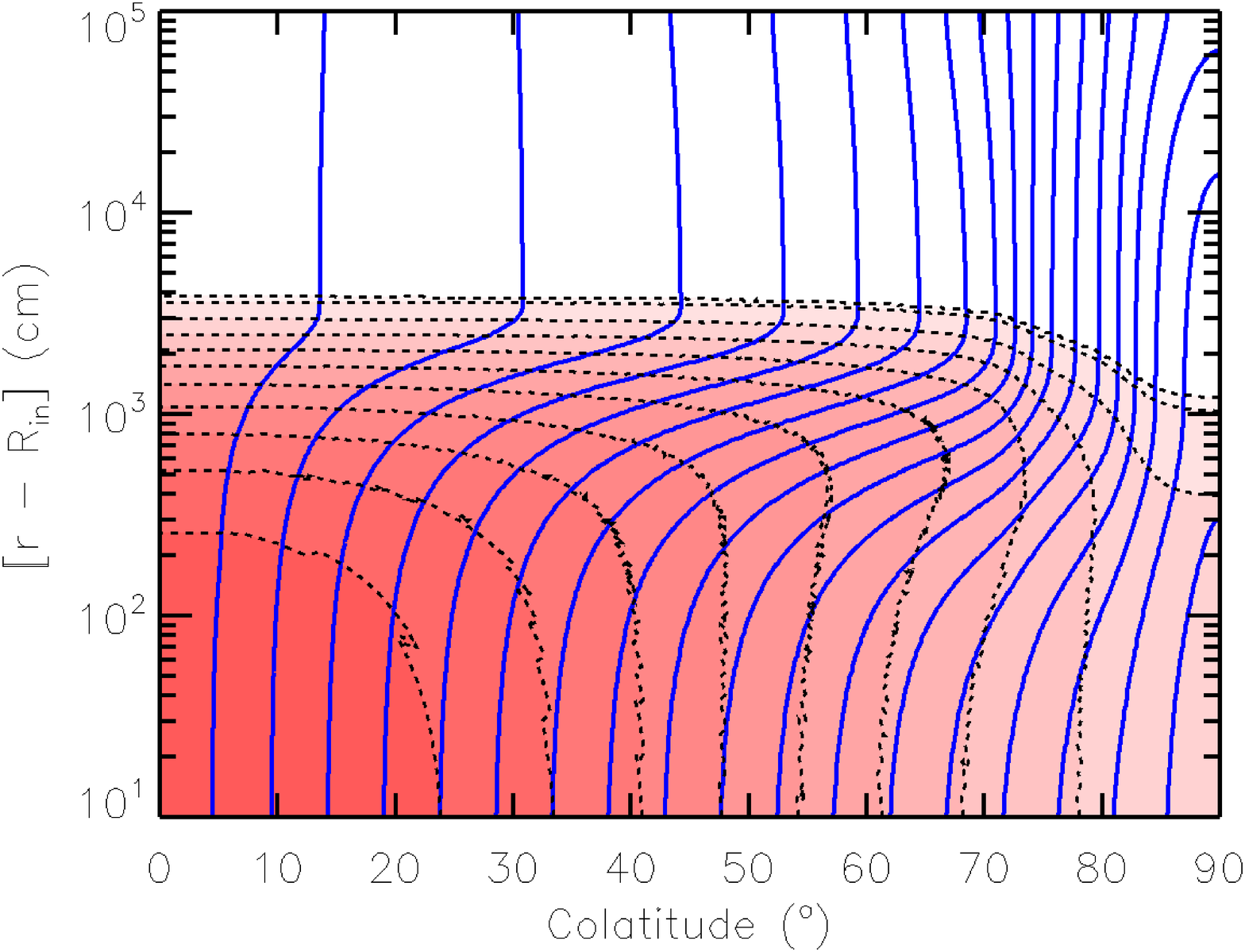}
}
\caption{Hydromagnetic structure of axisymmetric mountain equilibria at $M_{\mathrm{a}} =
M_{\mathrm{c}}$, showing magnetic field lines (solid blue curves) and isodensity contours
(dashed black curves) for model A with $M_{\mathrm{a}} = 5.2\times10^{-4} \mathrm{M}_{\sun}$ (top-left
panel), model B with $M_{\mathrm{a}} = 2.8\times10^{-8} \mathrm{M}_{\sun}$ (top-right panel),
model C with $M_{\mathrm{a}} = 1.2\times10^{-7} \mathrm{M}_{\sun}$ (bottom-left panel) and model
D with $M_{\mathrm{a}} = 2.0\times10^{-6} \mathrm{M}_{\sun}$ (bottom-right panel). Density
contours are drawn for $\eta \rho_{\mathrm{max}}$ (maximum at the pole), with
$\rho_{\mathrm{max,A}} = 3.9\times10^{15} \ \mathrm{g} \ \mathrm{cm}^{-3}$,
$\rho_{\mathrm{max,B}} = 6.4\times10^{8} \ \mathrm{g} \ \mathrm{cm}^{-3}$,
$\rho_{\mathrm{max,C}} = 7.4\times10^{9} \ \mathrm{g} \ \mathrm{cm}^{-3}$,
$\rho_{\mathrm{max,D}} = 3.8\times10^{11} \ \mathrm{g} \ \mathrm{cm}^{-3}$,
$\eta_{\mathrm{A}} = 0.9, \ 0.8, \ 0.7, \ 0.6, \ 0.5, \ 0.4, \ 0.3, \ 0.2, \
0.1, \ 0.05, \ 10^{-2}, \ 10^{-3}, \ 10^{-4}, \ 10^{-6}, \ 10^{-8}, \ 10^{-10},
\ 10^{-14}$, and $\eta_{\mathrm{B,C,D}} = 0.9, \ 0.8, \ 0.7, \ 0.6, \ 0.5, \
0.4, \ 0.3, \ 0.2, \ 0.1, \ 10^{-2}, \ 0$.}
\label{fig:magnetic_mountain}
\end{minipage}
\end{figure*}

A meridional cross-section of the magnetic mountain produced by models A--D in
Table \ref{table:eos} is displayed in Fig. \ref{fig:magnetic_mountain}, for
$M_{\mathrm{a}} = M_{\mathrm{c}}$. The magnetic field lines and isodensity contours are graphed as
solid and dashed curves, respectively; the shading also represents the density
and is included to guide the eye. Note that the vertical scale changes
dramatically from panel to panel. Adiabatic mountains stand $10^{1}-10^{2}$
times higher than an isothermal mountain for $M_{\mathrm{a}} = M_{\mathrm{c}}$ (see also Section
\ref{section_4:dipole_moment}). Moreover, one finds $\rho \to 0$ as $r \to
\infty$ in an isothermal mountain, whereas an adiabatic mountain drops to $\rho
= 0$ at a finite altitude. Mountains with an ideal degenerate electron gas EOS
(models B and C in Table \ref{table:eos}) are approximately 1 order of
magnitude taller than those with an ideal degenerate non-relativistic neutron
gas EOS (model D).

The polar ($r_{\mathrm{p}}$) and equatorial ($r_{\mathrm{e}}$) mountain heights
as well as their ratio $S$ can be estimated analytically in the small-$M_{\mathrm{a}}$
approximation developed in Appendix \ref{appendix:gs_analytic}. For models B, C
and D, and fiducial neutron star values (see Section \ref{section_3}), we obtain

\begin{equation}
r_{\mathrm{p}}|_{\mathrm{B}} = 5.2\times10^{4} \ \mathrm{cm}, \ r_{\mathrm{e}}|_{\mathrm{B}} = 7.2\times10^{3} \
\mathrm{cm}, \ S_{\mathrm{B}} = 7.2\times10^{0},
\end{equation}
\begin{equation}
r_{\mathrm{p}}|_{\mathrm{C}} = 2.8\times10^{4} \ \mathrm{cm}, \ r_{\mathrm{e}}|_{\mathrm{C}} = 8.3\times10^{3} \
\mathrm{cm}, \ S_{\mathrm{C}} = 3.3\times10^{0},
\end{equation}
\begin{equation}
r_{\mathrm{p}}|_{\mathrm{D}} = 6.2\times10^{3} \ \mathrm{cm}, \ r_{\mathrm{e}}|_{\mathrm{D}} = 2.1\times10^{2} \
\mathrm{cm}, \ S_{\mathrm{D}} = 2.9\times10^{1}. 
\end{equation}
Comparing with Fig. \ref{fig:magnetic_mountain}, we see that the analytic
formula in Appendix \ref{appendix:gs_analytic} generally overestimates
$r_{\mathrm{p}}$ and underestimates $r_{\mathrm{e}}$. This discrepancy arises
because the small-$M_{\mathrm{a}}$ approximation assumes the magnetic field is nearly
dipolar, whereas, at $M_{\mathrm{c}}$, the dipole is significantly deformed. For $M_{\mathrm{a}}
\ll M_{\mathrm{c}}$, there is better agreement between the numerical and analytical
solutions.

\section{Crustal equation of state}
\label{section_5}

A realistic crustal EOS is not a simple polytrope. It includes various pressure
contributions from thermal electrons, relativistic/non-relativistic degenerate
electrons, non-relativistic degenerate neutrons and the ionic lattice \citep{brown2000}. 
These partial pressures depend on the composition of the
crust; accreted matter undergoes nuclear reactions (e.g. electron captures and
pycnonuclear fusion) as the mass density and electron Fermi energy of the
compressed matter increases with depth \citep{chamel2008}. Our models of magnetic mountains in accreting X-ray systems necessitate 
the inclusion of a realistic accreted EOS of the neutron star crust. In this section, we
start from the realistic crustal EOS investigated by other authors \citep{negele1973, paczynski1983, brown2000} and derive an equivalent effective
adiabatic EOS ($K_{\mathrm{eff}}, \Gamma_{\mathrm{eff}}$) as a function of
$M_{\mathrm{a}}$. This EOS is labelled model E in Table \ref{table:eos}. The magnetic
mountains produced by this more realistic EOS are compared with the pure
adiabatic ones from Section \ref{section_4}.

We adopt the one-component plasma approximation for the accreted matter
\citep{haensel1990b, haensel1990a, brown2000, chamel2008}, together with the
nuclear composition proposed by \citet{haensel1990b, haensel1990a} at
temperature $T = 10^{8} \ \mathrm{K}$. This temperature is representative of the
steady-state thermal profile for $10^{6} \lesssim \rho/(\mathrm{g} \
\mathrm{cm}^{-3}) \lesssim 10^{14}$ in accreting neutron stars containing no
exotic matter such as a pion condensate or strange quarks in their interior
\citep{miralda_escude1990}. The foregoing assumptions hold for accretion rates
in the range $-11 < \log_{10}[\dot{M}/(\mathrm{M}_{\sun} \ \mathrm{yr}^{-1})] < -10$.

Recent work by \citet{read2009} [see also \citet{vuille1999}] produced a four-parameter fit to the set of candidates 
for high-density EOSs in order to systematize the study of various observational constraints on the EOSs. The low-density EOS was assumed to be that of ground-state cold matter given by 
\citet{douchin2001}, while the high-density candidate EOSs were parametrized by three free-parameter piecewise polytropes. Since the EOS of the accreted crust is stiffer 
than that of a cold-catalyzed one \citep{chamel2008}, making the radius of a $2$--$1 \mathrm{M}_{\sun}$ star $50$--$200 \ \mathrm{m}$ larger than that in the cold-catalyzed 
case \citep{zdunik2011}, the effective polytropic form for the accreted crust calculated in this section  
can be combined with observations of accreting neutron stars to constrain the parameters of the parametric EOS of \citet{read2009}.

\begin{figure*}
\begin{minipage}{170mm}
\subfigure
{
	\includegraphics[width=84mm, height=60mm]{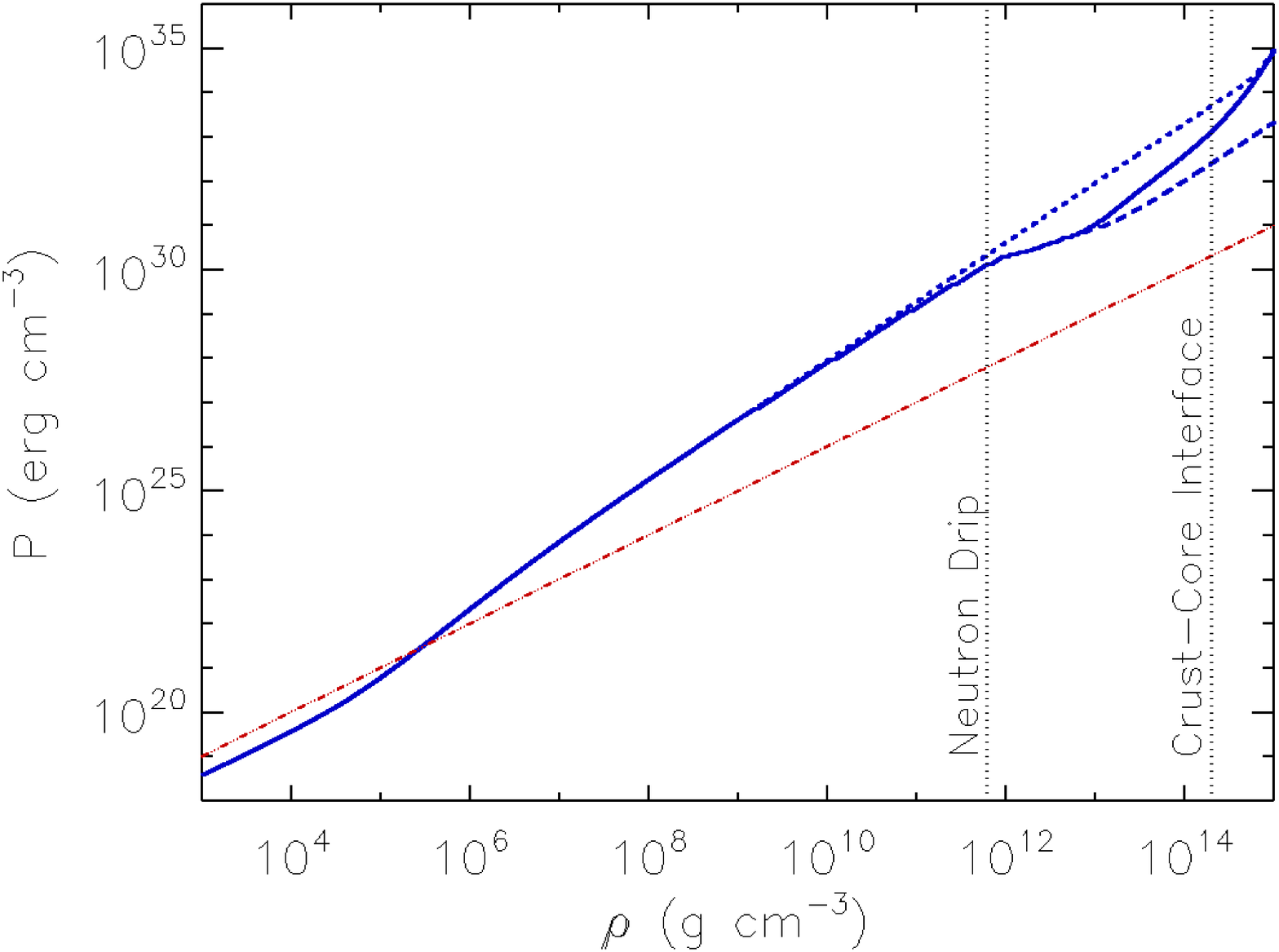}
}
\subfigure
{
	\includegraphics[width=84mm, height=60mm]{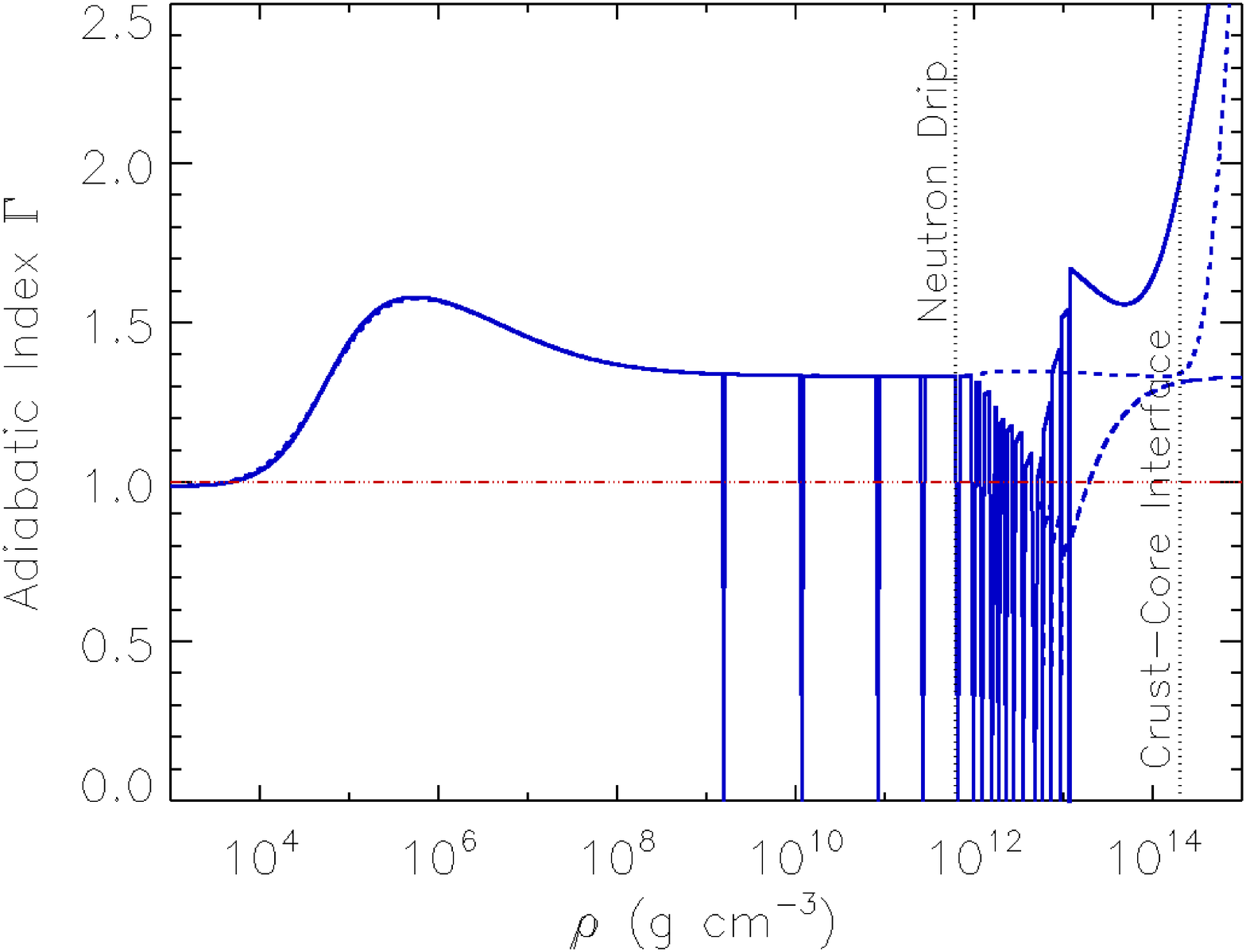}
}
\caption{Pressure as a function of density (left-hand panel) and adiabatic index as a
function of density (right-hand panel) for the full nuclear EOS (solid blue curve),
minus neutron pressure (long-dashed blue curve) and minus compositional
variations (short-dashed blue curve). For comparison, the isothermal EOS (triple-dot--dashed red
line) is also plotted. The neutron drip and crust--core interface densities are
marked with vertical dotted lines. Discontinuities in the adiabatic index for $\rho >
10^{9} \ \mathrm{g} \ \mathrm{cm}^{-3}$ are caused by density jumps between
compositional layers. The axes are log--log and linear--log in the left- and right-hand 
panels, respectively, with pressure and density measured in units of
$\mathrm{erg} \ \mathrm{cm}^{-3}$ and $\mathrm{g} \ \mathrm{cm}^{-3}$, 
respectively.} 
\label{fig:eos}
\end{minipage}
\end{figure*}

\subsection{Partial pressures}
\label{section_5:partial_pressures}

There are three principal contributions to the pressure in a mountain at
densities $\rho \lesssim \rho_{\mathrm{CC}}$. They are as follows. (i)
\textit{Electron pressure}: this is exerted by non-relativistic, relativistic 
and thermal electron populations \citep{paczynski1983}. (ii) \textit{Lattice
pressure}: the ionic lattice exerts negative pressure due to electrostatic
interactions within the Wigner--Seitz cells. It is calculated by fitting to the
free energy in Monte Carlo simulations of a one-component plasma
\citep{farouki1993}. (iii) \textit{Neutron pressure}: the effect is included in
a cold-catalyzed EOS which is parametrized to fit $11$ ground state nuclei above
the neutron drip line \citep{negele1973}. We sum the partial pressures (i)--(ii)
subject to pressure continuity across reaction surfaces, matching to the
cold-catalyzed EOS of \citet{negele1973} at densities above neutron drip.
Although the ground-state and accreted crusts contain different nuclei, their
respective EOS are indistinguishable for $\rho \gtrsim 10^{13} \ \mathrm{g} \
\mathrm{cm}^{-3}$, where the composition-insensitive neutron pressure dominates
\citep{chamel2008}.

The pressure and the adiabatic index $\Gamma = \mathrm{d}(\log P)/\mathrm{d}(\log \rho)$ of the
resultant EOS are graphed versus density in Fig. \ref{fig:eos}. Although some
parts of the EOS are piecewise adiabatic, other parts are not. At certain
densities where electron capture reactions occur rapidly, e.g. $\rho \gtrsim
10^{9} \ \mathrm{g} \ \mathrm{cm}^{-3}$, the density jumps discontinuously to
compensate for the sharp decline in electron pressure at a compositional
interface. This behaviour is accompanied by a sharp drop in the adiabatic index.
These discontinuities are an artefact of the one-component plasma approximation.
The presence of nuclear reactions softens the EOS for $10^{12} \lesssim
\rho/(\mathrm{g} \ \mathrm{cm}^{-3}) \lesssim 10^{13}$, relative to uniform
composition, whereas the addition of neutron pressure stiffens the EOS for $\rho
\gtrsim 10^{13} \ \mathrm{g} \ \mathrm{cm}^{-3}$.

\subsection{Effective polytrope}
\label{section5:effective_polytrope}

The realistic EOS [$K(\rho), \Gamma(\rho)$] in Fig. \ref{fig:eos} is transformed
into an effective adiabatic EOS, of the form $P = K_{\mathrm{eff}}
\rho^{\Gamma_{\mathrm{eff}}}$, by computing the mass-weighted averages
\begin{equation}
K_{\mathrm{eff}} = \frac{\int \mathrm{d}r \ r^{2} \rho(r) K(\rho)}{\int \mathrm{d}r \ r^{2}
\rho(r)}, \\
\Gamma_{\mathrm{eff}} = \frac{\int \mathrm{d}r \ r^{2} \rho(r) \Gamma(\rho)}{\int \mathrm{d}r \
r^{2} \rho(r)},
\end{equation}
for a spherically symmetric accreted layer of mass $M_{\mathrm{a}}$ whose density profile
$\rho(r)$ satisfies hydrostatic equilibrium. For simplicity, we ignore general
relativistic effects and assume the acceleration due to gravity to be uniform,
as in models A--D.

\begin{figure*}
\begin{minipage}{170mm} 
\subfigure
{
	\includegraphics[width=84mm, height=60mm]{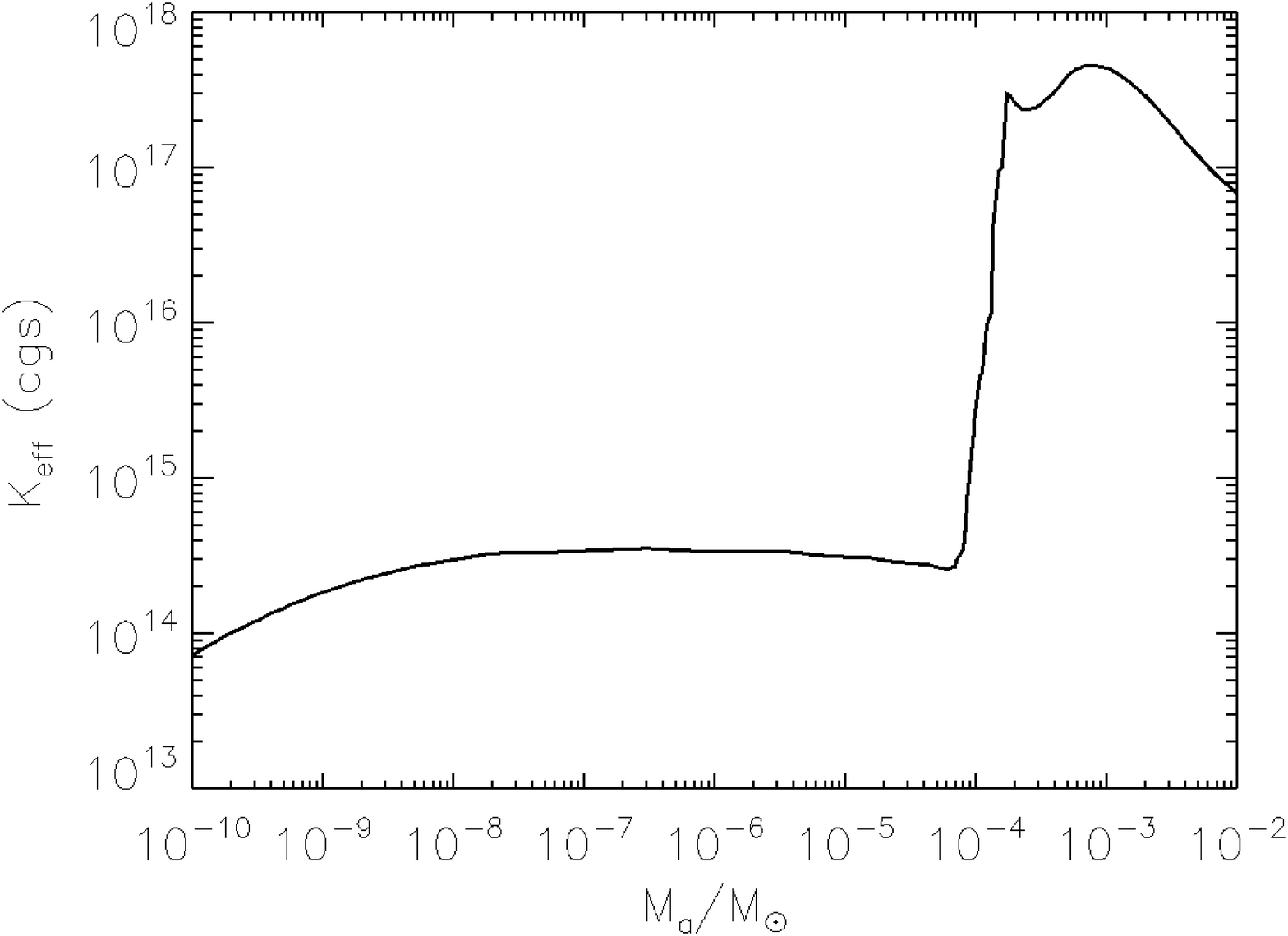}
}
\subfigure
{
	\includegraphics[width=84mm, height=60mm]{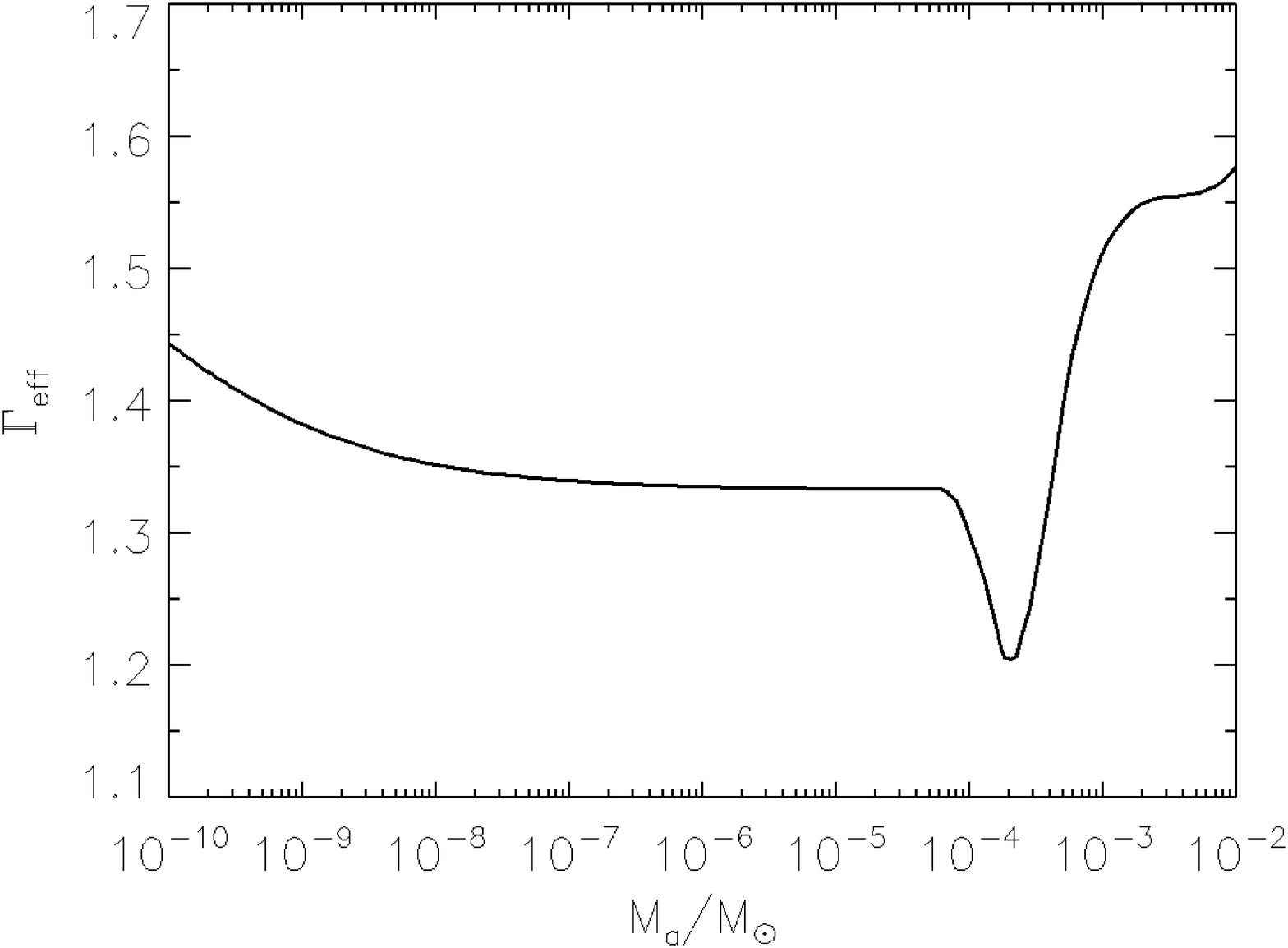}
}
\caption{Effective adiabatic EOS for mountains of different sizes (model E in
Table \ref{table:eos}). $K_{\mathrm{eff}}$ in cgs units (left-hand panel) and
adiabatic index $\Gamma_{\mathrm{eff}}$ (right-hand panel) as functions of accreted
mass $M_{\mathrm{a}}$ (in solar masses).}
\label{fig:K_Gamma_mass}
\end{minipage}
\end{figure*}

The scaling of $K_{\mathrm{eff}}$ and $\Gamma_{\mathrm{eff}}$ with $M_{\mathrm{a}}$ is
shown in Fig. \ref{fig:K_Gamma_mass}. The large radial variations of $K$ and
$\Gamma$ within the crust imply that $K_{\mathrm{eff}}$ and
$\Gamma_{\mathrm{eff}}$ depend strongly on the maximum achieved density and
hence $M_{\mathrm{a}}$. The mass-weighted averages are dominated by the base of the
mountain. Hence, the mass ellipticity and magnetic dipole moment of an adiabatic
mountain with a realistic nuclear EOS depend on $M_{\mathrm{a}}$, not just through the
weight of the accreted layer and the confining magnetic stresses but also
through the density-dependent thermodynamics at the mountain's base.

\begin{figure*}
\begin{minipage}{175mm}
\subfigure
{
	\includegraphics[width=84mm, height=65mm]{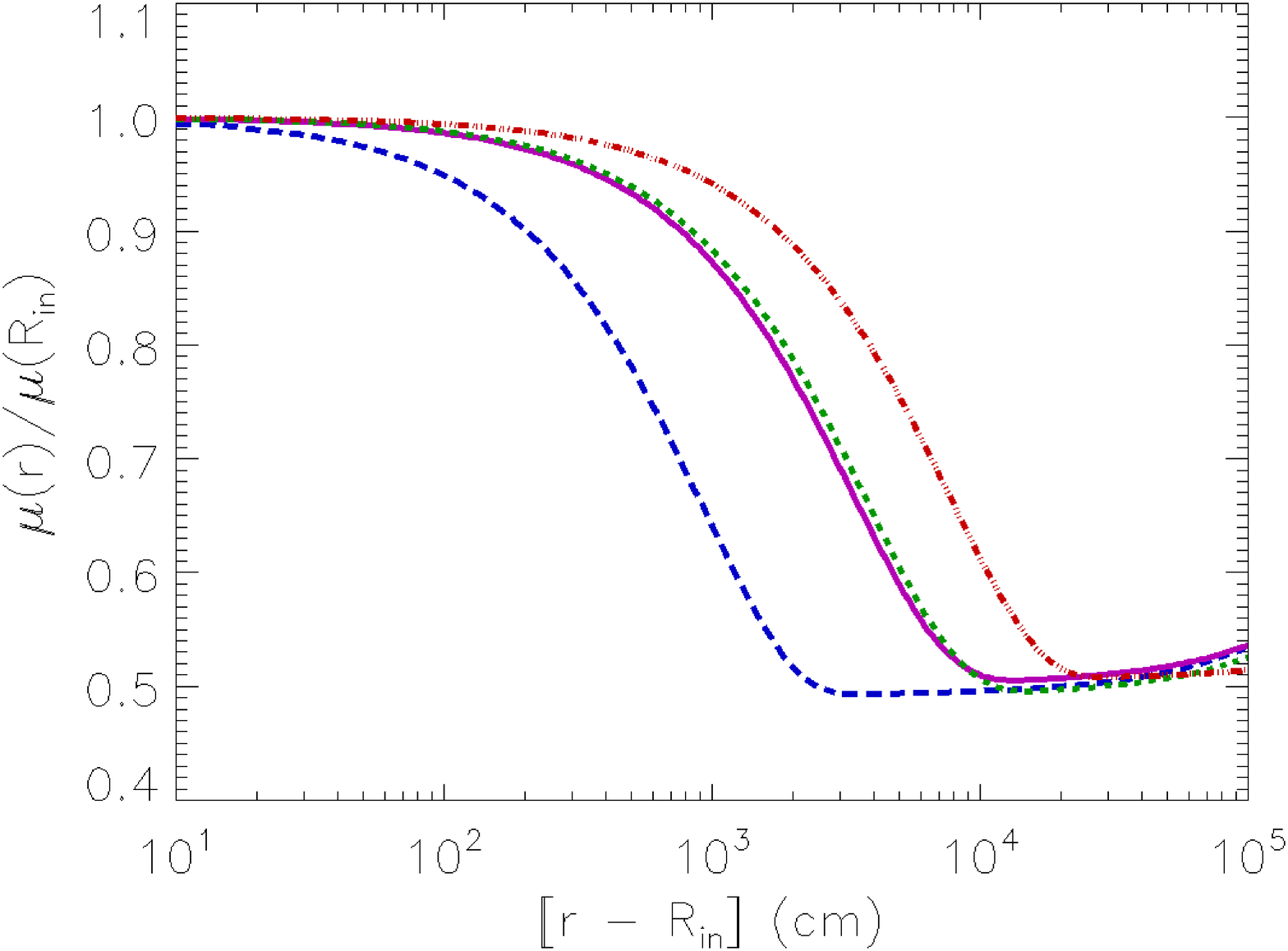}
}
\subfigure
{
	\includegraphics[width=84mm, height=65mm]{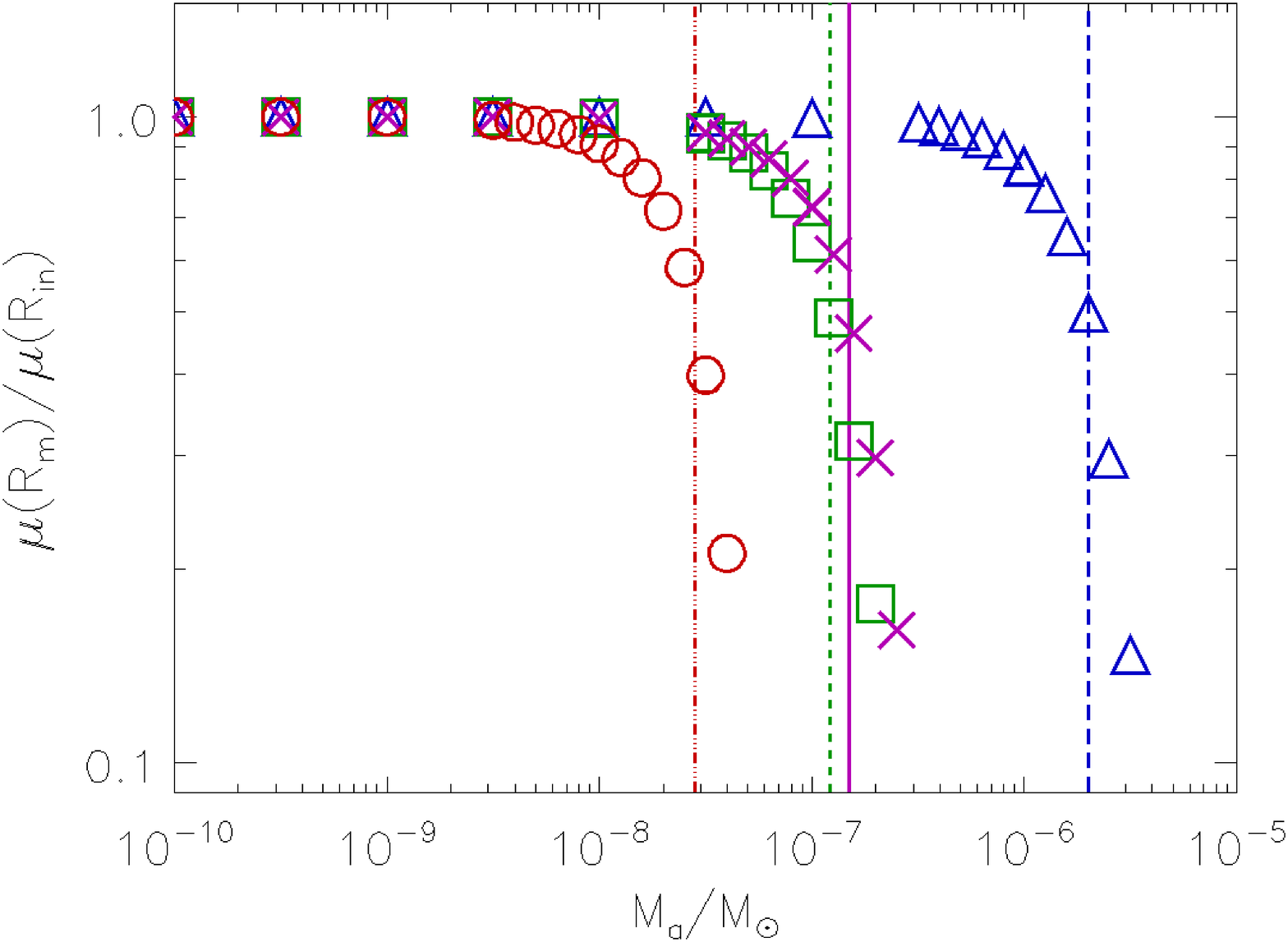}
}
\subfigure
{
	\includegraphics[width=84mm, height=65mm]{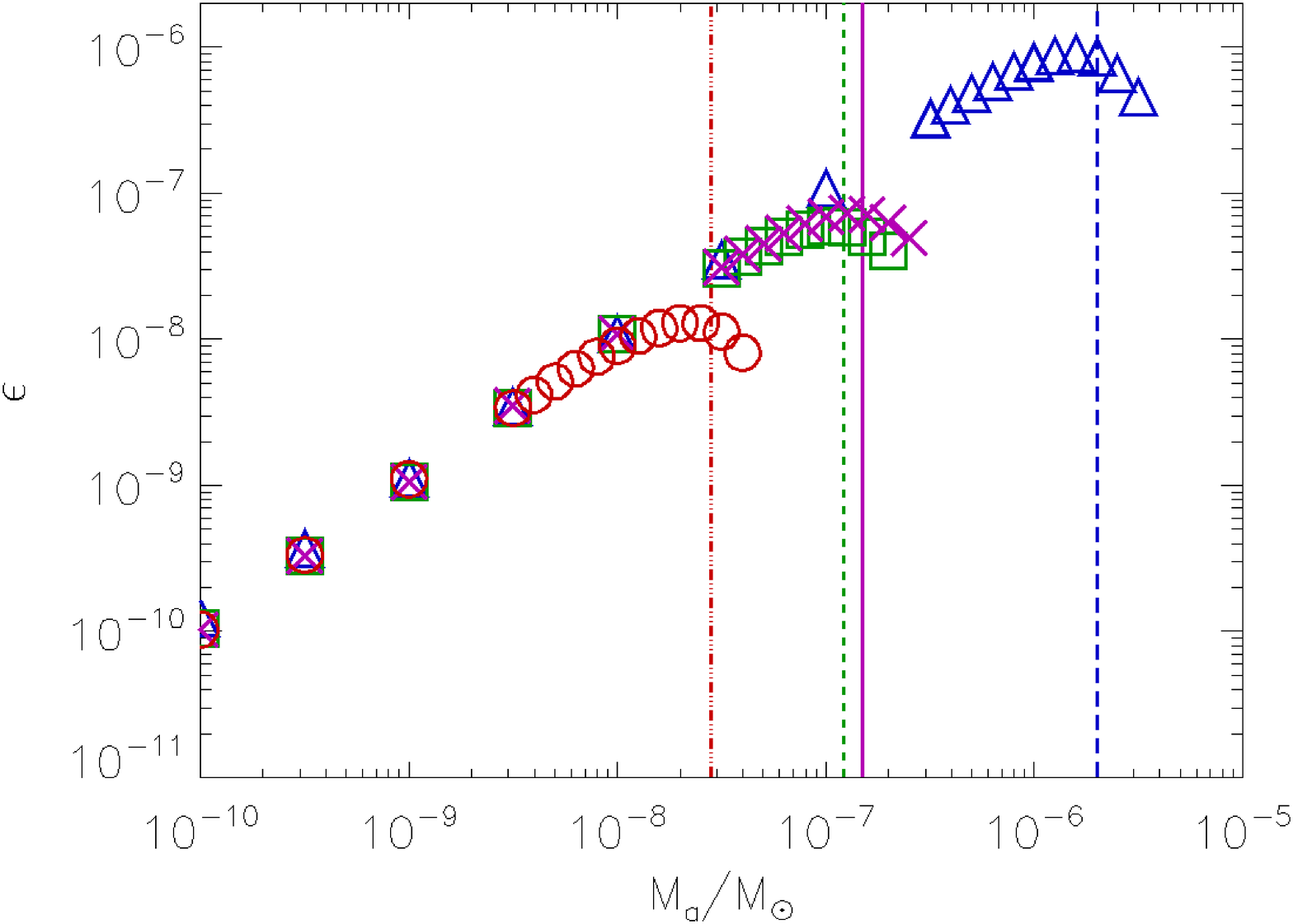}
}
\subfigure
{
	\includegraphics[width=84mm, height=65mm]{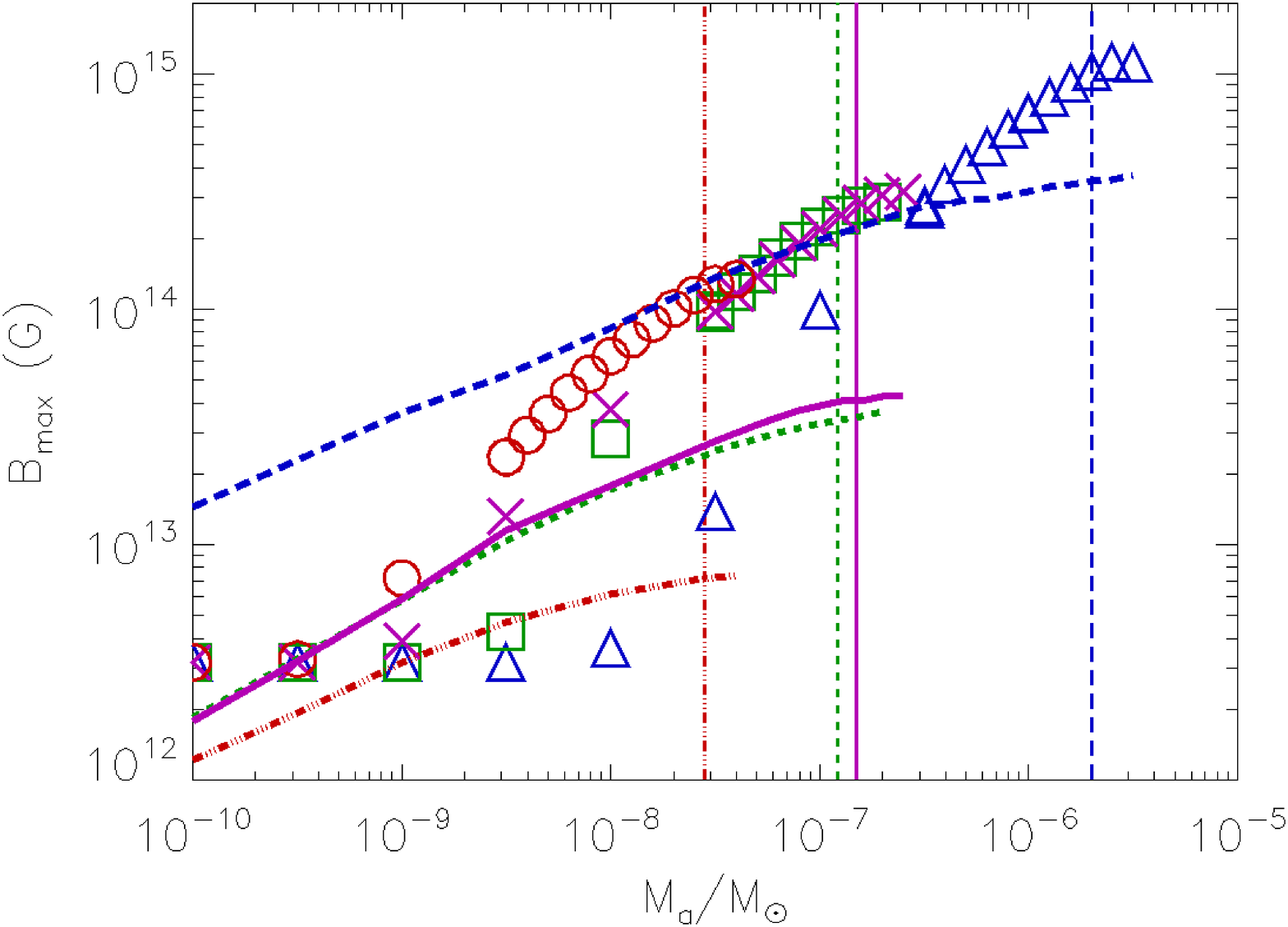}
}
\subfigure
{
	\includegraphics[width=84mm, height=65mm]{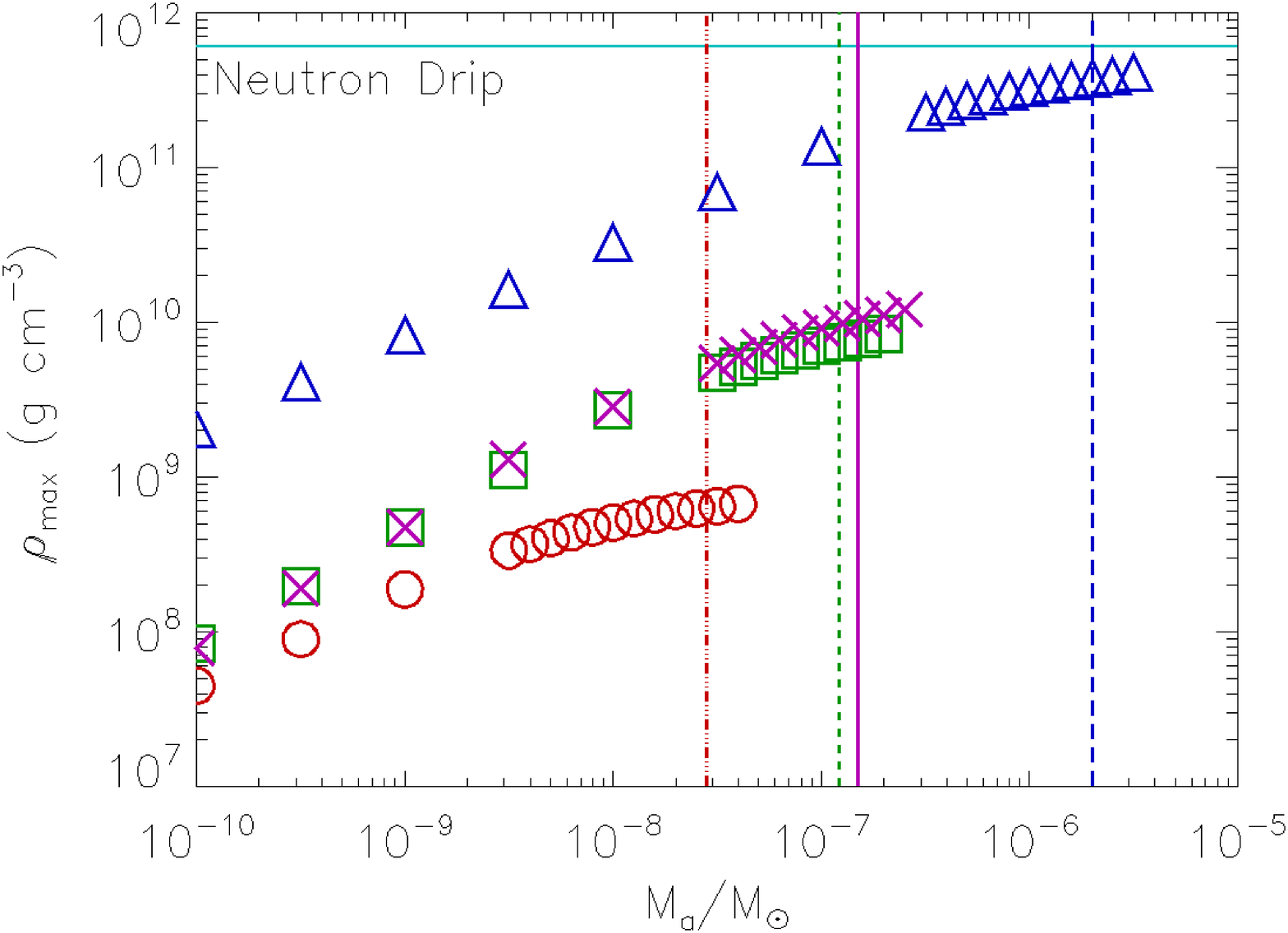}
}
\subfigure
{
	\includegraphics[width=84mm, height=65mm]{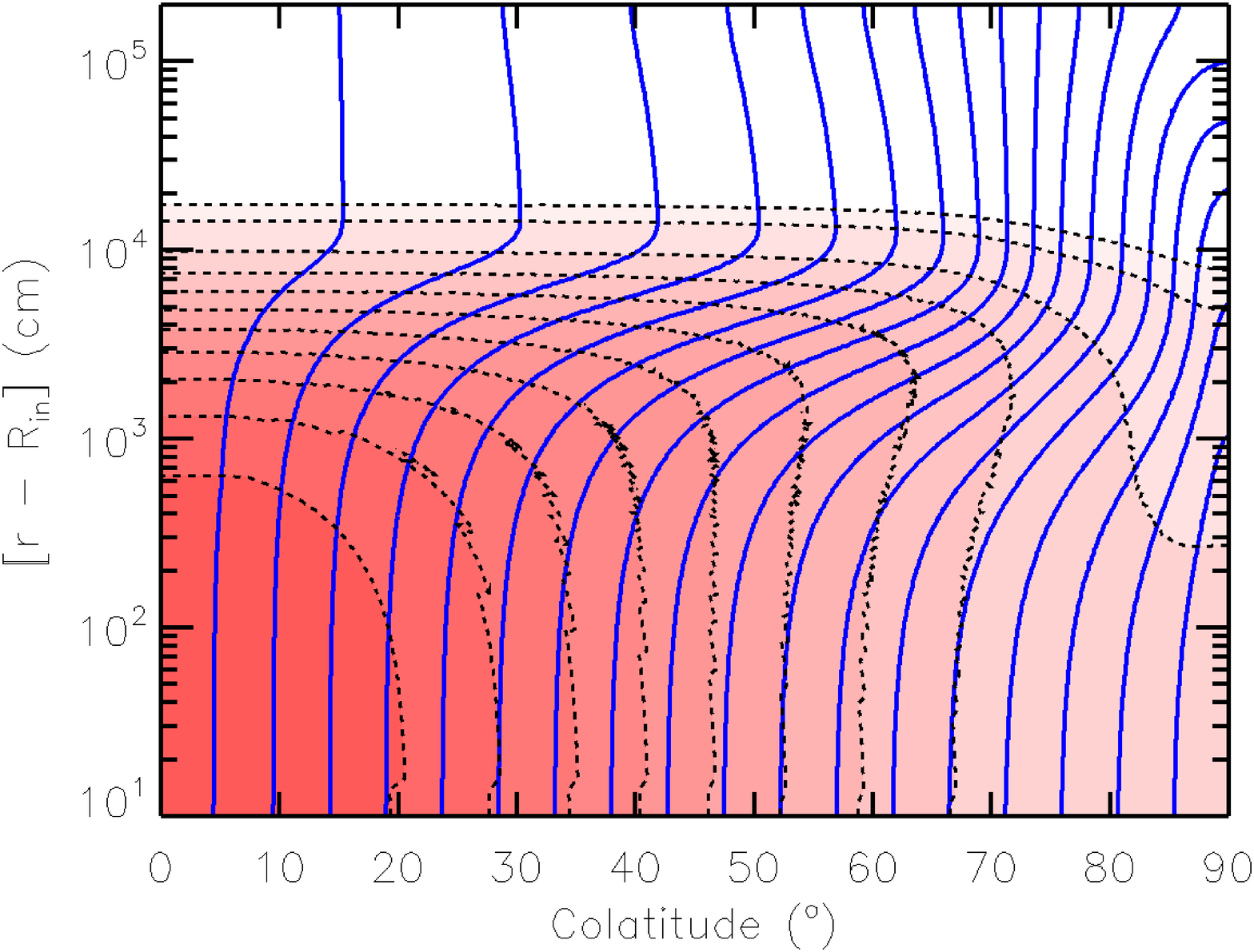}
}
\caption{Magnetic dipole moment $\mu$, calculated as a function of altitude (in
centimetres) from the inner boundary of the computational mesh $R_{\mathrm{in}}$ and
normalized to its surface value (top-left panel), for the characteristic masses
$M_{\mathrm{c}}$, measured in solar masses, for models B (triple-dot--dashed red curve), C (short-dashed green curve), D
(long-dashed blue curve) and E (solid purple curve). Also plotted are: normalized magnetic dipole
moment $\mu$ (top-right panel), mass ellipticity $\epsilon$ (middle-left panel),
maximum field strength $|\bmath{B}|_{\mathrm{max}}$ in gauss (middle-right panel) and
maximum density $\rho_{\mathrm{max}}$ in units of $\mathrm{g} \
\mathrm{cm}^{-3}$ (bottom-left panel), as a function of accreted mass $M_{\mathrm{a}}$,
measured in solar masses, for models B (red circles), C (green squares), D (blue
triangles) and E (purple crosses). Values of the characteristic masses $M_{\mathrm{c}}$ are plotted as vertical lines for models B (triple-dot--dashed line), C (short-dashed line), D (long-dashed line), 
E (solid line), and coloured accordingly. Overplotted in the bottom-left panel is the line of the neutron drip density $\rho_{\mathrm{ND}} = 6 \times
10^{11} \ \mathrm{g} \ \mathrm{cm}^{-3}$.
The hydromagnetic structure of model E is displayed in the bottom-right panel at
$M_{\mathrm{a}} = M_{\mathrm{c}} = 1.5 \times 10^{-7} \mathrm{M}_{\sun}$, showing magnetic field lines
(solid blue curves) and isodensity contours (dashed black curves). Density contours
are drawn for $\eta \rho_{\mathrm{max}}$ (maximum at the pole), with
$\rho_{\mathrm{max}} = 1.0 \times 10^{10} \ \mathrm{g} \ \mathrm{cm}^{-3}$, and
$\eta = 0.9, \ 0.8, \ 0.7, \ 0.6, \ 0.5, \ 0.4, \ 0.3, \ 0.2, \ 0.1, \ 10^{-2},
\ 0$.}
\label{fig:comparison}
\end{minipage}
\end{figure*}

We simulate magnetic mountains with the EOS of model E for $10^{-10} \lesssim
M_{\mathrm{a}}/\mathrm{M}_{\sun} \lesssim 10^{-7}$ by utilizing $K_{\mathrm{eff}}(M_{\mathrm{a}})$ and
$\Gamma_{\mathrm{eff}}(M_{\mathrm{a}})$ given in Fig. \ref{fig:K_Gamma_mass}. The dipole
moment $\mu$, ellipticity $\epsilon$, maximum magnetic field strength
$|\bmath{B}|_{\mathrm{max}}$ and maximum density $\rho_{\mathrm{max}}$ of model E are
compared with those of models B, C and D in Fig. \ref{fig:comparison}. The
hydromagnetic equilibrium for model E is also plotted in Fig. \ref{fig:comparison} for $M_{\mathrm{a}} = M_{\mathrm{c}}$ (cf.
Fig. \ref{fig:magnetic_mountain}). As the partial pressures are dominated by
relativistic degenerate electrons at $10^{-8} \lesssim M_{\mathrm{a}}/\mathrm{M}_{\sun} \lesssim
M_{\mathrm{c}}$, $\mu$, $\epsilon$, $|\bmath{B}|_{\mathrm{max}}$ and $\rho_{\mathrm{max}}$ in
model E behave like in model C at these accreted masses. At $M_{\mathrm{a}} \lesssim
10^{-8} \ \mathrm{g} \ \mathrm{cm}^{-3}$, $|\bmath{B}|_{\mathrm{max}}$ of model B
approaches that of model C due to the presence of non-relativistic degenerate
electron gas. Relativistic degenerate electrons dominate model E for $M_{\mathrm{a}} \geq
M_{\mathrm{c}}$, so model C can be used to approximate the realistic mountain for $M_{\mathrm{a}}
\lesssim 10^{-2} \mathrm{M}_{\sun}$. For $M_{\mathrm{a}} \gtrsim 10^{-2} \mathrm{M}_{\sun}$ (e.g. in
LMXBs), the dominant partial pressure comes from degenerate non-relativistic
neutrons (model D).

For the range of natal magnetic fields $10^{12} \leq B_{\ast}/(\mathrm{G})
\leq 10^{13}$ inferred by \citet{faucher-giguere2006}, $M_{\mathrm{c}}$ stays within the
range where model C applies [see equations (\ref{characteristic_mass_adiabatic})--(\ref{explicit_integrand}) of Appendix
\ref{appendix:gs_analytic}, which give the model-dependent scaling of $M_{\mathrm{c}}$ with respect to $B$ and $M_{\mathrm{a}}$]. If $B_{\ast}$ rises to $\sim 10^{15} \ \mathrm{G}$,
appropriate for a magnetar, $M_{\mathrm{c}}$ increases from $1.2\times10^{-7}$
to $2.6\times10^{-4} \mathrm{M}_{\sun}$ for model C, still below where degenerate neutron
pressure dominates (see Fig. \ref{fig:K_Gamma_mass}). We can therefore use model
C to calculate $M_{\mathrm{c}}$ under all plausible astrophysical scenarios.

\section{Application to gravitational-wave spin stalling}
\label{section_6}

Several mechanisms can brake the spin-up of an accreting neutron star: the
magnetospheric centrifugal barrier \citep{illarionov1975, ghosh1979},
GW emission \citep{wagoner1984, bildsten1998} and the
magnetic-dipole torque \citep{ostriker1969}. Every one of these mechanisms
eventually balances the accretion torque and stalls the spin-up process, when
the spin frequency $\nu_{\mathrm{s}}$ is large enough. We use equation (\ref{spin_equilibrium}) for spin balance which assumes the usual thin-disc accretion model \citep{bildsten1998}. It should be noted that this is not necessarily valid, 
as more refined accretion models weaken the spin-up torque or strengthen the propeller effect, thus obviating the need for a strong GW torque. 
The feedback provided by radiation pressure in rapidly accreting systems could lead to a thick and sub-Keplerian inner accretion disc, which modulates the accretion torque of the standard thin-disc model \citep{andersson2005}. 
Also, for weak accretors, if the magnetospheric radius becomes larger than the corotation radius, the star can exist in either a strong or weak `propeller' phase 
(see \citet{romanova2008} and references therein), with the transition between these phases being strongly dependent on the kinematic viscosity and magnetic diffusivity of the accreting matter \citep{romanova2004, romanova2005}.
Nevertheless, these improved accretion models do not invalidate any of the proposed GW-generating mechanisms.

In this section, we investigate how the stalling frequency depends on the EOS, if all the braking comes from gravitational radiation reaction. In this work, we do not consider radiation-pressure feedback on the accretion disc since we are interested in modelling moderately 
accreting LMXBs where this effect is small. Also, in the vicinity of the bottom magnetic field [$10^{7} - 10^{8} \ \mathrm{G}$; see \citet{van_den_Heuvel1995} and
\citet{zhang2006}], where the magnetosphere touches the stellar surface and the propeller effect can be neglected, the
GW torque dominates the magneto-centrifugal and magnetic-dipole torques. Clearly, this approach yields an upper bound on $\nu_{\mathrm{s}}$; the other mechanisms can lower
$\nu_{\mathrm{s}}$ further.

We synthesize five Monte Carlo populations of LMXBs, whose spins are such that
their gravitational radiation reaction torque exactly balances the accretion
torque. We assume that each simulated LMXB population undergoes magnetic burial
according to one of the five EOS in Table \ref{table:eos}. The number of neutron
stars in each population is chosen large enough $(\sim10^{5})$ to yield an
accurate cumulative spin distribution. We assume fiducial neutron star
parameters (see Section \ref{section_3}) and solve
\begin{equation}
\nu_{\mathrm{s}} = 2.09\times10^{1} \ \mathrm{Hz} \ \Bigg( \frac{\dot{M}}{10^{-10}
\mathrm{M}_{\sun} \ \mathrm{yr}^{-1}} \Bigg)^{1/5} \Bigg( \frac{\epsilon}{10^{-5}}
\Bigg)^{-2/5},
\label{spin_equilibrium}
\end{equation}
for the equilibrium spin frequency, assuming the wobble angle $\alpha$ tends to
$\alpha = \pi/2$ due to GW back reaction \citep{cutler2002} or
crust--core coupling \citep{alpar1988}. The accretion rates are selected from the
empirical luminosity function of Galactic LMXB sources \citep{grimm2002}, 
\begin{align}
N(>L) = 105 \Bigg[ \Bigg( \frac{L}{10^{36} \ \mathrm{erg} \ \mathrm{s}^{-1}} \Bigg)^{-0.26} - \Bigg( \frac{L_{\mathrm{max}}}{10^{36} 
\ \mathrm{erg} \ \mathrm{s}^{-1}} \Bigg)^{-0.26} \Bigg],
\end{align}
where $L$ is the apparent luminosity in the $2-10 \ \mathrm{keV}$ band, and
$L_{\mathrm{max}}$ is the cut-off luminosity, combined with the
luminosity-dependent mass fraction of the Galaxy which is visible to the \textit{RXTE}
All-Sky Monitor [see fig. 11 of \citet{grimm2002}]. The long-term average
bolometric luminosity is related crudely to the accretion rate by the familiar
expression.
\begin{equation}
\dot{M} \approx L R_{\ast} / G M_{\ast}.
\label{accretion_rate}
\end{equation}

\begin{figure*}
\begin{minipage}{175mm}
\subfigure
{
	\includegraphics[width=84mm, height=65mm]{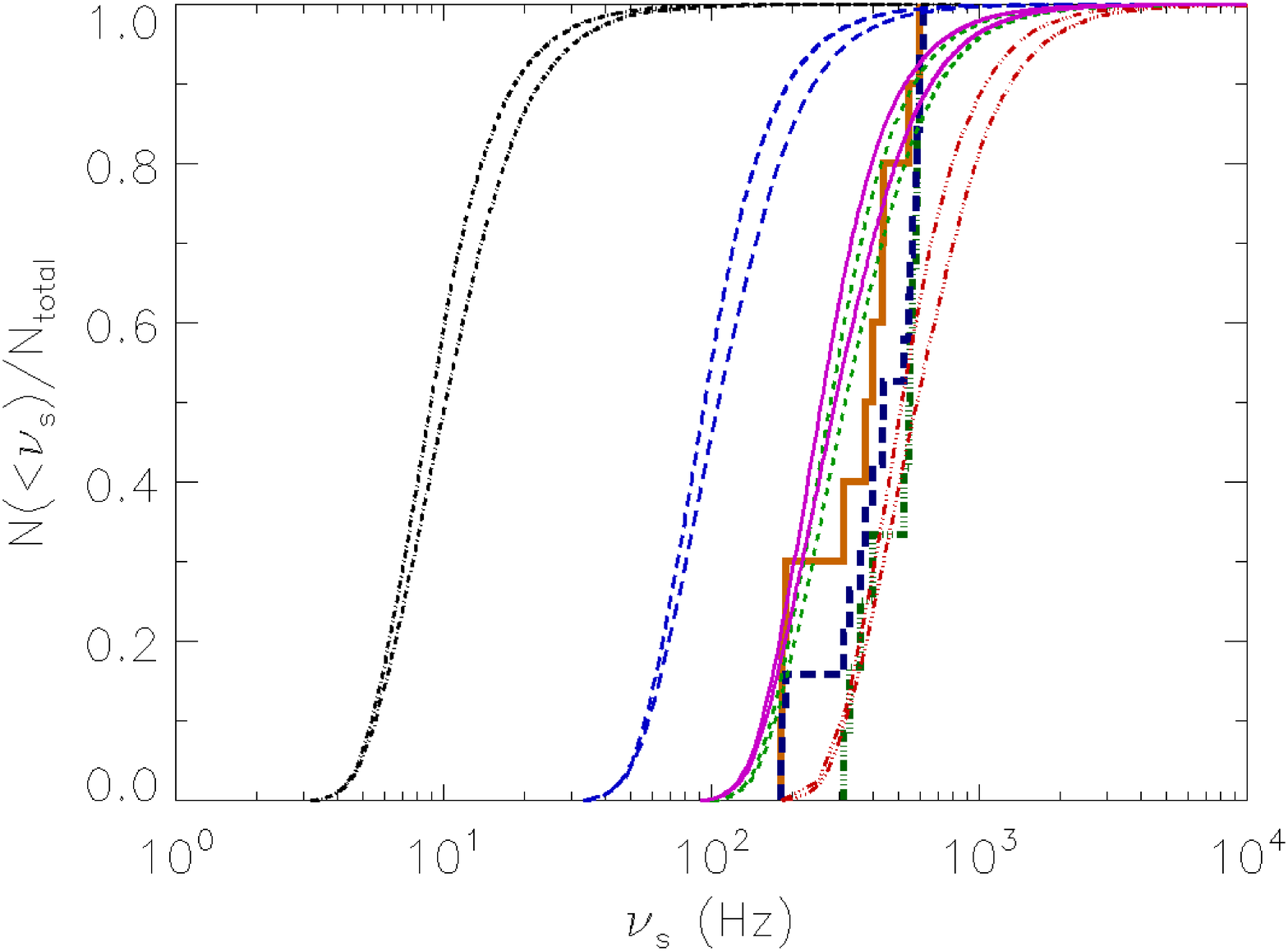}
}
\subfigure
{
	\includegraphics[width=84mm, height=65mm]{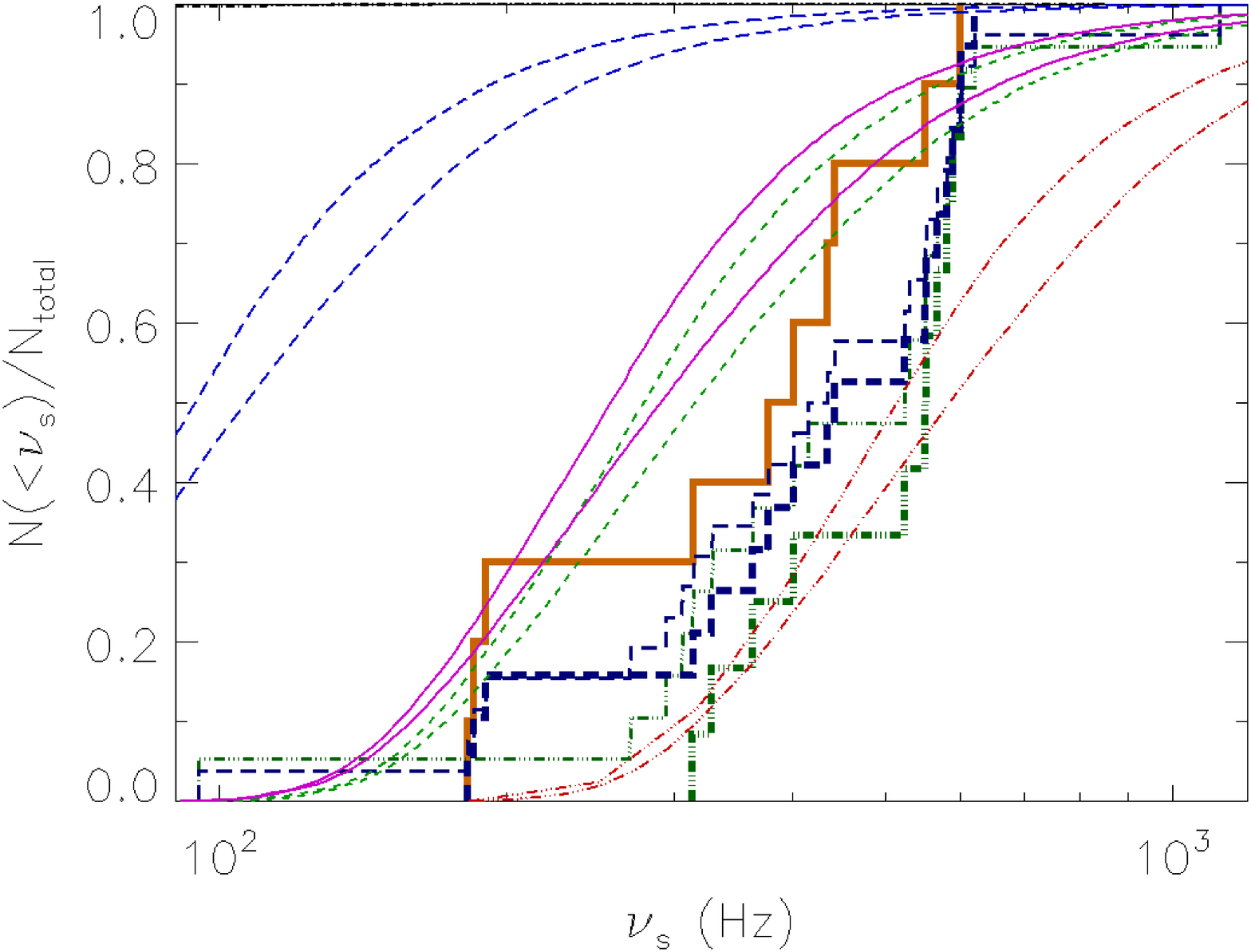}
}
\caption{Observed cumulative spin distributions of all NMPs (thin triple-dot--dashed green line), confirmed
NMPs (thick triple-dot--dashed green line), AMPs (thick orange line), all AMXPs (thin dashed blue line)
and confirmed AMXPs (thick dashed blue line), together with the theoretical
distributions predicted by model A (dot--dashed black curves), model B (triple-dot--dashed red curves),
model C (short-dashed green curves), model D (long-dashed blue curves) and model E (solid
purple curves). The theoretical curves are based on the GW 
stalling mechanism of Bildsten (1998), with the saturation ellipticities
$\epsilon(M_{\mathrm{a}} = M_{\mathrm{c}})$ for each of the models in Table 1. Two
luminosity cut-offs are considered: $L_{\mathrm{\mathrm{max}}} =
2.7\times10^{38} \ \mathrm{erg} \ \mathrm{s}^{-1}$ (rightmost theoretical curves) and
$L_{\mathrm{max}} = 3.2\times10^{37} \ \mathrm{erg} \ \mathrm{s}^{-1}$ (leftmost theoretical curves). The right-hand panel zooms into the range $90 \leq
\nu_{\mathrm{s}}/\mathrm{Hz} \leq 1200$ in the left-hand panel and displays the unconfirmed sources as well.} 
\label{fig:cumulative_spins}
\end{minipage}
\end{figure*}

The results of the Monte Carlo simulations are shown in Fig. \ref{fig:cumulative_spins}, where we compare the cumulative distribution function of our spin-equilibrium models with the observed distribution of
nuclear-powered millisecond pulsars (NMPs) (i.e. sources that show brightness oscillations in the tails of Type I X-ray bursts), accretion-powered
millisecond pulsars (AMPs) (i.e. sources that exhibit X-ray pulsations) and accreting millisecond X-ray pulsars (AMXPs) (i.e.
sources that exhibit either millisecond burst oscillations, X-ray pulsations or both). We obtain data on the spins of these objects from table 1 of \citet{watts2008}. To be consistent with contemporary 
literature on millisecond X-ray binaries \citep{chakrabarty2003, galloway2008a}, 
we adopt the following naming convention for these sources: accreting millisecond pulsars are AMPs, burst oscillation sources are NMPs, and we combine 
these two populations into AMXPs\footnote{By contrast, the naming convention used by \citet{watts2008} reflects how the spins are inferred observationally: accreting millisecond pulsars, burst oscillation sources and quasi-periodic oscillation sources.}. 
To distinguish between the confirmed and unconfirmed sources, we plot all/confirmed NMPs (thin/thick triple-dot--dashed green lines), AMPs (thick orange line) and all/confirmed AMXPs (thin/thick dashed blue lines). 
Curves represent cumulative distribution functions of models A (dot--dashed black curves), B (triple-dot--dashed red curves), C (short-dashed green curves), D (long-dashed blue curves) and E (solid purple curves). We
update the spin of EXO 0748--676 from $45$ to $552 \ \mathrm{Hz}$ \citep{galloway2010}, and we do not discriminate between intermittent pulsars
and AMPs (i.e. those sources which exhibit intermittent or persistent X-ray pulsations during outburst, respectively).

The luminosity function is defined for the \textit{RXTE} All-Sky Monitor catalogue ($2-10
\ \mathrm{keV}$ band), which is flux-limited below $\sim 10^{35} \ \mathrm{erg}
\ \mathrm{s}^{-1}$ \citep{grimm2002}. Two maximum luminosity cut-offs are
investigated, namely $L_{\mathrm{max}} = 2.7\times10^{38} \ \mathrm{erg} \
\mathrm{s}^{-1}$ (to include the most luminous LMXB Sco X-1) and
$3.2\times10^{37} \ \mathrm{erg} \ \mathrm{s}^{-1}$ (most luminous AMP Aql
X-1), encompassing the luminosity range of all confirmed and unconfirmed AMXPs.
All sources are assumed to follow the same power-law scaling of the luminosity
function. 

The All-Sky Monitor underestimates the true bolometric luminosity, and
hence the accretion rate, due to the presence of significant hard X-ray tails
$\gtrsim 10 \ \mathrm{keV}$ in LMXB X-ray spectra \citep{barret2001}. Although
this can be corrected \citep{galloway2008b}, we do not attempt to do so here,
because equation (\ref{accretion_rate}) is approximate anyway, equation
(\ref{spin_equilibrium}) depends weakly on $\dot{M}$, and the bolometric
correction factors differ by up to $\approx 40$ per cent between sources. 

Considering typical LMXB lifetimes of $\sim 10^{8} \ \mathrm{yr}$ \citep{podsiadlowski2002}, 
the accreted masses in these systems are evaluated to be in the range of $10^{-4} \lesssim
M_{\mathrm{a}}/\mathrm{M}_{\sun} \lesssim 10^{1}$. Therefore, enough matter has been transferred
in these systems to reach the characteristic masses and saturation ellipticities
for the models in Table \ref{table:eos}, given initial magnetic fields of
$10^{12.5} \ \mathrm{G}$. Hence, for each simulated LMXB population, we assign
the ellipticities of the neutron stars to be the saturation values for the
respective EOS in Table \ref{table:eos}.

From Fig. \ref{fig:cumulative_spins}, we see that an isothermal magnetic
mountain (model A) stalls the star at $\nu_{\mathrm{s}} \sim 1 \ \mathrm{Hz} \
(B_{\ast}/10^{12.5} \ \mathrm{G})^{-4/5}$, where the $B_{\ast}$ scaling follows
from $M_{\mathrm{c}} \propto B_{\ast}^{2}$ of equation (30) in PM04 and equation
(\ref{spin_equilibrium}). One would therefore need $B_{\ast} \approx 10^{10} \ \mathrm{G}$ to fit 
the observed spin distribution, contradicting population synthesis studies of isolated pulsars
\citep{hartman1997, arzoumanian2002, faucher-giguere2006}. Adiabatic magnetic
mountains (models B--E) are generally in better agreement with the observed spin
distribution. In fact, models B, C and E produce a good fit to all of the
observed spin distributions. Equation (\ref{characteristic_mass_adiabatic}) in
Appendix \ref{appendix:gs_analytic} for $M_{\mathrm{c}}(B_{\ast})$ implies
$\nu_{\mathrm{s}} \propto B_{\ast}^{-4/9}$ for models B and D, and
$\nu_{\mathrm{s}} \propto B_{\ast}^{-8/15}$ for model C. Thus, a better fit to 
the empirical spin distributions can be obtained for models B and C if the fiducial magnetic field 
in the range of $10^{12}-10^{13} \ \mathrm{G}$, rather than $10^{12.5} \ \mathrm{G}$, is considered. 
Although model D cannot match the observed spin distribution in this range, it is possible that Ohmic 
diffusion can improve the agreement by allowing the mountain to spread, resulting in a lower saturation 
ellipticity and hence higher equilibrium spin frequencies.

It appears that the equilibrium spin frequencies of confirmed NMPs are systematically
higher than those of AMPs; their cumulative distributions are offset to the
right and left of the AMXP distribution, respectively (see Fig.
\ref{fig:cumulative_spins}). This is qualitatively consistent with the GW spin
stalling mechanism, as the median time-averaged accretion
luminosities of NMPs are $\sim 20$ times higher than
those of AMPs, resulting in higher equilibrium spin frequencies by a factor of
$\approx (20)^{1/5} \approx 1.8$ (under the assumption of similar ellipticities in these systems).
This roughly corresponds to the frequency separation between the observed NMP and AMP distributions 
in Fig. \ref{fig:cumulative_spins}, supporting the GW spin stalling hypothesis. On the other hand, 
if outburst luminosities of these objects are considered instead, the separation in predicted equilibrium spin 
frequencies becomes negligible.

Another noteworthy feature of Fig. \ref{fig:cumulative_spins} is the steep
gradient of the observed distribution at $\nu_{\mathrm{s}} \approx 500 \
\mathrm{Hz}$ (D. K. Galloway, private communication). The theoretical curves for models A--E can reproduce the shape of the 
distribution for $\nu_{\mathrm{s}} \lesssim 400 \ \mathrm{Hz}$. For the range of $L_{\mathrm{max}}$ investigated, 
theoretical curves do not rise steeply enough to fit the higher-frequency ($\nu_{s} \gtrsim 400 \ \mathrm{Hz}$) end of the distribution. This is a problem for stalling models in
general, not just magnetic mountains; the $\epsilon^{-2/5}$ scaling in equation (\ref{spin_equilibrium}) 
is too gentle. The observed steepening could be caused by differences between the luminosity functions of Galactic LMXB sources and AMPs/NMPs. Allowing for a realistic distribution of saturation
ellipticities (e.g. due to a lognormal natal magnetic field distribution of
isolated pulsars, predicted by population synthesis studies) worsens the
steepening problem, if the luminosity function is assumed to be independent of
the magnetic field. It is possible that another mechanism (such as the
`propeller' effect) sets $\nu_{\mathrm{s}}$, but its dependence on underlying
variables (i.e. $\nu_{\mathrm{s}} \propto B_{\ast}^{-6/7} \dot{M}^{3/7}$) is
even gentler than gravitational radiation reaction. 
We defer a full investigation of this puzzle to a future paper.

\section{Discussion}
\label{section_7}

Magnetic burial in accreting neutron stars has several important astrophysical
consequences. It creates a significant mass quadrupole moment, which potentially
stalls the spin-up of an LMXB by gravitational radiation reaction. It also
reduces the magnetic dipole moment, in accord with the observed $\mu$ versus $M_{\mathrm{a}}$
relation in neutron star binaries presented in fig. 2 in \citet{taam1986}. In the
context of the statistical evidence against field decay over $10^{6} - 10^{7} \
\mathrm{yr}$ in isolated pulsars \citep{bhattacharya1992, lorimer1997}, magnetic
burial can be invoked to explain both the low magnetic fields in LMXBs and
millisecond pulsars \citep{chanmugam1992, lamb2005, zhang2006} and the observed spin
distribution of LMXBs \citep{chakrabarty2003}. However, before magnetic burial
is deemed a viable explanation for the above phenomena, the effect of the EOS on the burial process must
be quantified.

In this paper, we show that the effect of the EOS is large. Magnetic burial is
more effective for $4/3 \leq \Gamma \leq 5/3$ than for $\Gamma = 1$, in the
sense that less matter must be accreted in the former case than in the latter in
order to achieve the same amount of magnetic dipole screening. For the EOS
listed in Table \ref{table:eos}, $M_{\mathrm{c}}$ decreases from $5.2\times10^{-4}
\mathrm{M}_{\sun}$ (model A) to $2.8\times10^{-8} \mathrm{M}_{\sun}$ (model B), $1.2\times10^{-7}
\mathrm{M}_{\sun}$ (model C), $2.0\times10^{-6} \mathrm{M}_{\sun}$ (model D) and
$1.5\times10^{-7} \mathrm{M}_{\sun}$ (model E), for $B = 10^{12.5} \ \mathrm{G}$.
Likewise, the saturation ellipticities decrease from $3.2\times10^{-4}$ (model
A) to $1.3\times10^{-8}$ (model B), $6.0\times10^{-8}$ (model C),
$9.0\times10^{-7}$ (model D) and $7.3\times10^{-8}$ (model E). This is a
general result, applicable to a variety of scenarios where magnetic confinement
of accreted matter can occur, such as T Tauri stars \citep{bertout1988,
hartmann1998}, young neutron stars accreting from a fallback disc
\citep{chatterjee2000, wang2006} and magnetic white dwarfs \citep{king1979, wickramasinghe2000}. 
The characteristic mass scales quadratically with the magnetic field strength in all models but with different powers of the
accreted mass: we have $M_{\mathrm{c}} \propto B^{2} M_{\mathrm{a}}^{\beta}$, with $\beta = 0$
(model A), $-4/5$ (model B), $-1/2$ (model C) and $-4/5$ (model D) (see
Appendix \ref{appendix:gs_analytic}). The maximum density at the base of an
adiabatic mountain satisfies $\rho_{\mathrm{max}} \ll 10^{14} \ \mathrm{g} \
\mathrm{cm}^{-3}$, unlike for isothermal mountains, where it is unrealistically
high. We find that crustal cracking occurs as burial proceeds, because the yield magnetic field strength
is typically surpassed in non-isothermal models.

A Monte Carlo analysis of neutron stars in LMXBs, with $\dot{M}$ drawn from an empirical distribution 
and $B_{\ast}$ set to the fiducial $10^{12.5} \ \mathrm{G}$, shows that models B, C and E yield 
$100 \lesssim \nu_{\mathrm{s}}/(\mathrm{Hz}) \lesssim 600$ within the gravitational 
spin-equilibrium scenario \citep{bildsten1998}. This is in accord with the $\approx 180$--$620 \ \mathrm{Hz}$ 
confirmed spins of AMXPs. Model D predicts $50 \lesssim \nu_{\mathrm{s}}/(\mathrm{Hz}) \lesssim 300$, 
slightly too low to explain the data. In comparison, the isothermal magnetic mountain (model A) does not agree with 
the data at all, yielding $\nu_{\mathrm{s}}$ values $1$ order of magnitude lower than those from model D.

\begin{figure}
\includegraphics[width=84mm]{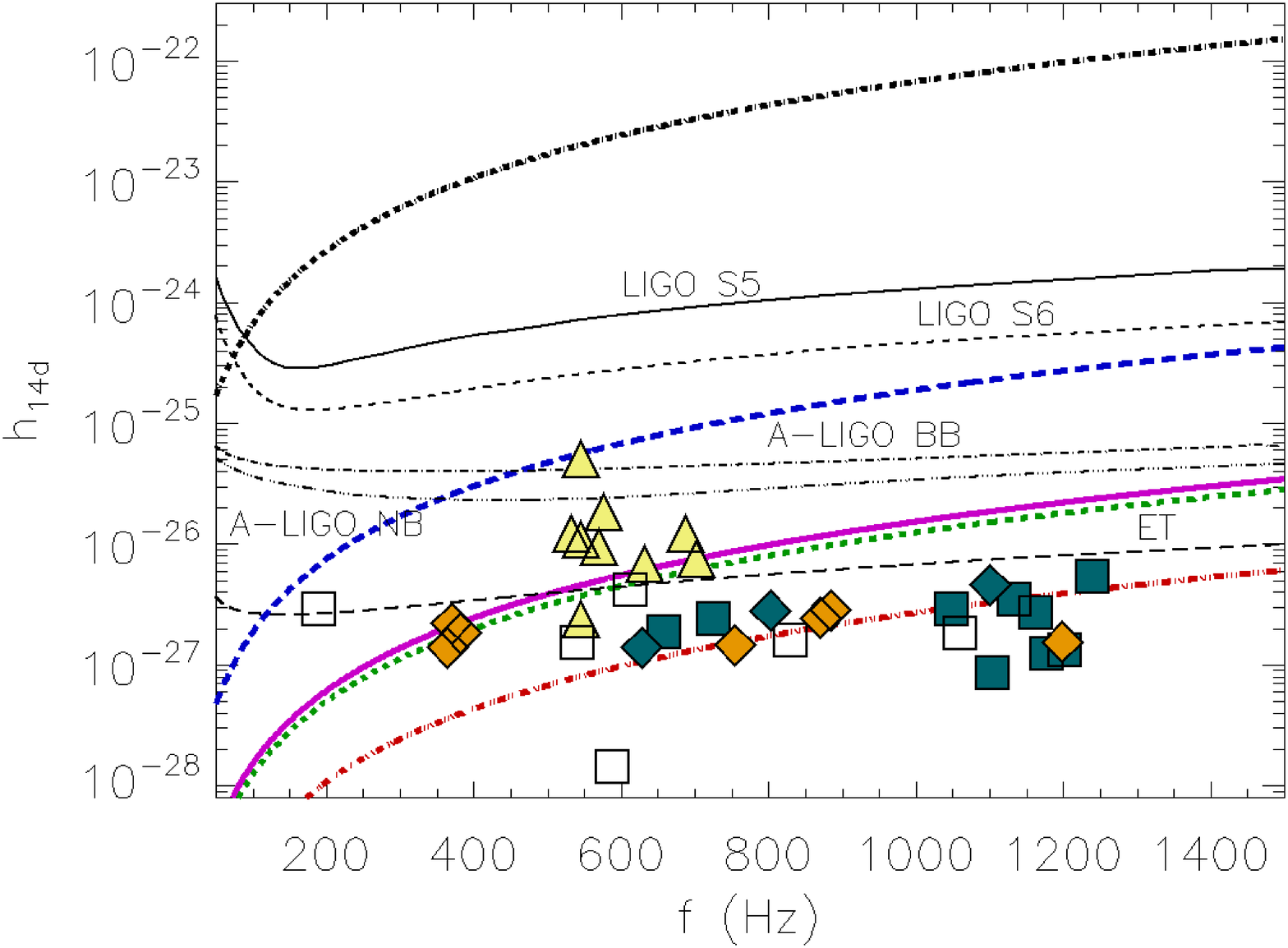}
\caption{Amplitude of the GW signal $h_{0}$ (dimensionless)
versus signal frequency $f (\mathrm{Hz})$ for magnetic mountains with saturation ellipticities $\epsilon = 3.2\times10^{-4}$ (model A) (thick
dot--dashed black curve), $1.3\times10^{-8}$ (model B) (thick triple-dot--dashed red curve), $6.0\times10^{-8}$
(model C) (thick short-dashed green curve), $9.0\times10^{-7}$ (model D) (thick long-dashed blue curve),
$7.3\times10^{-8}$ (model E) (thick solid purple curve), computed for natal magnetic fields
of $10^{12.5} \ \mathrm{G}$. Also plotted as points are the wave strain upper limits from Bildsten (1998) for AMPs (orange diamonds), confirmed NMPs (teal
squares), unconfirmed NMPs (unfilled squares), QPO sources (yellow triangles),
and sources exhibiting both pulsations and burst oscillations (teal diamonds), computed from observed X-ray fluxes. The sensitivities of LIGO S5 (thin solid curve), LIGO S6 (thin short-dashed
curve), Advanced LIGO in the broad-band configuration (thin dot--dashed curve), lower
envelope of Advanced LIGO in the narrow-band configuration (thin triple-dash-dotted
curve) and the proposed conventional Einstein Telescope (thin long-dashed curve)
configuration, assuming a feasible $14$ d coherent integration and a signal-to-noise ratio of 11.4, are overplotted.}                   
\label{fig:detectability}
\end{figure}

We compute the magnitude of the GW strain $h_{0}$ of the AMXPs and quasi-periodic
oscillation (QPO) sources by applying the gravitational spin-equilibrium
argument of \citet{bildsten1998} to the sources in table 1 in \citet{watts2008}.
Here, we differentiate between the confirmed and unconfirmed sources, as well as
AMPs, NMPs and sources that exhibit both persistent pulsations and burst
oscillations. The results for AMPs (orange diamonds), confirmed NMPs (teal
squares), unconfirmed NMPs (unfilled squares), QPOs (yellow triangles) and
sources exhibiting both pulsations and burst oscillations (teal diamonds) are
shown on a wave strain $h_{0}$ versus wave frequency $f$ plot in Fig. \ref{fig:detectability}, where $f=2\nu_{\mathrm{s}}$.
The highest $f$ value considered here corresponds to $2\nu_{\mathrm{s,max}}$, where $\nu_{\mathrm{s,max}}=760\ \mathrm{Hz}$ 
is the maximum inferred spin in NMPs via Bayesian analysis \citep{chakrabarty2003}. When computing $h_{0}$, 
we assume that the transient sources are in torque balance during outburst. This is in accord with \citet{hartman2008}, 
who argued that SAX J1808.4--3658 is secularly spinning down between outbursts and is thus likely to
be in spin equilibrium during outburst\footnote{The transition between
the spin-up and spin-down episode within the 2002 outburst of SAX J1808.4--3658
found by \citet{burderi2006} is probably due to pulse shape changes.}.

The characteristic GW strain $h_{0}$ [defined in
\citet{jaranowski1998}] detectable by Laser Interferometer Gravitational Wave Observatory (LIGO) and the proposed Einstein Telescope
from a periodic source at a distance of $3$ kpc [representative of Sco X-1; see
\citet{bradshaw1999}] with a false alarm rate of $1$ per cent and a false dismissal
rate of $10$ per cent for a computationally feasible integration time of $14$ days is
overplotted in Fig. \ref{fig:detectability} for LIGO S5 (thin solid curve), LIGO S6
(thin short-dashed curve), Advanced LIGO in the broad-band configuration (thin dot--dashed
curve), lower envelope of Advanced LIGO in the narrow-band configuration
(thin triple-dot--dashed curve) and the proposed conventional\footnote{The xylophone configuration of the Einstein
Telescope closely matches the sensitivity of the conventional configuration at
frequencies $\gtrsim 30 \ \mathrm{Hz}$ \citep{hild2010}.} Einstein Telescope
(thin long-dashed curve) \citep{hild2011, watts2008, smith2009}. We also plot $h_{0}$ versus $f$ for neutron stars with
magnetic mountains at a distance of $5 \ \mathrm{kpc}$, with magnetic field of
$10^{12.5} \ \mathrm{G}$, for models A (thick dot--dashed black curve), B (thick triple-dot--dashed red curve), C (thick short-dashed green curve), D (thick long-dashed blue curve) and E
(thick solid purple curve).

Model A significantly overestimates $h_{0}$ with respect to both the interferometer sensitivity curves and the 
inferred \citet{bildsten1998} limits. In contrast, model E undercuts the \citet{bildsten1998} limit for QPO sources, 
implying either the natal magnetic fields of these sources are $\sim 10^{13.5} \ \mathrm{G}$, or that these objects are 
not in GW spin equilibrium. All the confirmed AMXPs and most of the unconfirmed AMXPs are consistent with 
model E. They lie below the model E curve either because they have $B_{\ast} < 10^{12.5} \ \mathrm{G}$ or because 
Ohmic diffusion prevents the ellipticity from saturating. We note that the current magnetic mountain models are still preliminary. Effects that have not yet been modelled faithfully 
in the context of magnetic burial may modify the saturation ellipticities. Therefore, it is still premature to quantify the
absolute detectability of magnetic mountains as GW sources.

There have been two directed searches for GWs from the accreting
neutron star Sco X-1 \citep{abbott2007a, abbott2007b}, which is expected to be
the strongest emitter of its class in the GW spin stalling
scenario \citep{bildsten1998}. The first, coherent search computed the
\textit{F}-statistic on 6 h of LIGO S2 data, coincident between the Hanford and
Livingston interferometers. Assuming a non-eccentric orbit, it placed a $95$ per cent
confidence upper limit on the GW strain from Sco X-1 of
$h_{0}=1.7\times10^{-22}$ in the $464$--$484 \ \mathrm{Hz}$ frequency band, and
$h_{0}=2.2\times10^{-22}$ in the $604$--$624 \ \mathrm{Hz}$ frequency band
\citep{abbott2007a}, which corresponds to an upper limit on the ellipticity of
the neutron star of $\epsilon \approx 4\times10^{-4}$. The second, semicoherent
search performed a radiometer analysis of $20$ days of triple-coincidence LIGO
S4 data. It yielded a $90$ per cent confidence upper limit of $h^{90\%}_{\mathrm{RMS}}
\approx 3.4\times10^{-24} (f/200 \ \mathrm{Hz})$ \citep{abbott2007b}. As
required by the non-detection of gravitational emission from accreting neutron
stars \citep{abbott2007a, abbott2007b}, adiabatic EOS reduce the GW detectability of magnetic mountains below the current detection threshold
of $h_{0} \approx 10^{-23}$. In comparison, the saturation ellipticities of
ideal isothermal magnetic mountains of model A are above this threshold and
should have already been detected.

The models in this paper are not the final word on magnetically confined
mountains. The range of accreted masses investigated here is well below $M_{\mathrm{a}}
\sim 10^{-1} \mathrm{M}_{\sun}$, the typical value for an LMXB \citep{burderi1999}, due
to numerical breakdown. If $\epsilon$ truly saturates for $M_{\mathrm{a}} \gg M_{\mathrm{c}}$,
then this failing is less serious for the GW applications than
for understanding $\mu(M_{\mathrm{a}})$, but it should be noted that the saturation
hypothesis has not been tested rigorously for $M_{\mathrm{a}} \gtrsim 10 M_{\mathrm{c}}$
\citep{payne2004, vigelius2009b}. A precise calculation of mountain equilibria
for an exact, depth-dependent nuclear EOS cannot be carried out within our
Grad--Shafranov formulation, although a relativistic degenerate electron EOS
(model C) is a fair approximation for $M_{\mathrm{a}} \approx M_{\mathrm{c}}$. The models in this paper are constructed on an impenetrable and EOS- and $M_{\mathrm{a}}$-dependent surface $R_{\mathrm{in}}$ within the crust, 
which prevents sinking past this boundary. \citet{wette2010} showed that, for isothermal mountains, sinking reduces
$\epsilon$ by up to $60$ per cent. In the presence of Ohmic diffusion, a balance is
achieved after a mass $M_{\mathrm{d}}$ is accreted ($M_{\mathrm{d}}$ depends on magnetic field, temperature, accretion rate and EOS), in which the rate of cross-field mass transport
equals the accretion rate \citep{melatos2005}. As our model is not time-dependent, 
the Hall effect is also missing. Hall drift acts to break down the
magnetic field to shorter scales \citep{hollerbach2002, hollerbach2004} and may
operate in isolated neutron stars \citep{rheinhardt2002, rheinhardt2004} but is
thought to be relatively unimportant in accreting neutron stars, where it is
dominated by Ohmic diffusion \citep{cumming2004}. The crystalline lattice of the
crust is thought to melt in thin layers where electron captures have
significantly reduced the nuclear charge \citep{brown2000}. This is expected to
have non-negligible effects on magnetic burial, as the boundary condition on the
magnetic field becomes a function of density rather than radius (line-tying
where solid, free where liquid). Finally, the three-dimensional stability of MHD equilibria
depends on the EOS \citep{kosinski2006}. We leave the investigation of these phenomena to future work.

\section*{Acknowledgements}

The authors are grateful for computing time allocated by the Victorian
Partnership for Advanced Computing (http://www.vpac.org/). MP was
supported by an Australian Postgraduate Award.

\bibliographystyle{mn2e}
\bibliography{bibliography}

\appendix

\section{Numerical Algorithm}
\label{appendix:pseudocode}

The iterative solver described in appendix B of PM04 must be modified to handle
an adiabatic EOS. Equations (\ref{gs_adiabatic}) and (\ref{F_adiabatic}) show
that two quantities, $\psi(r, \theta)$ and $F(\psi)$, must be solved for at each
iteration step.

$F(\psi)$ appears implicitly in both the left- and right-hand sides of equation
(\ref{F_adiabatic}) for $\Gamma \neq 1$, whereas it is derived explicitly from
$\psi$ in one go via equation (\ref{F_isothermal}) for $\Gamma = 1$. Below we
outline briefly the major steps necessary to calculate hydromagnetic equilibria
in the general case.

\begin{enumerate}
\item A two-dimensional mesh is defined, of size $N_{X} \times N_{Y}$ (typically
$256 \times 256$). The \textit{X}-axis scales proportionally to $\log r$ to increase
the resolution close to the surface, where most of the screening currents lie;
we write $\tilde{X} = \log[\tilde{x} \exp(L_{x}) + 1]$, with $\tilde{x} = (r -
R_{\mathrm{in}})/x_{0}$. The \textit{Y}-axis scales as $\cos \theta$, with $\tilde{Y} = \cos
\theta$.

\item The flux function $\psi^{(0)}(\tilde{X}, \tilde{Y})$ is initialized across
the mesh, with $\psi^{(0)}(r, \theta) = \psi_{\ast} R_{\mathrm{in}} \sin^{2} \theta/r$
(i.e. dipole).

\item $N_{Y} - 1$ contours are laid down (the number is chosen to minimize the
occurrence of grid crossings), with contour values spaced linearly in
$\tilde{Y}$ (i.e. linear in $\cos \theta$).

\item Equation (\ref{mass-flux}) is used to compute $\mathrm{d}M/\mathrm{d}\psi$ along each
contour.

\item \label{iteration_point}For $\Gamma = 1$, compute $F(\psi)$ via equation
(\ref{F_isothermal}). For $\Gamma \neq 1$, iteratively solve equation
(\ref{F_adiabatic}) using the previous iterate $F^{(n-1)}(\psi)$ as a starting
point [with $F^{(0)}(\psi)$ uniform]. The value of $F^{(0)}(\psi)$ is chosen to
guarantee that the term in square brackets within the integral in
(\ref{F_adiabatic}), which equals the density, vanishes at the edge of the
computational grid after every iteration, so that the integral in
(\ref{F_adiabatic}) is always well defined. The iteration is halted when the
grid-averaged fractional residual drops below a threshold, usually $\langle
[F^{(n+1)}(\psi) - F^{(n)}(\psi)]/F^{(n)}(\psi)\rangle < 1.0\times10^{-6}$.
Convergence is usually achieved after $\sim 20$ iterations. 

\item The resultant $F^{\ast}(\psi)$ is under-relaxed with the initial input
value $F^{(0)}(\psi)$ via $F(\psi) = \Lambda F^{(0)}(\psi) + (1 -
\Lambda)F^{\ast}(\psi)$. The under-relaxation parameter is usually set to
$\Lambda = 0.9$.

\item Fit $F(\psi)$ with a $\chi^{2}$ polynomial fit of the form
$\sum_{i=0}^{N_{p}} A_{i} \psi^{i}$, where the degree of the polynomial is
typically $N_{p} = 8$. $F^{\prime}(\psi)$ is computed analytically from the
coefficients of the polynomial.

\item Values of $F(\psi)$ and $F^{\prime}(\psi)$ are linearly interpolated
across the mesh and fed into the source terms in equations
(\ref{gs_isothermal}), (\ref{gs_adiabatic}), (\ref{pressure_isothermal}) and
(\ref{pressure_adiabatic}) as required.

\item The Grad--Shafranov equation is solved for the intermediate quantity
$\psi^{\star}(r, \theta)$ by an iterative Poisson solver that utilizes
successive over-relaxation with Chebyshev acceleration \citep{payne2004}. The
Poisson solver halts when the fractional residuals are $\lesssim 10^{-2}$
everywhere across the grid.

\item \label{end_iterations}The solution for $\psi^{(n)}(r, \theta)$ is
subsequently under-relaxed via $\psi^{(n)} = \Theta \psi^{(n-1)} + (1 -
\Theta)\psi^{\star}$. The under-relaxation parameter is typically given by
$\Theta = 0.999$ for $\Gamma = 1$ for $\Gamma = 1$ and $\Theta = 0.9995$ for
$\Gamma \neq 1$.

\item Repeat steps \ref{iteration_point} -- \ref{end_iterations} until
convergence is achieved.
\end{enumerate}

\section{Analytic approximation for the characteristic mass $M_{\mathrm{\lowercase{c}}}$}
\label{appendix:gs_analytic}

An analytic formula can be obtained for the characteristic mass $M_{\mathrm{c}}$ by
calculating $\mu(M_{\mathrm{a}})$ analytically in the small-$M_{\mathrm{a}}$ limit and looking for
the value of $M_{\mathrm{a}}$ where $\mu$ drops to half its unperturbed value.

To calculate $\mu(M_{\mathrm{a}})$ in the small-$M_{\mathrm{a}}$ limit, we follow appendix A3 in
PM04 and proceed in three steps. First, we pick a simple form of $F(\psi)$
which linearizes the Grad--Shafranov equation while approximating the exact
numerical result:
\begin{equation}
F(\psi) = Q_{0}(\psi_{\ast} - \psi).
\label{approximation_1}
\end{equation}
Secondly, we evaluate the right-hand side of the Grad--Shafranov equation
assuming that the flux function is approximately dipolar:
\begin{equation}
\psi(r, \mu) \approx \psi_{\mathrm{D}}(r, \mu) = \frac{\psi_{\ast} R_{\mathrm{in}} (1 -
\mu^{2})}{r}. 
\label{approximation_2}
\end{equation}
This is justified because the magnetic field is weakly distorted in the
small-$M_{\mathrm{a}}$ limit. Thirdly, we solve the Grad--Shafranov equation with the
above source term to obtain the leading-order correction to $\psi$.

We begin by re-expressing the radial coordinate in terms of the fractional
altitude $x$
\begin{equation}
r = R_{\mathrm{in}}(1 + x).
\label{approximation_3}
\end{equation}
In a typical mountain, with height $\lesssim 10^{5} \ \mathrm{cm}$ (see Section
\ref{section_4:hydromagnetic_structure}), one always has $x \ll 1$ within the
mountain. With equations (\ref{approximation_1})--(\ref{approximation_3}), the
Grad--Shafranov equation (\ref{gs_adiabatic}) in the small-$M_{\mathrm{a}}$ approximation
becomes
\begin{equation}
\Delta^{2} \psi = Q_{0} \Bigg[ 1 - \frac{\phi_{0} (\Gamma - 1) x }{\Gamma
K^{1/\Gamma} \{ Q_{0} \psi_{\ast}[1 - (1 - \mu^{2})(1-x)] \}^{(\Gamma -
1)/\Gamma}} \Bigg]^{1/(\Gamma - 1)}.
\label{gs}
\end{equation}
The Lorentz force vanishes when the right-hand side of equation (\ref{gs}) is
zero. Therefore, the maximum height of the magnetic mountain as a function of
latitude can be written as 
\begin{equation}
x_{\mathrm{max}} = \frac{\Gamma K^{1/\Gamma} (Q_{0}\psi_{\ast}\mu^{2})^{(\Gamma
- 1)/\Gamma}}{(\Gamma - 1) \phi_{0}},
\label{max_height_pole}
\end{equation}
for $\mu^{2} \gg x(1 - \mu^{2})$, and 
\begin{equation}
x_{\mathrm{max}} = \frac{K \Gamma^{\Gamma} [Q_{0}\psi_{\ast}]^{\Gamma -
1}}{[(\Gamma - 1) \phi_{0}]^{\Gamma}}
\end{equation}
for $\mu \ll x/(1-x)$ (i.e. near the magnetic equator). (It is easy to check
that one has $x_{m} \ll 1$ a posteriori for typical parameters.) Therefore, for
adiabatic magnetic mountains, the ratio of polar to equatorial heights is
\begin{equation}
[Q_{0} \psi_{\ast}]^{-(\Gamma - 1)^{2}/\Gamma} \Bigg[ \frac{\phi_{0} (\Gamma -
1)}{\Gamma K^{1/\Gamma}} \Bigg]^{\Gamma - 1}.   
\label{height_ratio}
\end{equation}

Equation (\ref{gs}) can be solved by the method of Green's functions. From
Section 3.1 in PM04, we write
\begin{equation}
\psi(r, \mu) = \psi_{\mathrm{D}}(r, \mu) \Bigg[ 1 + \frac{r}{\psi_{\ast} R_{\mathrm{in}}}
\sum_{l=0}^{\infty} N_{l}^{-1} C_{l}^{3/2}(\mu) D_{l}(r) \Bigg],
\label{analytic_psi}
\end{equation}
\begin{equation}
D_{l}(r) = \int_{-1}^{1} \mathrm{d}\mu^{\prime} \int_{R_{\mathrm{in}}}^{\infty} \mathrm{d}r^{\prime}
\ r^{\prime 2} g_{l+1}(r, r^{\prime}) C_{l}^{3/2}(\mu^{\prime}) Q(r^{\prime},
\mu^{\prime}),
\label{integral}
\end{equation}
\begin{equation}
g_{l}(r, r^{\prime}) = \frac{1}{(2l+1) r^{\prime 2}}
\frac{r_{<}^{l+1}}{r_{>}^{l}} \Bigg[ \Bigg( \frac{R_{\mathrm{in}}}{r_{<}} \Bigg)^{2l +
1} - 1 \Bigg],
\label{spherical_greens_functions}
\end{equation}
\begin{equation}
Q(r^{\prime}, \mu^{\prime}) = Q_{0} (1 - \mu^{\prime 2}) r^{\prime 2}
\Bigg\{ 1 - \frac{\phi_{0} (\Gamma - 1) (r^{\prime}/R_{\mathrm{in}} - 1)}{\Gamma
K^{1/\Gamma} [F(\psi^{\prime})]^{(\Gamma - 1)/\Gamma}} \Bigg\}^{1/(\Gamma - 1)},
\label{Q}
\end{equation}
\begin{equation}
N_{l} = \frac{2(l+1)(l+2)}{(2l+3)},
\end{equation}
with $r_{<} = \min(r, r^{\prime})$ and $r_{>} = \max(r, r^{\prime})$. The symbol
$C_{l}^{3/2}(\mu)$ denotes the \textit{l}th Gegenbauer polynomial. The first few are
listed for reference: $C_{0}^{3/2}(\mu) = 1$, $C_{1}^{3/2}(\mu) = 3 \mu$,
$C_{2}^{3/2}(\mu) = (3/2)(5 \mu^{2} - 1)$. Since we are interested in how
the dipole moment is screened at large $r$, we assume $r >
r^{\prime}_{\mathrm{max}}$ always, where $r^{\prime}_{\mathrm{max}}$ is the top
of the mountain. This simplifies the radial Green's function to
\begin{equation}
g_{l}(r, r^{\prime}) = \frac{r^{\prime (l - 1)}}{(2l+1) r^{l}} \Bigg[ \Bigg(
\frac{R_{\mathrm{in}}}{r^{\prime}} \Bigg)^{2l + 1} - 1 \Bigg]. 
\end{equation}
Away from the magnetic equator, i.e. $\mu^{\prime 2} > x^{\prime}(1-\mu^{\prime
2})$, equations (\ref{approximation_1})--(\ref{approximation_3}),
(\ref{max_height_pole}) and (\ref{analytic_psi})--(\ref{Q}) combine to give
\begin{equation}
\psi(r, \mu) = \psi_{\mathrm{D}}(r, \mu) \Bigg[ 1 + \frac{r}{\psi_{\ast} R_{\mathrm{in}}}
\sum_{l=0}^{\infty} N_{2l}^{-1} C_{2l}^{3/2}(\mu) D_{2l}(r) \Bigg], \\
\label{analytic_psi_even}
\end{equation}
\begin{align}
\label{integral_even}
D_{2l}(r) = & \frac{-2 Q_{0} R_{\mathrm{in}}^{2l+5}}{r^{2l+1}} \\
& \times \int_{x^{\prime 1/2}}^{1} \mathrm{d}\mu^{\prime} \int_{0}^{\mu^{\prime 2(\Gamma - 1)/\Gamma}/\Lambda_{0}} \mathrm{d}x^{\prime} \nonumber \\
& \quad \ C_{2l}^{3/2}(\mu^{\prime})(1 - \mu^{\prime 2}) \Bigg[ 1 - \frac{\Lambda_{0} x^{\prime}}{\mu^{\prime 2(\Gamma - 1)/\Gamma}} \Bigg]^{1/(\Gamma - 1)} \nonumber,
\end{align}
with
\begin{equation}
\Lambda_{0} = \frac{\phi_{0} (\Gamma - 1)}{\Gamma K^{1/\Gamma} (Q_{0}
\psi_{\ast})^{(\Gamma - 1)/\Gamma}}.
\label{lambda}
\end{equation}

\begin{figure}
\includegraphics[width=84mm]{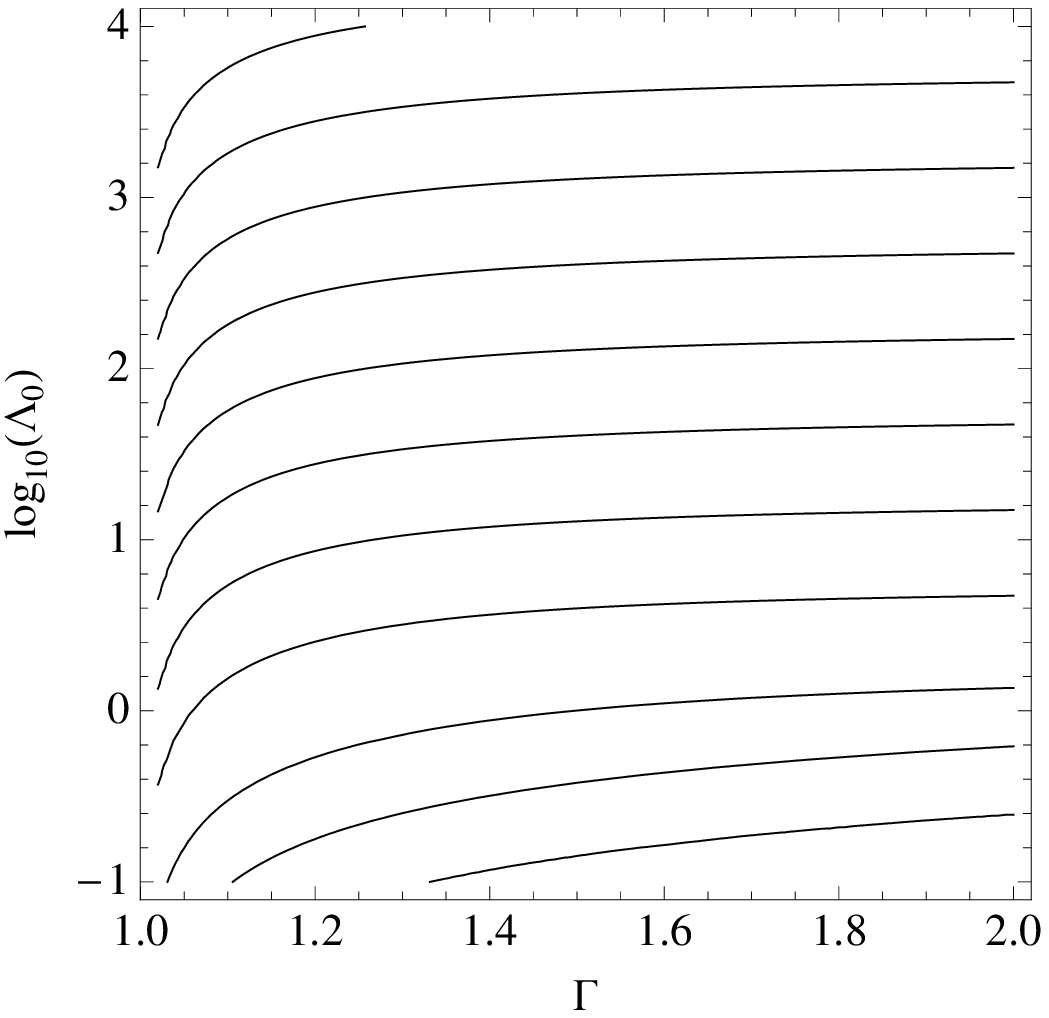}
\caption{Contour plot of the integral $I(\Lambda_{0}, \Gamma)$ defined by
equation (\ref{integral_plotted}), with $\Lambda_{0}$ defined by equation
(\ref{lambda}). $\Gamma$ and $\Lambda_{0}$ span the typical range expected in
magnetic mountains. The contours correspond to (from top to bottom)
$\log_{10}I(\Lambda_{0}, \Gamma) = -10, \ -9, \ -8, \ -7, \ -6, \ -5, \ -4, \
-3, \ -2, \ -1.7, \ -1.6$.}
\label{fig:lambda_gamma}
\end{figure}

Our goal is to calculate the dipole moment as a function of $M_{\mathrm{a}}$ given
(\ref{analytic_psi_even}) and (\ref{integral_even}). In the limit $r \to
\infty$, the $l \geq 1$ contributions to $\mu$ vanish, and equations
(\ref{analytic_psi_even}) and (\ref{integral_even}) reduce to
\begin{equation}
\psi(r, \mu) = \psi_{\mathrm{D}}(r, \mu) \Bigg[ 1 - \frac{3 Q_{0} R_{\mathrm{in}}^{4}}{2
\psi_{\ast}} I(\Lambda_{0}, \Gamma) \Bigg], 
\label{analytic_zeroth_order}
\end{equation}
with
\begin{align}
\label{integral_plotted}
I(\Lambda_{0}, \Gamma) = & \int_{x^{\prime 1/2}}^{1} \mathrm{d}\mu^{\prime} \ \int_{0}^{\mu^{\prime 2(\Gamma - 1)/\Gamma}/\Lambda_{0}} \mathrm{d}x^{\prime} \\
 & \ x^{\prime} (1 - \mu^{\prime 2}) \Bigg[ 1 - \frac{x^{\prime} \Lambda_{0}}{\mu^{\prime 2(\Gamma - 1)/\Gamma}} \Bigg]^{1/(\Gamma - 1)}. \nonumber 
\end{align}
Contours of $I(\Lambda_{0}, \Gamma)$ are plotted in Fig. \ref{fig:lambda_gamma}
for reference.

To express $Q_{0}$ in terms of the other variables, we substitute equations
(\ref{mass-flux}), (\ref{approximation_1}), (\ref{approximation_2}) and
(\ref{lambda}) into equation (\ref{F_adiabatic}) to give
\begin{align}
\label{q1}
[Q_{0} (\psi_{\ast} - \psi)]^{1/\Gamma} = & \frac{K^{1/\Gamma}}{2 \pi} \frac{M_{\mathrm{a}} \psi_{\ast} R_{\mathrm{in}}}{\psi_{\mathrm{a}}} \frac{\exp(-\psi/\psi_{\mathrm{a}})}{1-\exp(-b)} \\
 & \times \Bigg \{ \int_{C} \mathrm{d}s \ r^{3} \Bigg( 4 - \frac{3 \psi r}{\psi_{\ast} R_{\mathrm{in}}} \Bigg)^{-1/2} \nonumber \\
 & \quad \Bigg[ 1 - \frac{\Lambda_{0} (r/R_{\mathrm{in}} - 1)}{(1 - \psi/\psi_{\ast})^{(\Gamma - 1)/\Gamma}} \Bigg]^{1/(\Gamma - 1)} \Bigg\}^{-1}, \nonumber 
\end{align}
where $b=\psi_{\ast}/\psi_{\mathrm{a}}$ is a constant that parametrizes the lateral
extent of the accretion column. Equation (\ref{q1}) is not strictly an equality;
the linear ansatz $F(\psi) = Q_{0}(\psi_{\ast} - \psi)$ is not an exact solution
in the small-$M_{\mathrm{a}}$ limit (see fig. 6 in PM04, which presents a numerical
comparison). Hence, to evaluate $Q_{0}$ approximately, it is enough to integrate
equation (\ref{q1}) through the centre of the mountain, where most of the
mountain mass resides (i.e. along the polar flux line $\psi = 0$). This has the
added advantage that the resultant contour integral has no $\theta$ dependence
($\mathrm{d}s = \mathrm{d}r$, $\psi = 0$). Changing variables according to equation
(\ref{approximation_3}), substituting equation (\ref{max_height_pole}) for the
upper integration limit in $x^{\prime}$, and taking $(1 + x) \approx 1$ inside
the integral, we arrive at
\begin{align}
[Q_{0} \psi_{\ast}]^{1/\Gamma} = & \frac{K^{1/\Gamma}}{2 \pi} \frac{M_{\mathrm{a}}
\psi_{\ast}}{\psi_{\mathrm{a}} R_{\mathrm{in}}^{3} [1-\exp(-b)]} \\
& \times \Bigg[ \int_{0}^{1/\Lambda_{0}}
\mathrm{d}x \ (1 - \Lambda_{0} x)^{1/(\Gamma - 1)} \Bigg]^{-1} \nonumber.
\end{align}
The expression for $Q_{0}$ is therefore
\begin{equation}
Q_{0} = \frac{M_{\mathrm{a}} M_{\ast} G b}{2 \pi \psi_{\ast} R_{\mathrm{in}}^{4} [1 -
\exp(-b)]},
\label{Q0}
\end{equation}
and hence equation (\ref{lambda}) becomes
\begin{equation}
\Lambda_{0} = \frac{ (\Gamma - 1) G M_{\ast}}{\Gamma K^{1/\Gamma} R_{\mathrm{in}}}
\Bigg\{ \frac{b \exp(b) G M_{\mathrm{a}} M_{\ast}}{2 \pi [\exp(b)-1] R_{\mathrm{in}}^{4}}
\Bigg\}^{(1-\Gamma)/\Gamma}. 
\label{lambda_general}
\end{equation}
For fiducial neutron star parameters (e.g. Section \ref{section_3}) and the
adiabatic equations of state B--D in Table \ref{table:eos}, equation
(\ref{lambda_general}) reduces to 
\begin{align}
\Lambda_{0, \mathrm{B}} & = 1.8 \times 10^{-2} \Bigg( \frac{M_{\mathrm{a}}}{\mathrm{M}_{\sun}}
\Bigg)^{-2/5}, \\
\Lambda_{0, \mathrm{C}} & = 6.8 \times 10^{-1} \Bigg( \frac{M_{\mathrm{a}}}{\mathrm{M}_{\sun}}
\Bigg)^{-1/4}, \\
\Lambda_{0, \mathrm{D}} & = 8.4 \times 10^{-1} \Bigg( \frac{M_{\mathrm{a}}}{\mathrm{M}_{\sun}}
\Bigg)^{-2/5}. 
\end{align}

Upon substituting equations (\ref{analytic_zeroth_order}) and (\ref{Q0}) into
equation (\ref{multipole_moments}) (with $l = 1$, and $r$ instead of
$R_{\mathrm{m}}$) and comparing with the phenomenological burial law $\mu =
\mu_{i}(1 - M_{\mathrm{a}}/M_{\mathrm{c}})$ postulated by \citet{shibazaki1989} in the small-$M_{\mathrm{a}}$ limit, we obtain
\begin{align}
\label{characteristic_mass_adiabatic}
M_{\mathrm{c}} & = \frac{4 \pi [1 - \exp(-b)] \psi_{\ast}^{2}}{3 M_{\ast} G b I(\Lambda_{0}, \Gamma)} \\
& \approx \frac{2.8 \times 10^{-9} [1 - \exp(-b)]}{b} \Bigg[ \frac{B}{10^{12.5} \ \mathrm{G}} \Bigg]^{2} \nonumber \\
& \quad \; \times \Bigg[ \frac{I(\Lambda_{0}, \Gamma)}{10^{-9}} \Bigg]^{-1} \mathrm{M}_{\sun}. \nonumber
\end{align}

\begin{figure}
\includegraphics[width=84mm]{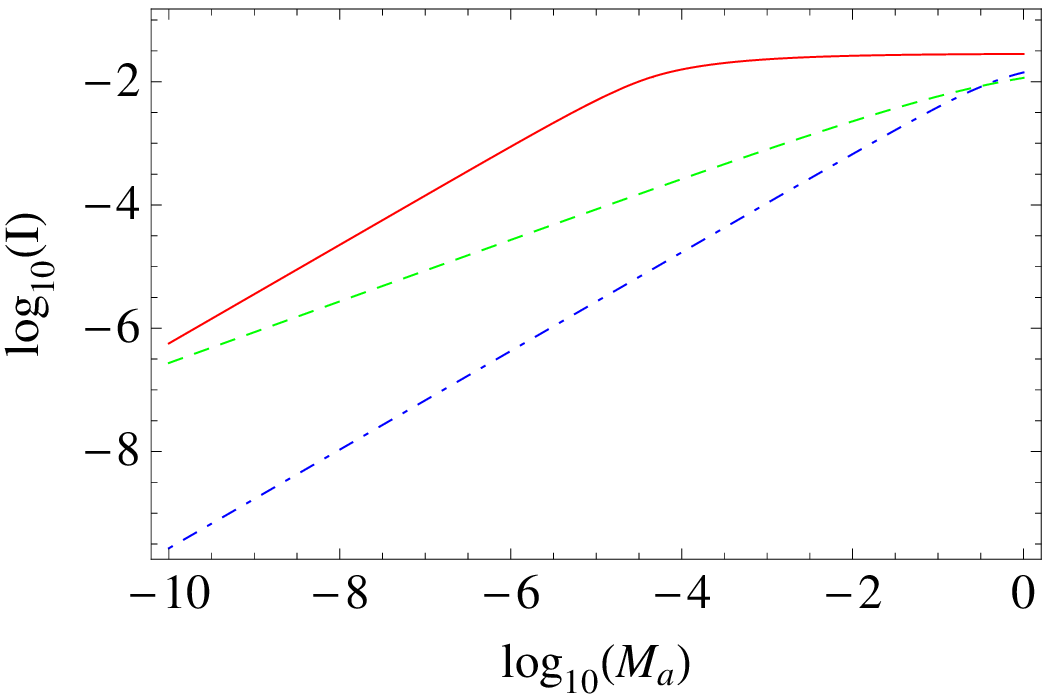}
\caption{Plot of equation (B18) for models B (solid red curve), C
(dashed green curve) and D (dot--dashed blue curve), as a function of accreted mass $M_{\mathrm{a}}$
(measured in solar masses). In the regime where the maximum height of the
magnetic mountain is $x<1$ (i.e. $\Lambda_{0} > 1$), $I(M_{\mathrm{a}})$ is polynomial.
$I(M_{\mathrm{a}})$ saturates at $\approx 10^{-1.5}$ in the case where $\Lambda_{0} < 1$,
which is an artefact of the approximations used in the small-$M_{\mathrm{a}}$ limit.}
\label{fig:I_Ma}
\end{figure}

The integral in equation (\ref{integral_plotted}) is computed for models B, C and D with fiducial neutron star parameters and plotted as a function of $M_{\mathrm{a}}$
in Fig. \ref{fig:I_Ma}. For the case $\Lambda_{0} > 1$, equation
(\ref{integral_plotted}) reduces to the following expressions
\begin{align}
I_{\mathrm{B}}(M) & = A_{\mathrm{B}} M^{7/3} + B_{\mathrm{B}} M^{5/3} +
C_{\mathrm{B}} M^{4/5},\\
I_{\mathrm{C}}(M) & = A_{\mathrm{C}} M^{7/6} + B_{\mathrm{C}} M^{5/6} +
C_{\mathrm{C}} M^{1/2},\\ 
I_{\mathrm{D}}(M) & = A_{\mathrm{D}} M^{7/3} + B_{\mathrm{D}} M^{5/3} +
C_{\mathrm{D}} M^{4/5},
\label{explicit_integrand}   
\end{align}
with $M = M_{\mathrm{a}}/\mathrm{M}_{\sun}$, $A_{\mathrm{B}}=3.5\times10^{7},
B_{\mathrm{B}}=-1.7\times10^{5}, C_{\mathrm{B}}=57,
A_{\mathrm{C}}=6.9\times10^{-3}, B_{\mathrm{C}}=2.2\times10^{-2},
C_{\mathrm{C}}=2.7\times10^{-2}, A_{\mathrm{D}}=7.2\times10^{-3},
B_{\mathrm{D}}=2.0\times10^{-2}$ and $C_{\mathrm{D}}=2.7\times10^{-2}$.

In contrast to equation (\ref{characteristic_mass_adiabatic}), the scaling of
$M_{\mathrm{c}}$ for isothermal magnetic mountains [from equations (29) and (30) in PM04]
is 
\begin{align}
M_{\mathrm{c}} & = \frac{GM_{\ast}\psi_{\ast}^{2}}{4c_{\mathrm{s}}^{4}b^{2}R_{\mathrm{in}}^{2}} \\
& \approx \frac{5.8\times10^{-4}}{b^{2}} \Bigg[ \frac{B}{10^{12.5} \
\mathrm{G}} \Bigg]^{2} \mathrm{M}_{\sun} \nonumber.
\end{align}

\end{document}